\documentclass[12pt,oneside,onecolumn]{article}
\usepackage{epsfig}
\usepackage{amsmath,amssymb,amsfonts}
\usepackage{a4wide}
\usepackage[top=30pt,bottom=30pt,left=48pt,right=46pt]{geometry}
\usepackage{slashed, enumitem}
\usepackage{jheppub}
\usepackage{enumerate}
\usepackage{float}
\usepackage{subfigure}
\usepackage{multirow}
\usepackage{slashed}
\usepackage{wasysym} 
\usepackage{mathrsfs} 
\usepackage{caption}
\usepackage{amsmath}
\usepackage{amssymb}
\usepackage{slashed}
\newcommand{\met}{$\cancel E_T$}

\newcommand{\newc}{\newcommand}
\newc{\wt}{\widetilde}
\newc{\ra}{\rightarrow}

\def\beq {\begin{equation}}
\def\eeq {\end{equation}}
\def\bi {\begin{itemize}}
\def\ei {\end{itemize}}
\def\bea {\begin{eqnarray}}
\def\eea {\end{eqnarray}}

\def \met{\rm E{\!\!\!/}_T}

\newcommand{\br}{\begin{eqnarray}}
\newcommand{\er}{\end{eqnarray}}
\newcommand{\be}{\begin{equation}}
\newcommand{\ee}{\end{equation}}

\def\lum{{\cal L}}
\newcommand{\ifb} {{fb}^{-1}}

\usepackage{array}
\newcolumntype{L}[1]{>{\raggedright\let\newline\\\arraybackslash\hspace{0pt}}m{#1}}
\newcolumntype{C}[1]{>{\centering\let\newline\\\arraybackslash\hspace{0pt}}m{#1}}
\newcolumntype{R}[1]{>{\raggedleft\let\newline\\\arraybackslash\hspace{0pt}}m{#1}}

\def \beq{\begin{equation}}
\def \eeq{\end{equation}}
\def \bea{\begin{eqnarray}}
\def \eea{\end{eqnarray}}
\def \ba{\begin{array}}
\def \ea{\end{array}}
\title{{\Large Scope of strongly self-interacting thermal WIMPs in a minimal $U(1)_D$ extension and its future prospects}}

\author[a]{Rahool Kumar Barman}
\author[a]{Biplob Bhattacherjee}
\author[b]{Arindam Chatterjee}
\author[c]{Arghya Choudhury}
\author[d]{Aritra Gupta}
\affiliation[a]{Centre for High Energy Physics, Indian Institute of Science, Bangalore 560012, India}
\affiliation[b]{Indian Statistical Institute, 203 B.T. Road, Kolkata, India.}
\affiliation[c]{Department of Physics, Indian Institute of Technology Patna, Bihta, 801103, India} 
\affiliation[d]{Tata Institute of Fundamental Research, Homi Bhabha Road, Mumbai, 400005, India.}
\emailAdd{biplob@iisc.ac.in}
\emailAdd{rahoolbarman@iisc.ac.in}
\emailAdd{arindam.chatterjee@gmail.com}
\emailAdd{arghya@iitp.ac.in}
\emailAdd{aritragupta@theory.tifr.res.in}

\abstract
{In this work we have considered a minimal extension of Standard Model by a local $U(1)$ gauge group in order to accommodate a stable (fermionic) Dark Matter (DM) candidate. We have focussed on parameter regions where DM possesses adequate self interaction, owing to the presence of a light scalar mediator (the dark Higgs), alleviating some of the tensions in the small-scale structures. 		
We have studied the scenario in the light of a variety of data, mostly from dark matter direct searches, collider searches and flavour physics experiments, with an attempt to constrain the interactions of the standard model (SM) particles with the ones in the Dark Sector (DS). Assuming a small gauge kinetic mixing parameter, we find that for rather heavy DM 
the most stringent bound on the mixing angle of the Dark Higgs 
with the SM Higgs boson comes from dark matter direct detection experiments, while for lighter DM, LHC constraints become more relevant. Note that, due to the presence of very light mediators the usual realisation of direct detection constraints in terms of momentum independent cross sections had to be reevaluated for our scenario. 
In addition, we find that the smallness of the relevant portal couplings, as dictated by data, 
critically suppress the viability of DM production by the standard ``freeze-out" mechanism in such simplified scenarios. In particular, the viable DM masses are $\lesssim \mathcal{O}(2)$ GeV $i.e.$ in the regions where direct detection limits tend to become weak. For heavier DM with large self-interactions, we hence conclude that non-thermal production mechanisms are favoured. Lastly, future collider reach of such a simplified scenario has also been studied in detail.}

\preprint{TIFR/TH/18-45}

\keywords{Dark Matter, Self-interaction, Collider Physics.}

\begin{document}
\maketitle

\section{Introduction} 
\label{sec:intro}
The evidence for existence of non-relativistic non-luminous Dark Matter (DM) has been 
overwhelming. Several astrophysical and cosmological observations have indicated the 
presence of DM \cite{Bertonebook}. Within the paradigm of the standard model of cosmology 
($\Lambda$CDM) recent measurements of the Cosmic Microwave Background Radiation (CMBR) 
estimated that about 26\% of the energy budget of our Universe consists of DM 
\cite{Aghanim:2018eyx, Ade:2015xua}\footnote{Very similar estimates have been obtained for simple extensions of $\Lambda$CDM and for $\omega$CDM models.}. CMBR, together with Big Bang Nucleosynthesis (BBN) requires DM to be non-baryonic. Astro-physical objects, especially primordial black holes (PBHs) have been considered as DM candidates. While such PBHs, for certain windows of mass\footnote{According to \cite{Carr:2016drx}, the allowed regions span the following range $(10^{16}-10^{17}, 10^{20}-10^{24}, 10^{33}-10^{36})$ gm.}
can account for the entire energy budget of DM, it has been shown that adequate production 
of PBHs after (single-field slow-roll) inflation can be difficult \cite{Drees:2011yz,Drees:2011hb}. 
However, the standard model (SM) of particle physics, does not incorporate any suitable DM 
candidate. Various extensions of the SM has been considered in the literature \cite{Jungman:1995df,
Bertone:2004pz}. A generic aspect in DM models concerns about the stability of DM. Constraints 
from structure formation, indirect searches \cite{Acciari:2018sjn,Cohen:2016uyg,Cirelli:2012ut} 
and CMB \cite{Slatyer:2016qyl} require DM to be very long-lived (life-time $\tau\gtrsim 10^{26}$ 
s depending on the decay modes). The simplest models introduce a global symmetry to prevent the 
decay of DM. However, it has been argued that such global symmetries may be broken due to 
gravitational effects \cite{Kallosh:1995hi,Banks:2006mm,}. This, in turn, can induce Planck-scale 
suppressed effective terms contributing to the decay of the DM \cite{MAMBRINI2016807}.
In this article, therefore, we extend the SM with a simple continuous local symmetry $U(1)$. 
We further assume that the DM is charged under this symmetry, and therefore, is stable. 
Earlier works that also explored simplest extensions of Standard Model with similar phenomenological signatures can be found in \cite{Chu:2014lja,Ko:2016ala,Garcia-Cely:2013wda,Garcia-Cely:2013nin,Okada:2010wd,PhysRevD.44.2118,PhysRevD.43.R22,Matsumoto:2018acr}
 
Recent results from the Large Hadron Collider (LHC), so far, present no convincing 
evidence for new physics, see e.g. \cite{Sirunyan:2017hnk,Aaboud:2016obm,Sirunyan:2018wcm}
for implications on DM models. Further, no evidence for particle DM has emerged from direct \cite{Aprile:2018dbl,Akerib:2016vxi,Cui:2017nnn} (and indirect \cite{Fermi-LAT:2016uux,Aguilar:2016kjl}) 
searches of DM. The Standard Model (SM) of particle physics remains a good low energy 
effective description. Stringent constraints on most well-motivated new physics scenarios 
present two distinct possibilities: any new physics may be beyond the energy reach of the 
LHC, and perhaps can only be probed indirectly through their contribution to the higher 
dimensional effective operators; or new physics, if exists at (or below) the electroweak 
scale, may be very weakly coupled to the SM. In the subsequent discussion, we will assume 
the latter and consider the Dark Sector (DS) particles to be accessible at LHC. 

There is an additional motivation for this consideration. Note that in spite of the enormous 
success of $\Lambda$CDM in the cosmological scales, there have been concerns, in particular 
when it comes to the small scale structures (see e.g. \cite{Weinberg:2013aya,Bull:2015stt, 
DelPopolo:2016emo} for recent reviews). The most notable ones include {\it Core-vs-cusp} 
\cite{Moore:1994yx,Flores:1994gz} and {\it Too-big-to-fail} \cite{2011MNRAS.415L..40B,2012MNRAS4252817F, 2014MNRAS444222G, Papastergis:2014aba}\footnote{{\it Missing satellite} problem \cite{Klypin:1999uc,Moore:1999nt} has also been extensively 
discussed. However, the discovery of faint dwarfs has lead to dissolution of this issue
 \cite{Tollerud:2008ze,Walsh:2008qn}.}.
It has been argued that self-interacting DM can also simultaneously resolve these 
issues \cite{Spergel:1999mh,Dave:2000ar}, see also \cite{Tulin:2017ara,Tulin:2013teo}
for a review.  Although this requires a very large 
self-interaction cross-section, it remains consistent with the present constraints 
on DM self-interaction \cite{Rocha:2012jg,Peter:2012jh,2013MNRAS.431L..20Z,Elbert:2014bma}
 \footnote{See also \cite{Tulin:2017ara} 
and references there for a compilation of constraints from bullet-cluster \cite{Markevitch:2003at,Clowe:2003tk,Randall:2007ph}, ellipticity of the halos 
and substructure mergers.} 
\footnote{An observed separation between the stars of a galaxy and the DM halo, while the 
galaxy falls into the core of a galaxy cluster Abel 3827, has been recently observed 
\cite{Massey:2015dkw}. When interpreted as due to DM self-interaction, this leads to 
$\dfrac{\sigma}{m}\simeq 1.5 (3)~{\rm cm^2gm^{-1}}$ for contact (long-range) interactions 
\cite{Kahlhoefer:2015vua}. However, the data has been argued to be also consistent with 
standard collisionless DM \cite{Massey:2017cwf}.}
\footnote{While baryonic feedback can possibly address some of these issues, it has been 
argued to face difficulties in the context of field galaxies. For a review see \cite{Tulin:2017ara}.}. Further, it has been argued that a velocity dependent self-interaction cross-section is favoured \cite{Kaplinghat:2015aga}. 
In the presence of light mediators, it is possible to generate large (velocity dependent) 
self-interaction among DM particles, thanks to the Sommerfeld enhanced self-scattering cross-section \cite{Hisano:2004ds,Cirelli:2007xd,ArkaniHamed:2008qn,Feng:2009hw,Tulin:2013teo}. 
In the present work, we will ensure that the $U(1)$ extended DS consists of at least one light mediator which facilitate adequate self-interaction among the DM particles.
 Imposing these criteria, 
we will investigate the implications on this simple DS. Further, combining constraints from 
collider searches, flavour physics, beam dump experiments and direct detection of DM, we will try to put restrictions on its interaction with the SM sector. 

This article is organised as follows : In Sec.\,\ref{sec:model} we describe the minimal $U(1)_D$ model and its particle content in detail. In Sec.\,\ref{sec:dm_self} the parameter space relevant for the study of large self interaction of dark matter is motivated quantitatively. Next, concentrating on the light $H_1$ region as motivated from self-interactions, impact of several experiments (like beam dump, flavour physics and collider) on our parameter space of interest is described studied in Sec.\,\ref{sec:par_space} and Sec.\,\ref{sec:all_constraints}. Sec.\,\ref{sec:DM_aspect} deals with bounds from dark matter phenomenology and discusses its implications. In sec.\ref{subsec:future} we have presented a detailed study about the future prospects of our scenario from a collider perspective. Finally we summarize and conclude in Sec.\,\ref{sec:conclusion}.

\section{Description of the Model}
\label{sec:model}
As described in the introduction, we consider a simple extension of SM 
where the stability of DM has been attributed to a local symmetry. However, 
any unbroken local symmetry would imply the existence of a massless gauge 
boson. Cosmological observations together with the success of BBN constrain 
the presence of such a boson through the tight limits of extra relativistic 
{\it dof}. This demands the $U(1)_D$ to be broken, and we employ Higgs 
mechanism to achieve this\footnote{An Abelian gauge boson can also get 
massive via Stuckelberg mechanism \cite{Stueckelberg, Stueckelberg1, Stueckelberg2,
Ogievetskii}, which essentially assumes the neutral scalar component to be very 
heavy and therefore decoupled from the spectrum. However, we will not consider 
that possibility here.}.

We have assumed minimal particle content for a Dark Sector with a local 
$U(1)_D$ gauge symmetry. Two Weyl fermions needed to be introduced so that 
the gauge anomaly associated with $U(1)_D$ group cancels, and the theory 
remains anomaly free. We will further elaborate on the motivation behind this 
particular choice. The Dark Sector is assumed to have the following particle 
content:  
\begin{table}[h]
\begin{center}
  \begin{tabular}{|c||c|c|c|c|c|c|}
   \hline
 Particles (spin)  &  $\xi_1$ ($\frac{1}{2}$) & $\xi_2$ ($\frac{1}{2}$) 
 & $\varphi$ (0) & $Z_D$ (1)\\
   \hline
 $U(1)_D$ charge ($q_D$)  &  1 &  -1 &  2 & 0 \\
   \hline
  \end{tabular}
  \caption{Particle content (in the gauge eigenbasis) of the Dark Sector (DS) and their 
charges ($q_D$) under $U(1)_D$ gauge group.}
 \label{charges}
\end{center}

\end{table} 
In Table~\ref{charges}, $\xi_1$ and $\xi_2$ represents left-chiral Weyl spinors, while 
$Z_D$ and $\varphi$ represents a vector boson and a complex scalar field respectively. 
We consider the following lagrangian, which is invariant under $\mathcal{G}_{\rm SM} \times U(1)_D$ (where, $\mathcal{G}_{\rm SM}$ denotes the SM gauge group). 
\beq
\mathcal{L}= \mathcal{L}_{\rm SM}+ \mathcal{L}_{\rm DS} + \mathcal{L}_{\rm portal}; 
\eeq
where ``DS" and ``portal" denote Dark Sector and mediator respectively. 
The two component Weyl spinors $\xi_1$ and $\xi_2$ can be expressed as a four 
component  fermion $\chi = (\xi_1,\tilde{\xi}_2)^{T}$, where 
$\tilde{\xi}_j = -i \sigma_2 \xi_j^*$. 
With this notation, we have, 
\begin{equation}
{\cal L_\text{DS}} = i \overline{\chi}\gamma^\mu D_\mu \chi+ 
D_{\mu}\varphi^{*} D^{\mu}\varphi - V(\varphi)- M\overline{\chi}\chi -  
\left(\dfrac{f}{\sqrt{2}}\varphi \overline{\chi}  \chi^c + h.c.\right)\,.
\label{DiracL}
\end{equation}
where $\gamma_{\mu}$ in the Weyl representation has been assumed. Further, in the 
lagrangian above $D^{\mu} = \partial^{\mu} - i g_D q_D Z_D^{\mu}$  and 
$\chi^c = i \gamma_2 \gamma_0 \bar\chi^T = (\xi_2,\tilde{\xi}_1 )^T $.
Note that a $U(1)_D$ charge of $1 (-1)$ for $\chi ~(\tilde{\chi})$ together with a 
charge of $\mp 2$ for $\varphi$ ensures the invariance of the Yukawa term under 
the guage group $U(1)_D$. 

The Dark Sector interacts with the SM sector via the following gauge invariant terms.
\begin{eqnarray}
\mathcal{L}_{\rm portal} & = -\dfrac{\epsilon_g}{4}F^{\mu \nu}F_{D \mu \nu}-
\dfrac{\lambda_{\rm mix}}{4} (\varphi^* \varphi)(H^{\dagger} H),
\label{lport}
\end{eqnarray}
Although none of the particles in the low energy spectrum are charged under both $U(1)_Y$ 
and $U(1)_D$ gauge symmetry, we still keep the gauge invariant phenomenologically viable 
kinetic mixing term \cite{Holdom:1985ag} in the lagrangian, which can possibly be generated 
due to high-scale physics. We parametrize the mixing term with parameter $\epsilon_g$. 
The scalar field $\varphi$ can couple with the SM Higgs via the usual portal interaction term. 
We will elaborate on the constraints on both mixing terms in a subsequent discussion.    

\subsection{The Higgs Sector} 
The scalar potential is given by, 
\begin{equation}
V(\varphi, H) = \mu_{H}^{2} H^{\dagger} H + \mu_{\phi}^{2} \varphi^{\dagger} \varphi+ \frac{\lambda_{H}}{4} \left(H^{\dagger}\,H\right)^{2} +\frac{\lambda_{\varphi}}{4} \left(\varphi^{\dagger} \varphi\right)^{2}+ \frac{\lambda_{\rm mix}}{4}\left(H^{\dagger} H\right)\left(\varphi^{\dagger} \varphi\right)
\end{equation}
The stability conditions can be read off as follows : 
\begin{eqnarray}
\lambda_H > 0, \, \lambda_{\varphi} > 0, \, \lambda_{\rm mix}^2 < 4 \lambda_H \lambda_{\varphi}
\end{eqnarray}
 
Since $\mathcal{G}_{SM} \times U(1)_D$ is broken spontaneously, we require $\mu_H^2 < 0$ and 
$\mu_{\varphi}^2 < 0$. Let $v_h$ and $v_{\varphi}$ denote the $vev$s of the scalar 
fields responsible for the spontaneous breaking of  $\mathcal{G}_{SM} \times U(1)_D$. 

\footnote{Note that while $v_{\varphi}$, in general, may be complex; its phase is not 
physical and can be absorbed by suitable field redefinitions both in the Higgs sector and in the Dark Sector, as may be evident from the lagrangian. Therefore, $v_{\varphi}$ has been assumed 
to be real. } 
Around the minimum of the scalar potential, these scalar fields can be described as,  
\begin{equation}
\varphi = \dfrac{1}{\sqrt{2}} (v_{\varphi}+\varphi_R + i \varphi_I) \,
\label{pform}
\end{equation}
and 
\begin{equation}
 H= 
\begin{pmatrix}
 \dfrac{1}{\sqrt{2}} \sigma_+ \\ 
 \dfrac{1}{\sqrt{2}} (v_h+ h +  i \sigma)  ,\,
\end{pmatrix} 
\label{hform}
\end{equation}
Here, $\phi_R$ ($h$) and $\phi_I$
($\sigma$) are the CP-even (CP-odd) parts of $\phi$ and $H$ respectively and $\sigma_+$ is a complex scalar. In the unitary gauge, where the low energy $d.o.f$ constitutes only the physical 
fields, the form of the corresponding expressions may simply be obtained by 
setting  $\varphi_I,~\sigma_+ $ and $\sigma$ to 0 in equations 
(\ref{pform}) and (\ref{hform}).   

The squared mass matrix for the scalar fields, in the basis $\{\varphi_R, h\}$, 
is given by, 
\begin{equation}
\mathcal{M}_{\rm s}^{2} =  
\begin{pmatrix}
 v_{\varphi}^2 \lambda_{\varphi}  & v_h v_{\varphi} \dfrac{\lambda_{\rm mix}}{2} \\ 
 v_h v_{\varphi} \dfrac{\lambda_{\rm mix}}{2} & v_h^2  \lambda_H  ,\,
\end{pmatrix} 
\end{equation}
The orthonormal mixing matrix $\mathcal{N}_s$, is then given by, 
\begin{equation}
\mathcal{N}_s =   
\begin{pmatrix}
 \cos \theta & \sin \theta \\ 
 -\sin \theta & \cos \theta   \,
\end{pmatrix}, 
\end{equation}
where 
\begin{equation} 
\theta_{\rm mix} =  \dfrac{1}{2} \sin^{-1}\dfrac{v_{\varphi} v_h \lambda_{\rm mix}}{(v_h^4 \lambda_H^2 + v_{\varphi}^4 \lambda_{\varphi}^2 + 
    v_h^2 v_{\varphi}^2 (\lambda_{\rm mix}^2 - 2 \lambda_H \lambda_{\varphi}))^{\frac{1}{2}}}.
\label{thetaH}    
\end{equation} 
The mass eigenvalues can simply be obtained as the elements of the diagonal matrix 
$(\mathcal{M}_s^{D})^2 =  \mathcal{N}_s^T \mathcal{M}_s^2 \mathcal{N}_s $ and the 
corresponding eigenvalues give the physical masses of the CP-even neutral scalar 
particles, 
\begin{equation}
M^2_{H_1,H_2}= \dfrac{1}{4} \{v_h^2 \lambda_H + v_{\varphi}^2 \lambda_{\varphi} \mp 
(v_h^4 \lambda_H^2 + v{\varphi}^4 \lambda_{\varphi}^2 + 
    v_h^2 v_{\varphi}^2 (\lambda_{\rm mix}^2 - 2 \lambda_H \lambda_{\varphi}))^{\frac{1}{2}}\}
\label{hmass}
\end{equation}
The mass eigenstates are given by,  
\begin{equation}
\{H_1, H_2 \}^T   =   \mathcal{N}_{s}^{T} \{\varphi_R, h\}^T, \, 
\end{equation} 
or more explicitly, 
\begin{eqnarray}
H_1 & = & \cos\theta_{\rm mix} ~\varphi_R - \sin\theta_{\rm mix} ~h,\\
H_2 & = & \sin\theta_{\rm mix} ~\varphi_R + \cos\theta_{\rm mix}~ h.
\end{eqnarray}
We will denote the mass eigenstate corresponding to the lightest eigenvalue $M_{H_1}$  
as $H_1$ and that to the heavier one ($M_{H_2}$) as $H_2$. In this convention $\cos \theta_{\rm mix}$ simply 
denotes the $\varphi_R$ content in the lightest mass eigenstate. As we will consider 
a scenario with a light $H_1$, $H_2$ would be the SM-like Higgs boson. 
therefore, $\cos \theta_{\rm mix}$ denotes the SM Higgs content in $H_2$ as well as $\varphi_R$ content in $H_1$. 


\subsection{The Gauge Boson Sector}
The gauge group in the present context is : $ \mathcal{G}_{\rm SM} \times U(1)_D$, 
where $ \mathcal{G}_{\rm SM}$ denotes the SM gauge group. In this section we focus on 
the electro-weak gauge bosons and implication of the gauge kinetic mixing with the 
$U(1)_D$ gauge boson. The relevant lagrangian for the neutral gauge boson sector 
is given by,

\begin{equation}
\cal{L}_{\rm gauge} = \cal{L}_{\rm kin} + \cal{L}_{\rm mass}+\cal{L}_{\rm int}
\end{equation}
where, 
\begin{eqnarray}
 \cal{L}_{\rm kin} &=&  -\frac{1}{4} \hat{W}_{\mu\nu}^3
\hat{W}_3^{\mu\nu} - \frac{1}{4} \hat{B}_{\mu\nu}
 \hat{B}^{\mu\nu} - \frac{1}{4} \hat{Z}_{D \mu\nu}
 \hat{Z}_{D}^{\mu\nu}  + \frac{\epsilon_g}{2} \hat{B}_{\mu\nu}
 \hat{Z}_{D}^{\mu \nu } \nonumber \\
 &=&  - \frac{1}{4} \hat{V}_{a \mu \nu}^T \mathcal{K}^{a}_{\rm V} \hat{V}_{a}^{\mu \nu} \\ \,
  \mathcal{L}_{\rm mass} &=&  -\frac{1}{2} \left( m_w \hat{W}_\mu^3
 -m_B \hat{B}_\mu \right) \left( m_w \hat{W}_3^\mu
 -m_{B} \hat{B}^\mu \right) - \frac{m_D^2}{2}
 \, \hat{Z}_{D\mu} \hat{Z}_{D}^\mu \nonumber\ \\  
 & = & -\frac{1}{2}\hat{V}_{\mu}^T \mathcal{M}^{a}_{\rm_V} \hat{V}^{\mu}.
\end{eqnarray}
The above equations are written in the (non-canonically normalized) basis , 
$\hat{V}_{\mu} = (\hat{W}^3_{\mu}~\hat{B}_{\mu}~ \hat{Z}_{D \mu})^T$. 
Further,   
\begin{equation}
 \mathcal{K}^{a}_{\rm V} = 
\begin{pmatrix}
1 & 0 & 0  \\ 
0 & 1 & \epsilon_g \\
0 &  \epsilon_g & 1  \,  
\end{pmatrix},\,
\mathcal{M}^{a}_{\rm V} = 
\begin{pmatrix}
m_w^2 & -m_w m_B & 0  \\ 
-m_w m_B & m_B^2 & 0 \\
0 &  0 & m_{D}^{2}  \,  
\end{pmatrix}. 
\end{equation}
where $m_w = \dfrac{g_2  v }{2}, ~m_B=\dfrac{g_1  v }{2}$.
The kinetic term is canonical in the basis $V_{\mu} = \zeta \hat{V}_{\mu}$ , with 
\begin{equation}
\cal{\zeta} = 
\begin{pmatrix}
1 & 0 & 0  \\ 
0 & 1-\dfrac{\epsilon_g^2}{8} & \dfrac{\epsilon_g}{2} \\
0 & \dfrac{\epsilon_g}{2} & 1-\dfrac{\epsilon_g^2}{8}  \,  
\end{pmatrix}, 
\end{equation}
where we have kept only terms upto $\mathcal{O}(\epsilon_g^2)$.
The invariance of the gauge covariant 
derivatives imply that the gauge couplings $\mathcal{G}$ in the modified basis are given by, 
\begin{equation}
\mathcal{G} = \mathcal{\hat{G}}.\mathcal{\zeta}^{-1} =
\begin{pmatrix}
\hat{g}_2 & 0 & 0  \\ 
0 & \hat{g}_1 \left(1-\dfrac{3\epsilon_g^2}{8}\right) & g_{1D}  \\
0 & g_{D1} & \hat{g}_D \left(1-\dfrac{3 \epsilon_g^2}{8}\right)  \,  
\end{pmatrix}, 
\end{equation}
where $ g_{D1}= \hat{g}_D \dfrac{\epsilon_g}{2}$ and $ \hat{g}_{1D} = g_1 \dfrac{\epsilon_g}{2}$, 
and $\mathcal{\hat{G}} = Diagonal(g_2, g_1, g_D)$.
In this basis, the mass matrix takes the following form, 
\begin{equation}
\mathcal{M}^b_{V} = \zeta^{-1 T}\mathcal{M}^a_{V}\zeta^{-1}
\end{equation}
In this context, we have used the approach described in refs \cite{Fonseca:2011vn,OLeary:2011vlq}.
We have used \texttt{SARAH} and \texttt{SPheno}~\cite{Porod:2003um,Porod:2011nf} to numerically diagonalize 
the mass matrix and thus obtain the respective mass eigenstates and the corresponding mixing matrix. The mass eigenstates include a zero mass state, corresponding to the 
unbroken electro-magnetic gauge group $U(1)$, and two massive states representing $Z$ and 
$Z_D$ bosons respectively. Needless to mention, the mass of the $Z$ boson, and the electro-weak precision data in general, constraints $\epsilon_g$ \cite{Babu:1997st,Williams:2011qb}. For the present work, we will mainly consider small 
enough $\epsilon_g$ only ensuring prompt decay of $Z_D$ at high energy colliders.

\subsection{The Dark Sector}
After EWSB, $\varphi_R$ assumes a ${\it vev}$ ($v_{\varphi}$) generating a Majorana mass 
term in the dark fermionic sector. The mass term for the fermions, then, can be expressed 
as : 
\begin{equation}
-\mathcal{L}_{\rm mass}   = \dfrac{1}{2} \{\bar{\tilde{\xi}}_1~  \bar{\tilde{\xi}}_2 \} \mathcal{M}_{\rm DM} \{ \xi_1~ \xi_2\}^T +{\rm h.c.}
\end{equation} 
where  
\begin{equation}
\mathcal{M}_{\rm DM} = 
\begin{pmatrix}
 f v_{\varphi} & M \\ 
 M &  f v_{\varphi}    \,  
\end{pmatrix}. 
\end{equation}
 
This mass matrix can be diagonalized by an orthonormal matrix 
\begin{equation}
\mathcal{N}_{\rm DM} =  \dfrac{1}{\sqrt{2}} 
\begin{pmatrix}
  -1 & 1 \\ 
  1 & 1   \,
\end{pmatrix}. 
\end{equation}
The mass of the physical states can be obtained from the eigenvalues of $\mathcal{M}_{\rm DM}$ 
and are given by  $M_{\pm} =|M \pm f v_{\varphi}|$. The corresponding physical states (mass 
eigenstates) are $\chi_{\pm} = \dfrac{\chi^c \pm \chi}{\sqrt{2}}$. The state with the lower mass {\it i.e.} $M_-$ is our dark matter candidate and from now on its mass will be denoted by $M_{DM}$ for definiteness. \footnote{Note that we have 
used the absolute value of the smallest eigenvalue, since the sign of the fermion mass can simply 
be rotated away by a chiral rotation. For example, with $M > f v_{\varphi}$ and $M > 0$, the smallest eigenvalue $M-f v_{\varphi}$ is positive. The corresponding mass eigenstate is 
$ \dfrac{1}{\sqrt{2}} \left( (\xi_1-\xi_2)~~  (-\tilde{\xi}_1+\tilde{\xi}_2)  \right)^T = 
\gamma_5 \dfrac{\chi^c  - \chi}{\sqrt{2}}$.} 

The lagrangian ${\cal L_\text{DS}}$, in the mass basis, is given by :

\begin{eqnarray}
\cal{L}_{ } &=& \frac{1}{2} \left( i \overline{\chi}_+\gamma^\mu \partial_{\mu} \chi_+ + i\overline{\chi}_-\gamma^\mu \partial_{\mu} \chi_- -g_D Z_D^{\mu} \bar{\chi}_{+} \gamma_{\mu}\chi_{-}-g_D Z_D^{\mu} \bar{\chi}_{-}\gamma_{\mu} \chi_{+}  \right)  \nonumber \\
&&\,-\frac{1}{2}\left(M_+ \overline{\chi}_+ \chi_+  -  M_-\overline{\chi}_- \chi_{-}\right) -\dfrac{f}{2} \left((\cos\theta\, H_1 + \sin\theta\, H_2)(\overline{\chi}_+ \chi_+ - \overline{\chi}_- \chi_-)  \right)\,.
\label{LDM}
\end{eqnarray}
In the expression above $H_1$ and $H_2$ denote the $\varphi_R$-like and $h$-like states (only true for small mixing), respectively, in the mass eigenbasis as discussed before.

%
%


\section{Self-interaction of dark matter: Allowed regions} 
\label{sec:dm_self}
A study of self interaction in our dark matter scenario can give rise to novel features and lead to interesting results. As was discussed in the introduction, strongly self-interacting dark matter can help to alleviate many small scale structure problems. In our model there are two Majorana fermions, an extra gauge boson and an extra Higgs-boson. Hence, it is
generic to any such $U(1)$ gauge theories to have vertices like $\chi_+ \chi_+ H_1$, $\chi_- \chi_- H_1$ and $\chi_+ \chi_- Z_D$.

Here, the new gauge boson is not charged under the Standard Model gauge group. Hence it can be very light in principle. The new Higgs boson can also be quite light. A remarkable
consequence of this is the Sommerfeld Enhancement in the limit when $M_{DM} \gg M_{H_1}$ or $M_{Z_D}$, where $M_{DM}, M_{H_1}$ and $M_{Z_D}$ are the masses of the DM, light Higgs
and the extra gauge boson. Depending on the mass
of the mediator we can have the following three cases :

\begin{itemize}
\item When $M_{H_1}$ is much lighter than the dark matter mass, but $M_{Z_D}$ is heavier. Sommerfeld enhancement takes place by exchanging the light Higgs boson only by the usual ladder diagrams.

\item When $Z_D$ is lighter than the new Higgs, then enhancement proceeds by
exchanging only this light extra gauge boson.

\item When both $Z_D$ and the new Higgs are of comparable masses then enhancement should in principle occur due to both the mediators.
\end{itemize}
The case when both the new gauge boson and the extra Higgs are very light may be interesting from a theoretical point of view. But, by making both of these particles light (lighter than the DM) simultaneously, we will lose out on the robust collider signatures that we also intend to explore in this work. So we do not probe those parameter regions where both of the mediators are light.

Next, let us discuss about the second case. Since the new scalar is not very light, this implies that $\lambda_{\rm mix} \, \, v_{D} $ is not small (from the expression of the mass of the new scalar). But on the other hand, we demand that $M_{Z_D} \sim g_{D}\,v_{D}$ should be much smaller than that of $M_{H_{1}}$. To satisfy both of these conditions we have to take resort to very small dark gauge boson couplings. But the enhancement factor also depends on $g_{D}$ (through $\chi_{+} \,\chi_{-} \, Z_D$ vertices). Hence in this case, the enhancement that we were expecting by introducing light gauge boson mediator is nullified by the presence of small gauge couplings. So, practically speaking, Sommerfeld enhancement with only light $Z_D$ is not a good a choice either.

From these two points of view, it hence seems to take the first case as our preferred choice for studying strong self interactions of dark matters. 

In calculating the self-interacting cross section we closely followed the analytical expressions presented in \cite{Feng:2009hw}. In the Born limit ($\alpha_D M_{DM}/M_{H_1} \ll 1$), the cross section is given by :
\begin{align} 
\sigma_T^{\rm Born} &= \frac{8\pi \alpha_D^2}{M_{DM}^2 v^4} \Big( \log\big(1+M_{DM}^2 v^2/M_{H_1}^2\big) -\frac{M_{DM}^2 v^2}{M_{H_1}^2 + M_{DM}^2 v^2} \Big) \label{born2} \; ,
\end{align}
where, $\alpha_D = f^2 / 4\pi$ and $v$ is the virial velocity of galaxies.

Outside the Born regime ($\alpha_D M_{DM} /M_{H_1} \gtrsim 1$) non-perturbative effects become important. Analytical results can be obtained in the classical limit ($M_{DM} v/M_{H_1} \gg 1$). For an attractive potential \cite{Feng:2009hw, PhysRevLett.90.225002,PhysRevE.70.056405} :
\begin{equation}
\sigma_T^{\rm clas} = 
\left\{\begin{array}{lc}
	\frac{4 \pi}{M_{H_1}^2} \beta^2 \ln\left(1+\beta^{-1}\right) & \beta \lesssim 10^{-1} \\
	\frac{8 \pi}{M_{H_1}^2} \beta^2 / \left(1+1.5 \beta^{1.65}\right) & \; 10^{-1} \lesssim \beta \lesssim 10^3 \\
	\frac{\pi}{M_{H_1}^2} \left(\ln \beta+1-\frac{1}{2} \ln^{-1}\beta \right)^2 & \beta \gtrsim 10^3
\end{array} \right. \label{plasmaAppendix}
\end{equation}
where $\beta \equiv 2 \alpha_D M_{H_1} / (M_{DM} v^2)$.
Analytical results can also be obtained in the resonance region in between by approximating the Yukawa potential by {\it {Hulthen}} potential \cite{1302.3898}.
\begin{equation}
\sigma_T^{\rm Hulth\acute{e}n} = \frac{16\pi}{M_{DM}^2 v^2} \sin^2 \delta_0 \label{hulthen}
\end{equation}
where the $l=0$ phase shift is given in terms of the $\Gamma$-function by
\begin{equation}
\delta_0 = \arg\left( \frac{ i \,\Gamma\big( \frac{ i M_{DM} v}{\kappa M_{H_1}} \big)}{\Gamma(\lambda_+) \Gamma(\lambda_-)} \right)
\end{equation}
with
\begin{equation}
\lambda_\pm \equiv 1 + \frac{ i M_{DM} v}{2 \kappa M_{H_1}} \pm  \sqrt{ \frac{ \alpha_D M_{DM}}{\kappa M_{H_1}}  - \frac{ M_{DM}^2 v^2 }{4 \kappa^2 M_{H_1}^2} }
\end{equation}
and $\kappa \approx 1.6$ is a dimensionless number.
The bound on self-interacting dark matter is given in terms of $\sigma_{\rm self} / M_{DM}$. This should be $\lesssim (0.1-10)$ cm$^2$ gm$^{-1}$  when $v_0 \sim 10$ km/sec to alleviate the {\it Core-vs-cusp} problem of the dwarf spheroidal galaxies \cite{2012MNRAS.423.3740V,Loeb:2010gj}. From the X-ray and the lensing observations of Bullet cluster we further have $\sigma_{\rm self}/M_{DM} \lesssim 1$ cm$^2$ gm$^{-1}$ \cite{Randall:2007ph} when $v_0 = 1000$ km/sec. 

\begin{figure}
	\centering
	\includegraphics[scale=0.27]{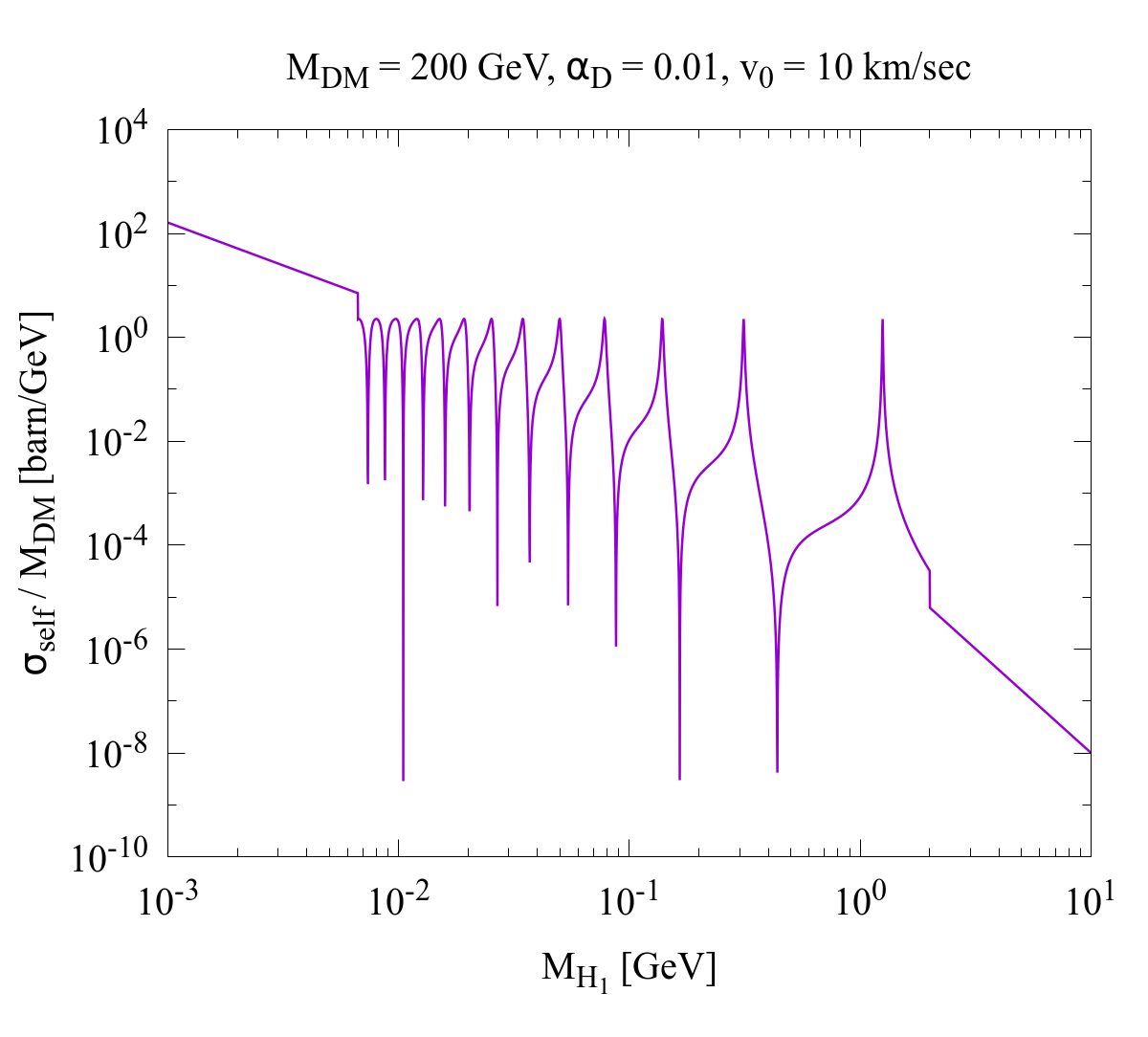}
	\caption{Sommerfeld Enhancement with respect to mediator mass for a fixed dark matter mass and fixed coupling strength $\alpha_D$.}
	\label{sigmaSelf}
\end{figure}

\begin{figure}[h!]
	\hspace*{-1.27cm}
	\includegraphics[scale=0.22]{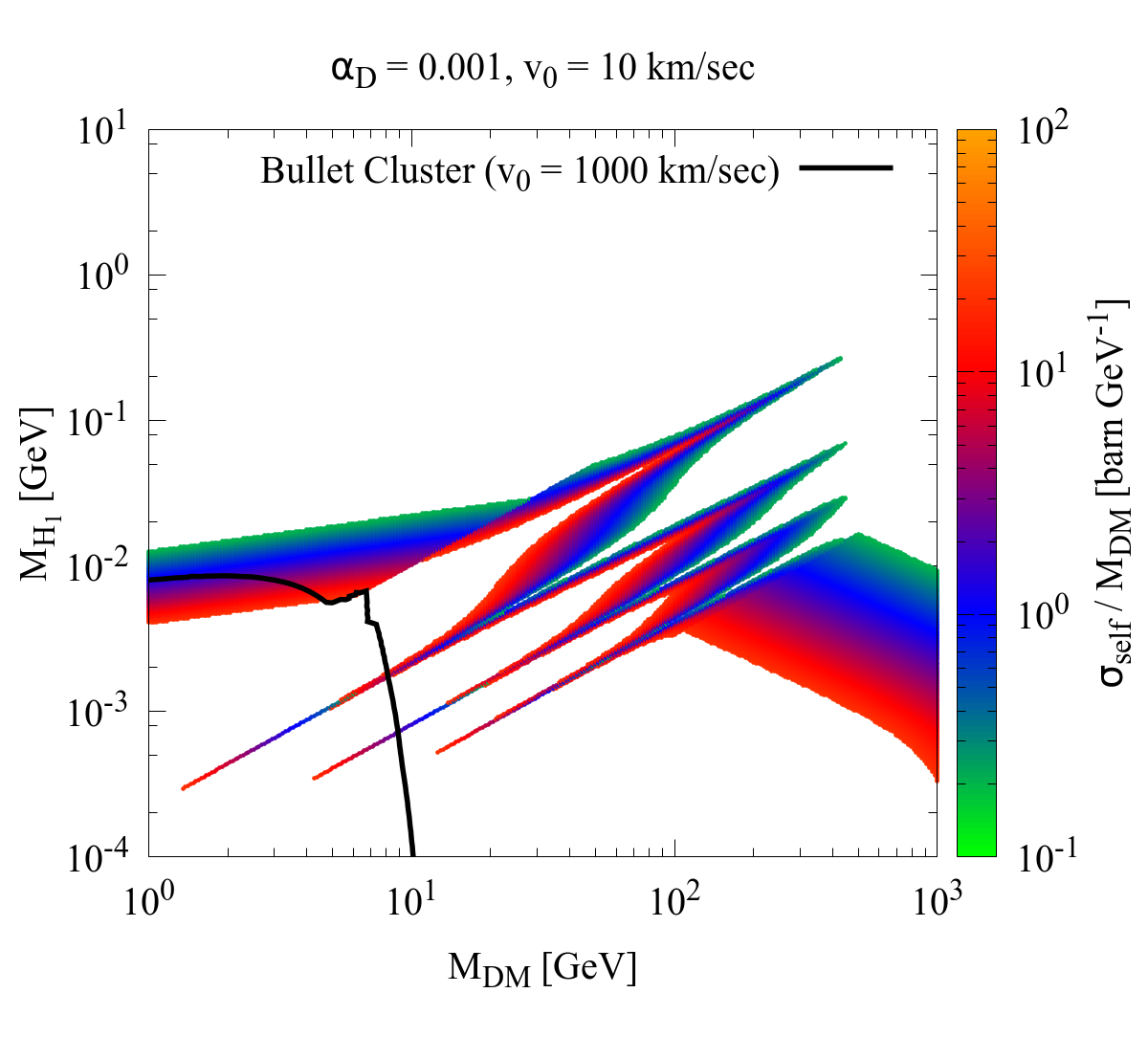}~
	\includegraphics[scale=0.24]{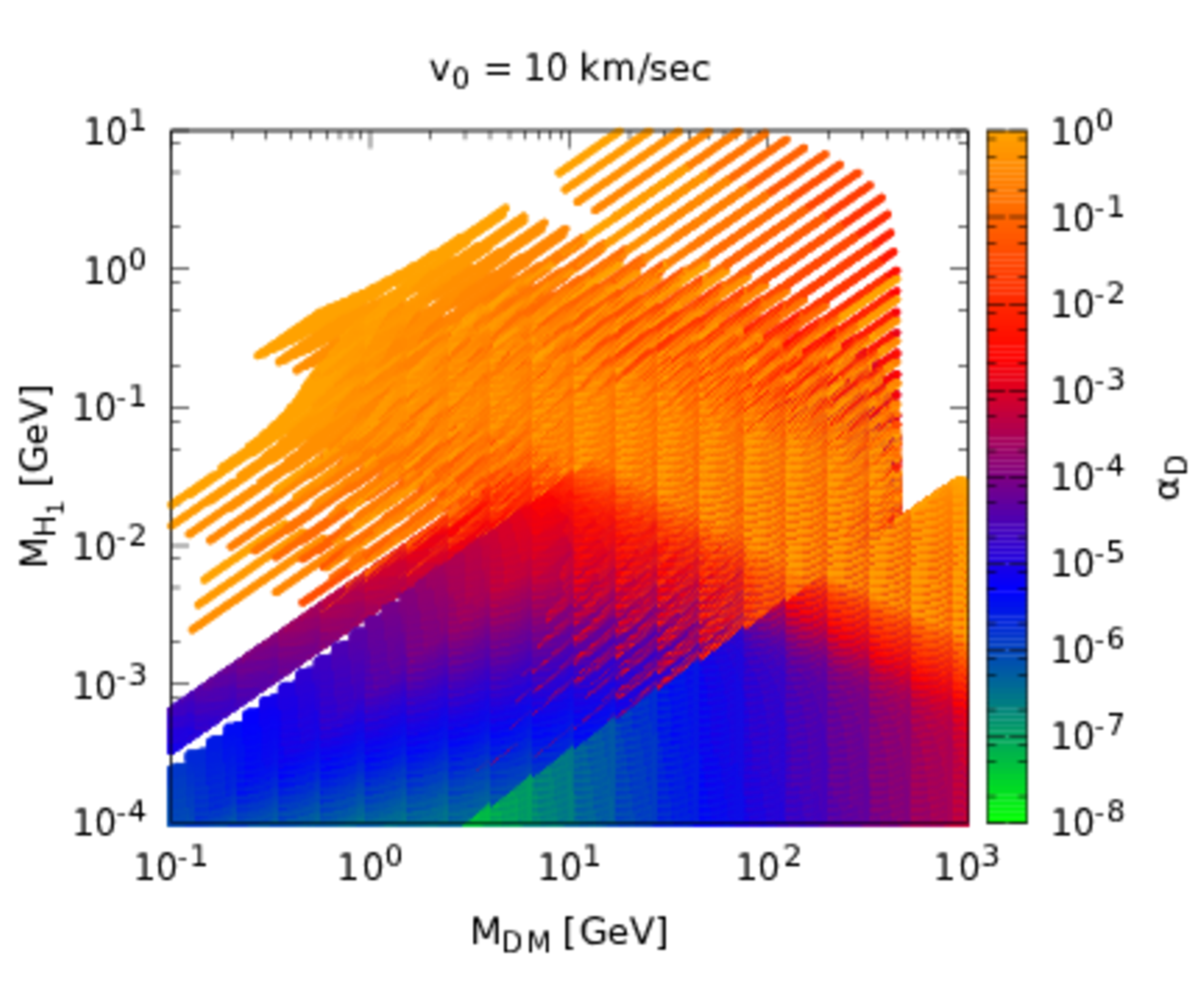}
	\caption{{\bf Left}: Region for allowed self-interaction with varying dark matter and mediator mass for a fixed $\alpha_D$. $v_0$, the virialized velocity is fixed to 10 km/sec. The region within the  black contour denotes the part of the parameter space ruled from Bullet cluster~\cite{Markevitch:2003at,Clowe:2003tk,Randall:2007ph} constraint (see text for more details).
	 {\bf Right}: Region of parameter space where moderate to strong self-interaction of dark matter is allowed to alleviate {\it Core-vs-cusp, Too-big-to-fail} problems, with varying $\alpha_D$ (0.1 cm$^2$ gm$^{-1}$ $\lesssim \sigma_{\rm self} / M_{DM} \lesssim 10$ cm$^2$ gm$^{-1}$). The variation of $\alpha_D$ is shown in the color bar. The allowed points shown in this figure satisfies the constraint arising from Bullet Cluster.}
	\label{scan_self1}
\end{figure}

The self-interaction cross section in units of the dark matter mass (for a fixed $\alpha_D$ and $M_{DM}$) is plotted with respect to the mediator mass in Fig.\,\ref{sigmaSelf}.
Since we are concentrating on the low mediator mass regime in this study, the mass of the light Higgs was varied up to 10 GeV throughout this work. If we now perform a scan by varying both the dark matter mass (0.1 GeV--1 TeV) and the mediator mass (0.1 MeV--10 GeV), and the allowed points ({\it i.e.} points with 0.1 cm$^2$ gm$^{-1}$ $\lesssim \sigma_{\rm self} / M_{DM} \lesssim 10$ cm$^2$ gm$^{-1}$) are plotted, then we arrive at Fig.\,\ref{scan_self1} (left). This was however done for a fixed $\alpha_D\, (\equiv f^2/4\pi) = 0.001$. The right hand panel of Fig.\,\ref{scan_self1} shows the more general case with varying $\alpha_D$. The values of $\alpha_D$ are shown in the colour bar. We see immediately that for light mediators (with mass reaching up to 10 GeV) and for a wide range of dark matter masses, the allowed range of $\alpha_D$ is quite large. Note that, self interaction can be achieved even for very small values of $\alpha_D$ ($\sim 10^{-7}$) when the mediator is extremely light ($\sim 100$ keV). Here, we have taken care of the bounds on the self-interaction cross-section arising from the Bullet Cluster.

So, to summarise, in this work we will be concentrating on self interaction of dark matter mediated by light scalar mediators alone. But, this mediator ($H_1$) cannot be extremely light in order to respect the constraints arising from BBN \cite{2011PhLB..701..296M,Cyburt:2015mya}. This is because, BBN places strong upper bounds on presence of relativistic particles when temperature of the universe is $\sim 1$ MeV. Hence, for all practical purposes, $M_{H_1} \gtrsim 10$ MeV would be admissible. Therefore, the mass range of the scalar $H_1$ for suitable self-interaction is expected to be : $10$ MeV $\lesssim M_{H_1} \lesssim 10$ GeV. We will concentrate on this range of $M_{H_1}$ throughout the rest of the work. The upper bound on self-interaction cross section from Bullet Cluster was also taken into consideration during our calculation.

\section{Survey of parameter space}
\label{sec:par_space}

The latest results from the SM Higgs coupling measurements still allow the SM Higgs to have non-standard couplings. In the context of the $U(1)_{D}$ model considered in this work, such non-standard decay modes of the SM-like Higgs boson could arise in three possible ways, viz., decay into a pair of light Higgs bosons ($H_{2} \to H_{1} H_{1}$), decay into a pair of dark gauge bosons ($H_{2} \to Z_{D} Z_{D} $) and decay into the Majorana fermions ($H_{2} \to \chi_{+} \chi_{+},\chi_{-}\chi_{-}$). The $H_{2} \to H_{1}H_{1}$ decay width is crucially controlled by $\sin\theta_{\rm mix}$, while the $H_{2} \to Z_{D}Z_{D}$ decay process also has a direct dependence on $\epsilon^{2}$. Throughout this analysis, we restrict the mass of $Z_{D}$ within $M_{Z_{D}} < 60~{\rm GeV}$. Thus, the $H_{2} \to H_{1}H_{1}$ and $H_{2}\to Z_{D}Z_{D}$ decay modes always remain kinematically feasible. 



We perform a random scan over the $9$-dimensional parameter space to capture the underlying physics of the $U(1)_{D}$ model. In this respect, we scan over the following parameters : $\lambda_{H}$, $\lambda_{\phi}$, $\lambda_{mix}$, $g_{D}$, $\epsilon_{g}$, $v_{D}$, $M$, and $f$. 

With the main focus on exploring the self-interaction motivated parameter region, we perform a random scan over the following range of input parameters. The scan range is inspired by the region of parameter space shown in Fig.~\ref{scan_self1} (b) which corresponds to those sectors which facilitate a dark matter with moderate to strong self-interactions. It can be observed from Fig.~\ref{scan_self1} that a self-interacting DM up to mass $1~{\rm TeV}$ could be accommodated with $M_{H_{1}} < 10~{\rm GeV}$. The mass of $H_{1}$ (Eqn.~\ref{hmass}) is determined by $\lambda_{\phi},~\lambda_{mix}$ and $v_{D}$, and the range of these parameters is chosen in such a way that a significant fraction of scanned points populate the $M_{H_{1}} \lesssim 10~{\rm GeV}$ region. Correspondingly, $M$ has also been varied from $1~{\rm GeV} - 1000~{\rm GeV}$ to obtain the DM mass in that range. Since we are interested in studying the collider prospects of $Z_{D}$, whose mass is given by $g_{D} \times v_{D}$, the choice of $g_{D}$ is governed primarily by the requirement of $M_{Z_{D}}$ within $2~{\rm GeV} - 60~{\rm GeV}$. 
\begin{eqnarray}
\label{eq:inp_par_range}
-0.10 < \lambda_{H} < -0.16, \quad ~ -10^{-8} < \lambda_{\phi} < -0.7, \quad ~ 10^{-12} < \lambda_{mix} < 0.3 \quad \quad \\
~10^{-6} < g_{D} < 0.6,\quad ~ 10^{-4} < \epsilon_{g} < 10^{-10} \quad \quad  \\
1~ {\rm GeV} < v_{D} < 1000~ {\rm GeV}, \quad~ 10^{-6} < f < 10^{-1},\quad ~1~{\rm GeV} < M < 500 ~{\rm GeV} \quad 
\label{eqn:par_space}
\end{eqnarray}

\begin{figure}
\begin{center}
\includegraphics[scale=0.23]{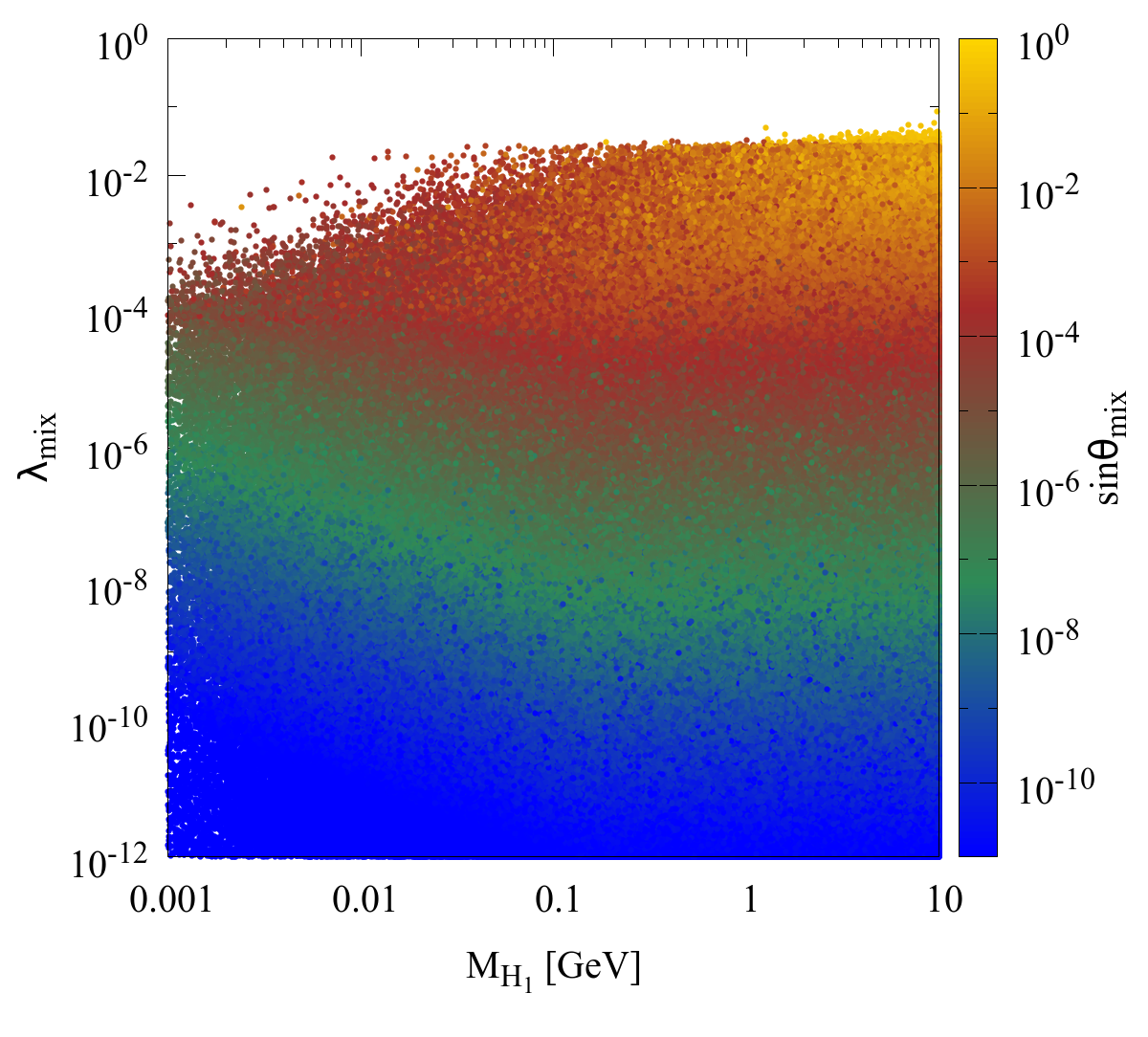}
\caption{Parameter space points with $M_{H_{1}} \leq 10~{\rm GeV}$, $M_{Z_{D}} \leq 60~{\rm GeV}$ and $ 124.4~{\rm GeV} \leq M_{H_{2}} \leq 125.8~{\rm GeV}$, 
in the $M_{H_{1}}-\lambda_{mix}$ plane with $\sin{\theta}_{mix}$ presented in the color palette.}
\label{fig:inp_par}
\end{center}
\end{figure}


\section{Constraints}
\label{sec:all_constraints}

The $U(1)_{D}$ parameter space discussed so far gets constrained by a multitude of experimental search results. Combined measurement of the Higgs boson mass by the ATLAS and CMS collaborations confines it within $124.4~{\rm GeV}$ - $125.8~{\rm GeV}$~\cite{Aad:2015zhl} at 3~${\rm \sigma}$. Consequently, the mass of $H_{2}$ is required to lie within $124.4~{\rm GeV} < M_{H_{1}} < 125.8~{\rm GeV}$. The parameter space points obtained by imposing the Higgs mass constraint, $M_{Z_{D}} < 60~{\rm GeV}$ and $M_{H_{1}} < 10~{\rm GeV}$ are shown in Fig.~\ref{fig:inp_par} in the $\lambda_{mix} - M_{H_{1}}$ plane. The color palette represents the value of $\sin\theta_{mix}$, and exhibits its direct proportionality with $\lambda_{mix}$. 


The effect of constraints on $\sin\theta_{mix}$ from light $H_{1}$ searches at LEP has been analyzed in this section. We also evaluate the implications from Higgs signal strength measurements through a global $\chi^{2}$ analysis, by combining, both, $8$ and $13$ TeV LHC results. The status of our parameter space in light of direct $Z_{D}$ searches at the LHC is studies as well. The $U(1)_{D}$ parameter space under study also receives strong constraints from measurements at B-factories and various beam dump experiments. In the remainder of this section, we will analyse in detail, the implications on our parameter space from each of these constraints.

\subsection{Constraints from LEP}
\begin{figure}
\begin{center}
\includegraphics[scale=0.20]{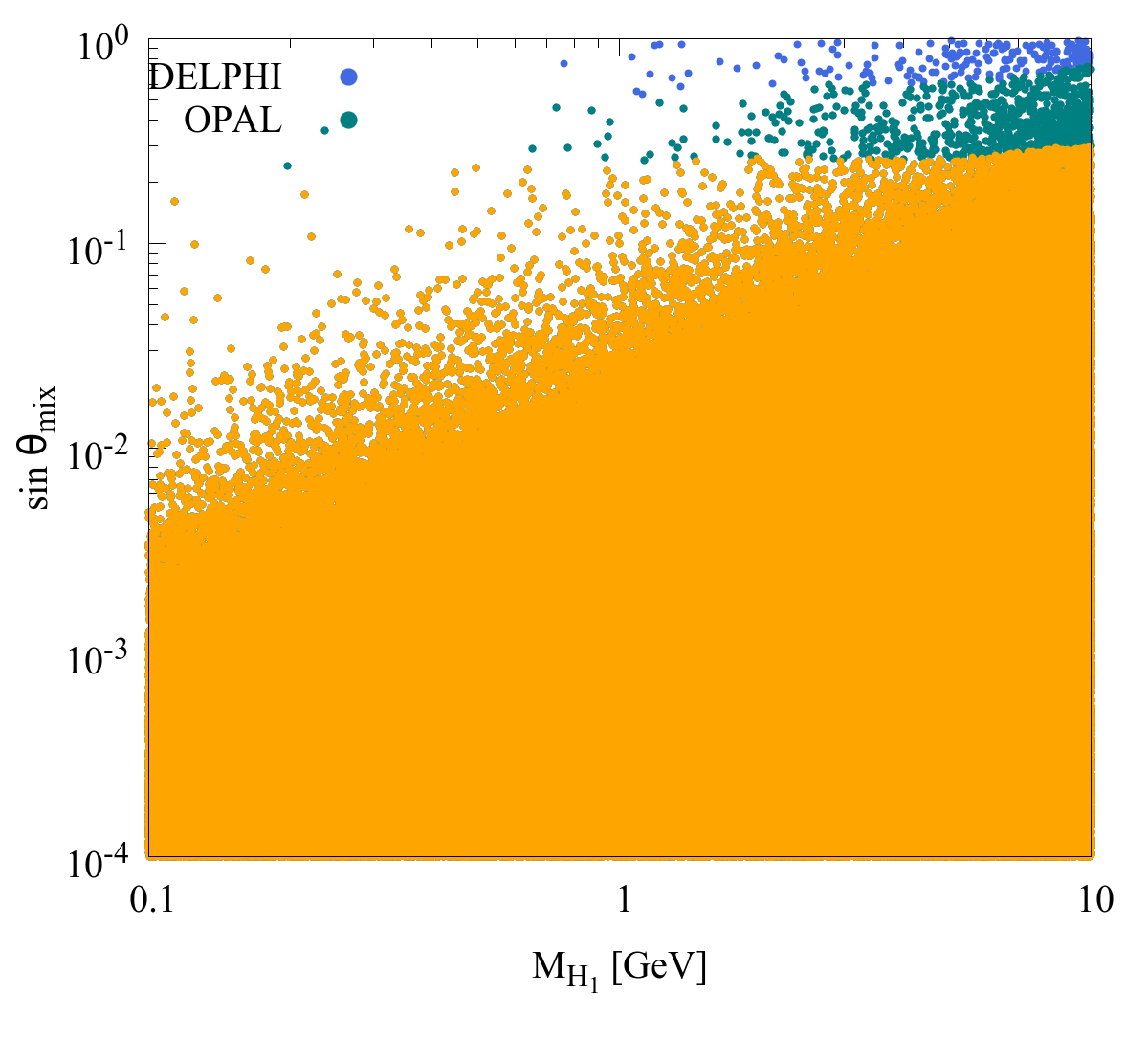}
\caption{Scatter plot in the $M_{H_{1}}$ - $\sin\theta_{mix}$ plane showing the implications of imposing the LEP constraints on our parameter space. All parameter space points satisfy the Higgs mass constraints and have $M_{H_{1}} < 10~{\rm GeV}$ and $M_{Z_{D}} < 60~{\rm GeV}$. The blue  colored parameter space points are ruled out by upper limits derived on $\zeta^{2}$ by DELPHI~\cite{Abdallah:2001ux}, where $\zeta$ is the normalized $H_{1}VV$ coupling with respect to SM. The green colored points are excluded by the upper limits on $\kappa$ derived by OPAL~\cite{Abbiendi:2002qp}, where $\kappa$ is the ratio of production cross-section of $H_{1}$ in the Higgs strahlung process to its SM value.}
\label{fig:LEP}
\end{center}
\end{figure}

The two LEP collaborations, DELPHI and OPAL, have performed numerous searches for the Higgs boson using data collected from $e^{+}e^{-}$ collisions at $\sqrt{s}$ ranging from $\sim 91~{\rm GeV}- 209~{\rm GeV}$, with no observation of signal like events. DELPHI has performed searches for light $H_{1}$ using the $e^{+}e^{-} \to ZH_{1} \to (Z \to e^{+}e^{-}, \mu^{+}\mu^{-})H_{1}$~\cite{Abreu:1990bq} and $Z \to e^{+}e^{-}$,
$\mu^{+}\mu^{-}$, $\tau^{+}\tau^{-}$,
$\nu \bar{\nu}$, $q\bar{q}$~\cite{Abreu:1990zc} processes. The di-Higgs ($e^{+}e^{-} \to H_{2}H_{1}$) and triple Higgs ($e^{+}e^{-} \to H_{2}H_{1} \to H_{1}H_{1}H_{1}$) final states have been also analyzed by LEP in $4\tau,~4b,~2b2\tau,~6\tau,~6b$ channel~\cite{Schael:2006cr}.


Upper bounds on $H_{1}VV(V=W^{\pm},Z)$ coupling normalized to that for SM, at $95\%$ C.L., taking into account, both, LEPI and LEPII data has been derived in \cite{Abdallah:2001ux}. In the context of our analysis, the normalized $H_{1}VV$ coupling ($\zeta$) is directly proportional to $\sin\theta_{\rm mix} $, and the upper limits derived in \cite{Abdallah:2001ux} exclude the blue colored points in Fig.~\ref{fig:LEP}. It can be observed from Fig.~\ref{fig:LEP} that the upper limits on $H_{1}VV$ coupling excludes $\sin\theta_{\rm mix} \gtrsim 0.5$, for all values of $M_{H_{1}}$ obtained in our parameter space scan. The study by OPAL collboration~\cite{Abbiendi:2002qp}, using the LEPI and LEPII dataset, has derived upper bounds on $\kappa$ at $95\%$ C.L., for $M_{H_{1}} = 10^{-6} - 100 ~{\rm GeV}$, where $\kappa$ is the ratio of production cross-section of the new light scalar in the Higgsstrahlung process to that of SM Higgs production in the Higgsstrahlung process, under the assumption that the mass of the new light scalar is equal to the SM Higgs. Within the $U(1)_{D}$ model considered here, $\kappa$ is proportional to ${\sin^2\theta_{\rm mix}}$, and the corresponding upper bounds, upon being implemented on our parameter space, excludes the green  colored region in Fig.~\ref{fig:LEP}. We find that the upper limits from \cite{Abbiendi:2002qp} exerts a relatively stronger constraint on the parameter space and excludes $\sin\theta_{\rm mix} \gtrsim 0.2$, over the entire range of $M_{H_{1}}$\footnote{It may be noted that decay channels originating from the light scalar Higgs ($M_{H_{1}} \lesssim 2~{\rm GeV}$) require a careful treatment owing to the uncertainties associated with partial decay widths of $H_{1}$ into hadronic channels. 
Within $2m_{\pi}\lesssim M_{H_{1}}\lesssim 2~{\rm GeV}$, the uncertainties associated with theoretical calculations remain significant~\cite{Gunion:425736}, and the corresponding partial decay widths are computed using low energy effective theories of QCD. In our analysis, we have considered the branching ratios of $H_{1} \to \gamma \gamma, ~e^{+}e^{-},~\mu^{+}\mu^{-}$ from \cite{Berger:2016vxi}.} obtained in our parameter space scan. 

\subsection{Constraints from LHC Higgs signal strength measurements} The ATLAS and CMS collaborations have performed numerous measurements of the coupling of the $125~{\rm GeV}$ Higgs boson using LHC Run-I and Run-II datasets. Results from these measurements are usually presented through signal strength observables, which considers the most important Higgs production modes at LHC : gluon fusion mode ($ggF$), vector boson fusion ($VBF$), associated production with vector bosons ($VH_{2},~V=W^{\pm},Z$) and associated production with a top-antitop pair ($t\bar{t}h$). The relevant Higgs boson decay modes are $H_{2} \to b\bar{b}$, $\tau^{+}\tau^{-}$, $WW$, $ZZ$, $\gamma\gamma$. The signal strength variable is defined as 
\begin{eqnarray}
\mu = \frac{({\sigma_{H_{2}}^{i} \times Br_{H_{2}}^{j}})_{Model}}{({\sigma_{H_{2}}^{i} \times Br_{H_{2}}^{j}})_{SM}}
\label{eqn:signal_strength}
\end{eqnarray} 

\begin{figure}
\begin{center}
\includegraphics[scale=0.2]{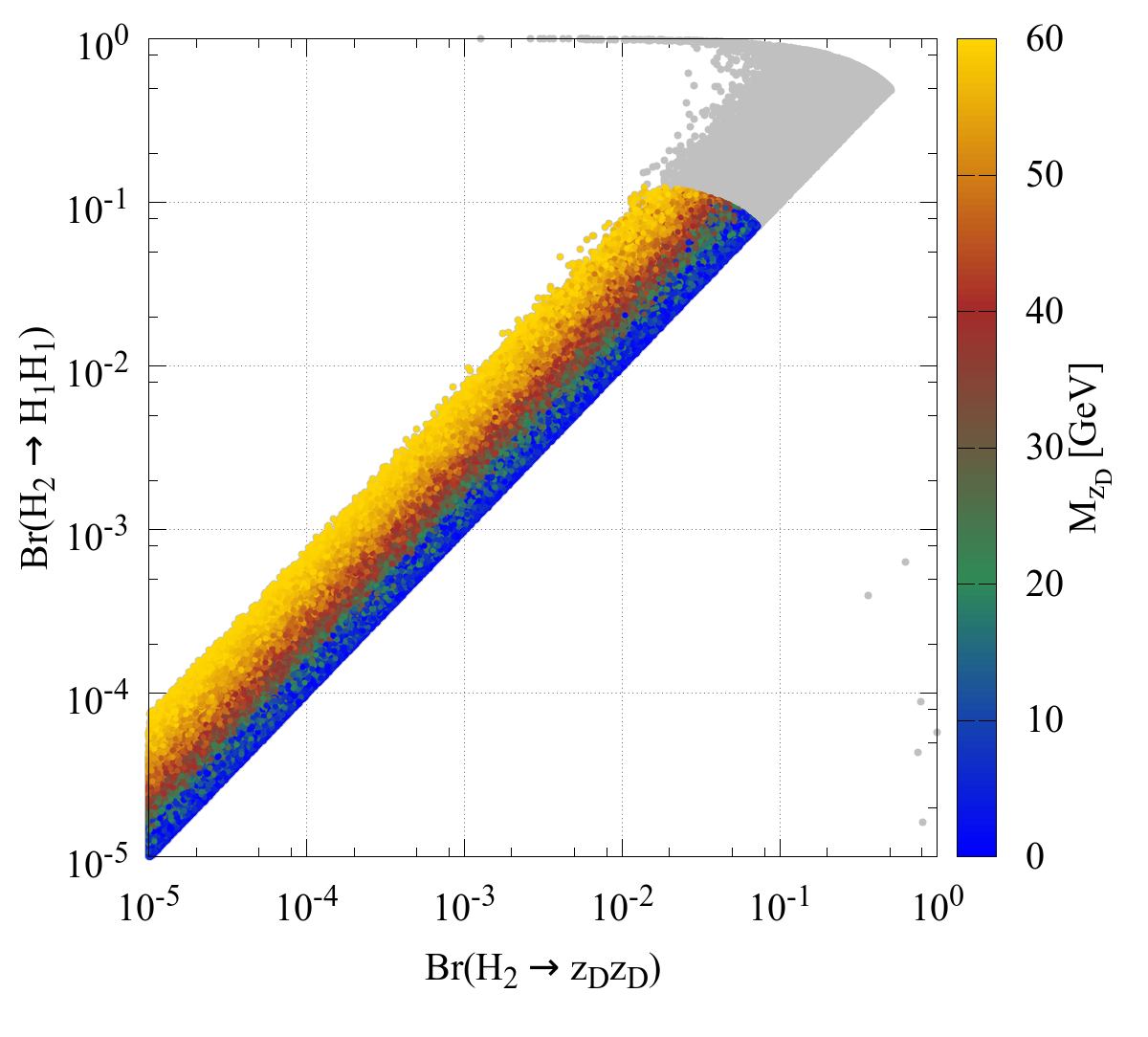}\includegraphics[scale=0.2]{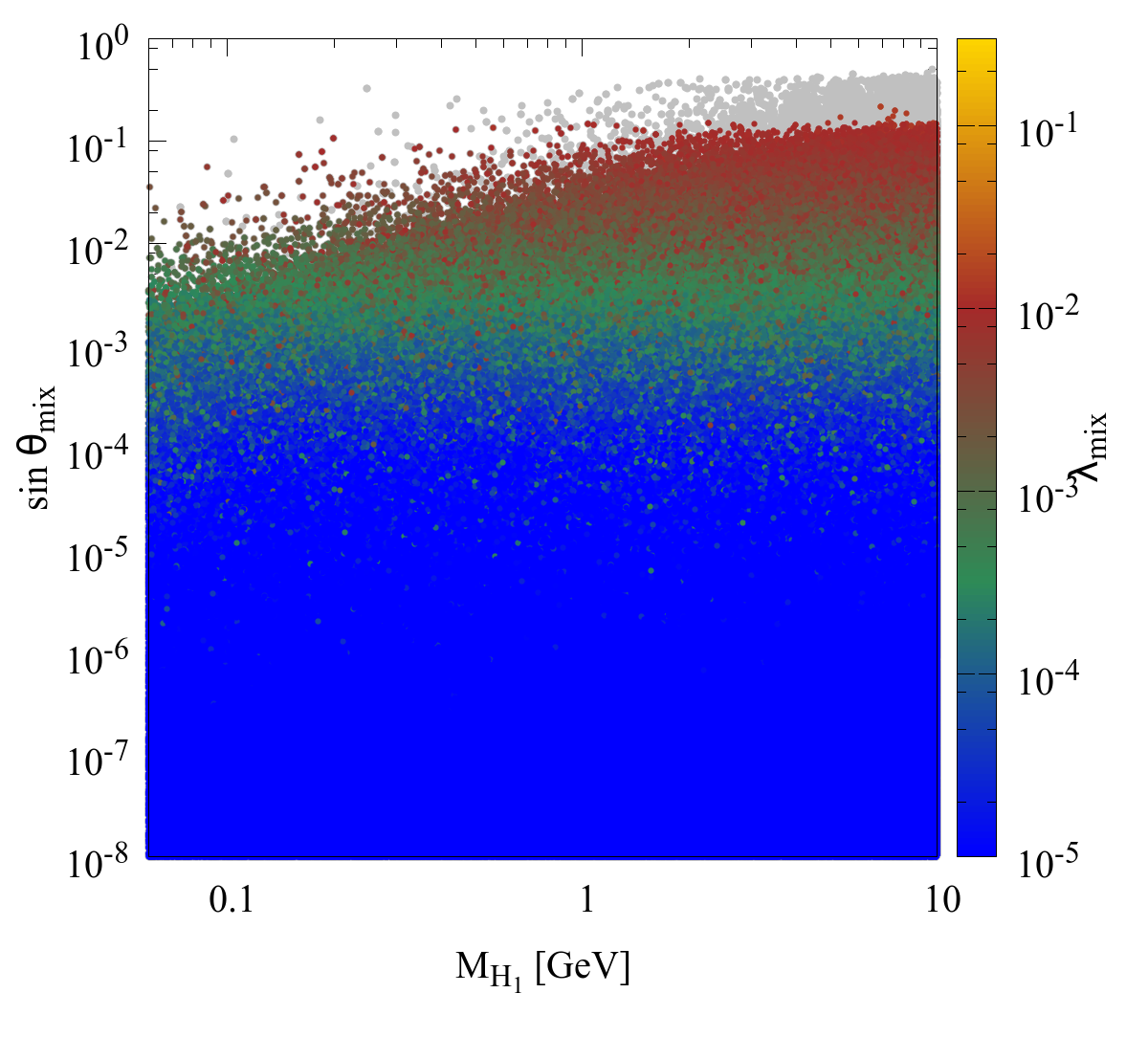}
\caption{The grey colored points are excluded by the global $\chi^{2}$ analysis. \textbf{Left}: Correlation between $Br(H_{2} \to H_{1}H_{1})$ and $Br(H_{2} \to Z_{D}Z_{D})$. The color palette represents the value of $M_{Z_{D}}$. \textbf{Right}: Scatter plot in the $M_{H_{1}}$ - $\sin\theta_{\rm mix}$ plane. The color palette represents the value of $\lambda_{mix}$. The grey colored points are excluded by the signal strength constraints.}
\label{fig:signal_strength}
\end{center}
\end{figure}

Here, $\sigma_{H_{2}}^{i}$ corresponds to the Higgs production cross-section in the $i^{th}$ mode ($i=~ggF$, $VBF$, $VH_{2}$ and $t\bar{t} H_{2}$), and $Br_{H_{2}}^{j}$ corresponds to the branching fraction of the Higgs in the $j^{th}$ decay mode $(j=~b\bar{b}$, $\tau^{+}\tau^{-}$, $WW$, $ZZ$, $\gamma \gamma)$. $(\sigma_{H_{2}}^{i})_{\rm SM}$ and $(Br_{H_{2}}^{i})_{\rm SM}$ corresponds to the SM counterparts. In the current analysis, the heavier scalar Higgs boson, $H_{2}$, is required to be consistent with the SM $125~{\rm GeV}$ Higgs boson, thereby requiring it to be dominantly doublet-like. However, mixing between the doublet and singlet Higgs fields renders a small singlet admixture in $H_{2}$ as well. The coupling of $H_{2}$ with the SM particles remain similar to that of the case of SM Higgs boson, except for an additional suppression (by ${\cos{{\theta}_{mix}}}$). Consequently, the $t\bar{t}H_{2}$ vertex in the $ggF$ production mode of Higgs, and the $VVH_{2}$ vertex in the $VBF$, $VH_{2}$ and $t\bar{t}H_{2}$ production modes of Higgs, gets an additional factor of $\cos{{\theta}_{\rm mix}}$, resulting in the Higgs production cross-section acquiring a $\cos^2\theta_{\rm mix}$ suppression. As a result, the ratio of ${\sigma_{H_{2}}^{i}}_{Model}/{\sigma_{H_{2}}}_{SM}$ in Eqn.~\ref{eqn:signal_strength} becomes equal to $\cos^2\theta_{\rm mix}$. 
The branching fraction of $H_{2}$ to SM final states also gets affected by the presence of new non-SM decay modes. As specified in the previous section, $H_{2}$ has three possible non-SM decay modes, and the relative interplay of input parameters determine the partial decay width in each of the SM channel. 

In this study, the signal strength constraints have been imposed upon the scanned parameter set through a global $\chi^{2}$ analysis\footnote{ This analysis set up has been validated in \cite{Bhattacherjee:2015sga,Barman:2016jov}.} performed by taking into account the most recent Higgs signal strength constraints, tabulated in Table~II and Table~III of \cite{Barman:2016jov}. The value of $\chi^{2}$ is computed as 
\begin{eqnarray}
\chi^{2} = \Sigma_{i} \frac{(\bar{x}_{i}-x_{i})^{2}}{\Delta{x_{i}^{2}}}
\label{eqn:chi2}
\end{eqnarray}
where, $x_{i}$ corresponds to the best-fit value of the observable derived through experimental measurements,  $\bar{x}_{i}$ corresponds to the value of the observable computed for the current model, and $\Delta{x_{i}}$ refers to the error associated with the experimental measurement. In the context of this study, $x_{i}$ represents the best-fit value of the signal strength observables. The value of $\chi^{2}$ was computed for all scanned parameter space points by combining $28$ signal strength observables from LHC Run-I data and $18$ observables from LHC Run-II data ($\sim 15~fb^{-1}$ and $\sim 36~fb^{-1}$)\footnote{For details see Table II and III of \cite{Barman:2016jov}.}, and the lowest value of $\chi^{2}$ was determined (represented by $\chi^{2}_{min}$). Allowing $2\sigma$ uncertainty, we choose parameter space points which lie within $\chi^{2}_{min} + 6.18$. The implications of the global $\chi^{2}$ analysis are shown in Fig.~\ref{fig:signal_strength}. We would like to note that the parameter space points corresponding to Fig.~\ref{fig:signal_strength}~(left) have been generated from the parameter space specified in Eqn.\ref{eqn:par_space}, except for $f$ and $M$, whose values were fixed in order to generate $M_{DM} = 500~{\rm GeV}$. Thus, $H_{2} \to H_{1}H_{1}$ and $H_{2}\to Z_{D}Z_{D}$ are the only non-SM decay modes for $H_{2}$. We show the correlation between these two non-SM decay modes in Fig.~\ref{fig:signal_strength}~(left), where the grey points (color palette points) have been excluded (allowed) by the global $\chi^{2}$ analysis. The $H_{2}Z_{D}Z_{D}$ coupling emerges from the $D_{\mu}\phi^{\dagger} D^{\mu}\phi$ term, when one of the singlet Higgs field receives a vacuum 
expectation value. The covariant derivative contains a term $ \propto g_{D} Z_{D}$, and the $\phi$ field yields a term proportional to $\sin\theta_{\rm mix}\, H_{2}$, resulting in the $H_{2}Z_{D}Z_{D}$ coupling to become proportional to $g_{D}^{2}\times\sin\theta_{\rm mix}$. Another contribution arises from the SM term ${D_{\mu}H}^{\dagger}D^{\mu}H$ through $Z - Z_{D}$ mixing. However, this term is proportional to $\epsilon_{g}^{2}$ and since we have restricted ourselves to small values of $\epsilon_{g}$, contributions from this term can be safely ignored. The $H_{2}H_{1}H_{1}$ coupling manifests from the quadratic Higgs mixing term in the scalar potential, $\dfrac{\lambda_{mix}}{4} (H^{+}H)(\phi^{\dagger}\phi)$, and is therefore, proportional to $\lambda_{mix} v_{H}$. It can be observed from Fig~\ref{fig:signal_strength}~(left) that the current constraints from Higgs signal strength measurements allow the SM-like Higgs ($H_{2}$) to have $\lesssim 15\%$ of non-SM branching fraction. In the low $\epsilon_{g}$ limit, ratio of $Br(H_{2} \to H_{1} H_{1})$ to $Br(H_{2} \to Z_{D}Z_{D})$ is directly proportional to $M_{Z_{D}}$, and the same can be visualized in Fig.~\ref{fig:signal_strength}~(left), where the color palette represents the value of $M_{Z_{D}}$. 

In Fig.~\ref{fig:signal_strength}~(right), we show the parameter space points allowed by the global $\chi^{2}$ analysis (color palette points) in the $M_{H_{1}} - \sin\theta_{\rm mix}$ plane, while the points excluded by the same have been shown in grey color. The generic parameter space of Eqn.~\ref{eqn:par_space} has been represented in Fig.~\ref{fig:signal_strength}~(right). Within this scenario, $H_{2}$ can also decay into the additional non-SM decay mode : $H_{2} \to DM~DM$, depending on the values of $f$ and $M$. It can be visualized from Fig.~\ref{fig:signal_strength}~(right) that the signal strength constraints wield an approximately flat exclusion on $\sin{\theta_{\rm mix}} \gtrsim 0.1$ over the entire range of $H_{1}$ mass. Hence limits on $\sin\theta_{\rm mix}$ from LHC signal strength measurements are two times stronger compared to the results from LEP (shown in Fig.~\ref{fig:LEP}).

\subsection{Constraints from LHC direct searches} 
\label{sec:LHC_direct_constraints}

\begin{figure}
\begin{center}
\includegraphics[scale=0.20]{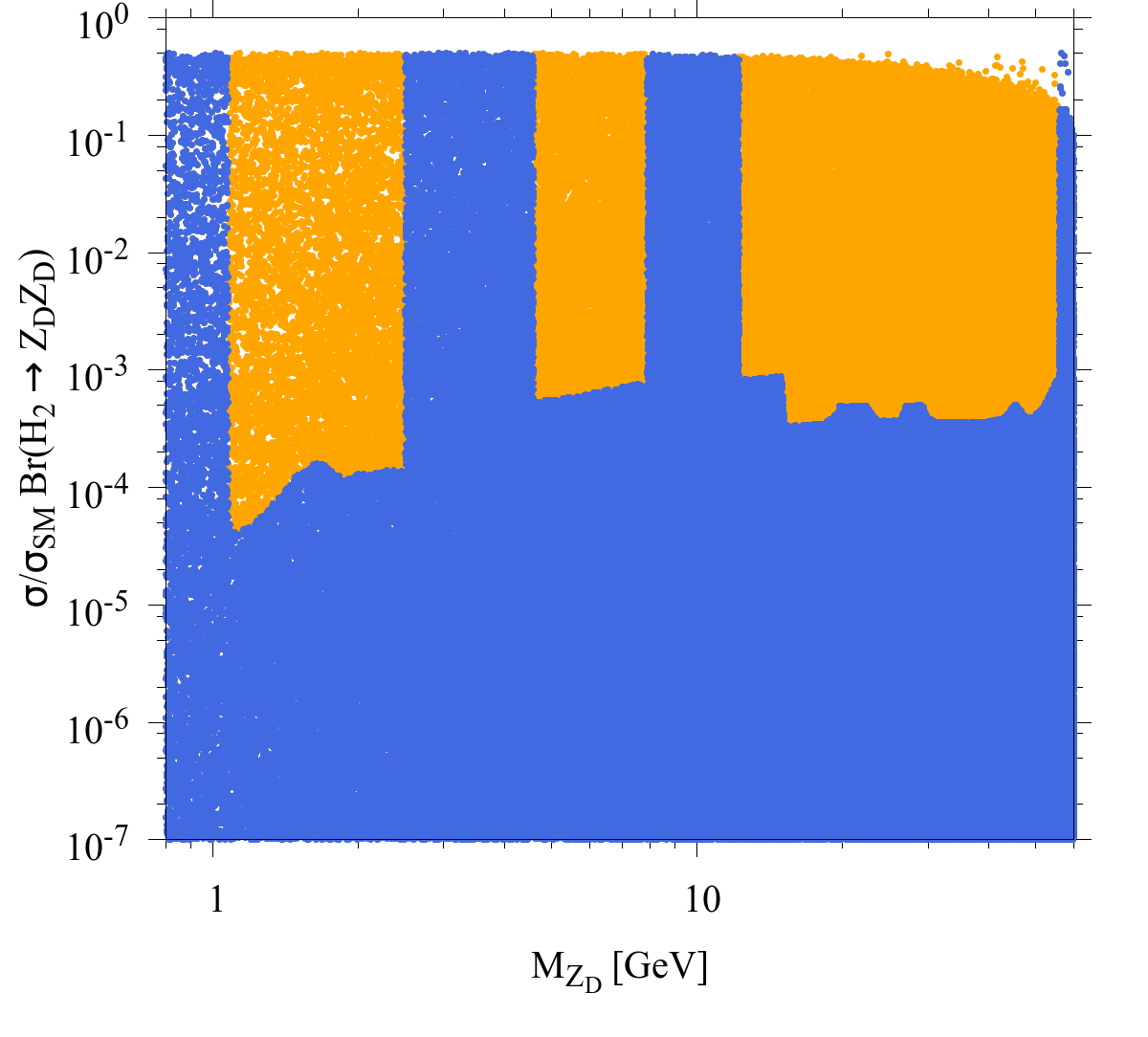}
\caption{Scatter plot in the $M_{Z_{D}}-\sigma/\sigma_{SM} \times Br(H_{2} \to Z_{D}Z_{D})$ plane exhibiting the implications from application of upper limits derived by LHC from search in the $H_{2} \to Z_{D}Z_{D} \to 4l$ channel~\cite{Aaboud:2018fvk}. All parameter space points in this figure satisfy the Higgs mass constraints. The blue colored points are the ones which are still allowed by the $H_{2} \to Z_{D}Z_{D}$ search limits. }
\label{fig:LHC_zdzd}
\end{center}
\end{figure}

ATLAS and CMS have performed different searches for the Higgs boson with a mass of 125 GeV decaying 
into two spin-zero particles, $H_2 \to AA (SS)$, in various final state using Run-I and Run-II datasets. The ATLAS collaboration has analysed the $4l~(l=e,\mu)$ final state originating from the decay of $125~{\rm GeV}$ Higgs boson via an intermediate $ZZ_{D}$, $Z_{D}Z_{D}$ and $AA$ pair production using the $13~{\rm TeV}$ dataset collected at $\lum \sim 36.1~\ifb$ \cite{Aaboud:2018fvk}. This search probed the mass range of $1~{\rm GeV} < M_{A} < 60~{\rm GeV}$. The same dataset was also used by ATLAS to probe the $2b2\mu$ final state produced via $H_{SM} \to SS \to 2b2\mu$~\cite{Aaboud:2018esj} and $4b$ final state originating via $H_{SM} \to SS \to 4b$~\cite{Aaboud:2018iil}, where S is a spin-zero boson, in the mass range $18~{\rm GeV} < M_{S} < 62~{\rm GeV}$. The CMS collaboration has also searched the $4\mu$ final state produced via $H_{\rm SM} \to SS \to 4\mu$, and derived upper limits in the mass range $0.25~{\rm GeV} < M_{S} < 8.5~{\rm GeV}$. This search used the $13~{\rm TeV}$ dataset collected at $\lum \sim 35.9~\ifb$~\cite{CMS-PAS-HIG-18-003}. The same dataset has also been used by CMS to search for exotic decays of the Higgs boson to a pair of light $A$ in the $2b2\tau$~\cite{Sirunyan:2018pzn} and $2\mu 2\tau$~\cite{Sirunyan:2018mbx} final state, focussing on the mass range $15~{\rm GeV} < M_{A} < 62~{\rm GeV}$. ATLAS and CMS have also performed similar searches using the LHC Run-I dataset for a multitude of final states : $4\tau$ \cite{Khachatryan:2017mnf}, $2\mu 2b$ \cite{Khachatryan:2017mnf}, $2\mu 2 \tau$ \cite{Khachatryan:2017mnf,Aad:2015oqa}, $4\mu$ \cite{Aad:2015sva,Khachatryan:2015wka}, $4\tau$ \cite{Khachatryan:2015nba,CMS-PAS-HIG-14-022}, $2\tau 2b$ \cite{CMS-PAS-HIG-14-041}.



The ATLAS collaboration has searched for the dark vector boson ($Z_{D}$) in the mass range of $1~{\rm GeV} \lesssim M_{Z_{D}} \lesssim 60~{\rm GeV}$, where $Z_{D}$ 
can be produced as pair ($Z_{D}Z_{D}$) or in association with SM $Z$ boson and eventually the $Z_{D}Z_{D}/ZZ_{D}$ pair decays into $4l(l=e,\mu)$ final state~\cite{Aaboud:2018fvk}. Upper limits were obtained on the quantity $\sigma/\sigma_{SM} \times Br(H_{2} \to ZZ_{D})$ or $\sigma/\sigma_{SM} \times Br(H_{2} \to Z_{D}Z_{D})$, where $\sigma$ and $\sigma_{SM}$ are the production cross-sections of the SM-like Higgs boson in the NP and SM scenarios. 
It is to be noted that the limits obtained in \cite{Aaboud:2018fvk} assume $Br(Z_{D} \to e^{+}e^{-}) \sim 50\%$ and $Br(Z_{D} \to \mu^{+}\mu^{-}) \sim 50\%$, resulting in $4l = 4e(25\%),~2e2\mu(50\%),~4\mu(50\%)$, and therefore, a correct scaling is required while evaluating the implication of these constraints on the NP model under consideration. 
The decay processes : $H_{2} \to ZZ_{D}$ and $H_2 \to Z_{D}Z_{D}$, depend on the kinetic mixing factor ($\epsilon_{g}$), and are independent of the mixing between the Higgs doublet from the SM and the singlet Higgs from the dark sector. As a result, these decay channels serve an excellent probe of $\epsilon_{g}$. However, in the case of $H_{2} \to ZZ_{D} \to 4l$ search channel, the SM $H_{SM} \to ZZ^{*} \to 4l$ process offers a strong irreducible background, and eventually dilutes the resolution between the signal and background, rendering these search limits insensitive to the low $\epsilon_{g}$ region which is relevant to our parameter space. On the other hand, the search channel, $H_{2} \to Z_{D} Z_{D} \to 4l$ stands on an advantageous ground owing to the possibility of application of SM Z-boson veto, which significantly improves the signal sensitivity as compared to the earlier case. We show the implications from the current $Z_{D}$ limits from $H_{2} \to Z_{D}Z_{D}$ searches (from \cite{Aaboud:2018fvk}) on our parameter space in Fig.~\ref{fig:LHC_zdzd}. The vertical axis corresponds to the $\sigma/\sigma_{SM} \times Br(H_{2} \to Z_{D}Z_{D})$ and the horizontal axis corresponds to $M_{Z_{D}}$. The yellow colored points in Fig.~\ref{fig:LHC_zdzd} are excluded by the current direct search constraints  while the blue colored points are still allowed.

\subsection{Constraints from B-factories and beam dump experiments}
\label{sec:beam_dump_current}
The LHCb collaboration~\cite{Aaij:2017rft} has derived upper limits on the kinetic mixing factor at 90$\%$ C.L., covering $214~{\rm MeV}\lesssim  M_{Z_{D}} \lesssim 70~{\rm GeV}$, using the LHC data set collected at an integrated luminosity of $1.6~ fb^{-1}$ at $\sqrt{s}=13~{\rm TeV}$. For $M_{Z_{D}} \gtrsim 10~{\rm GeV}$, this search offers the strongest upper limits on $\epsilon_{g}$, among all other contemporary dark photon experiments, and excludes $\epsilon_{g} \gtrsim 10^{-3}$ for $M_{Z_{D}}\sim 10.6~{\rm GeV}$. The limit becomes slightly weaker towards higher $M_{Z_{D}}$ and excludes $\epsilon_{g} \gtrsim 2\times 10^{-2}$ at $M_{Z_{D}} \sim 70~{\rm GeV}$. At $M_{Z_{D}}$ below $10~{\rm GeV}$, the most stringent constraints are offered by BaBar~\cite{Lees:2017lec}, which exclude $\epsilon_{g} \gtrsim 10^{-3}$ in the mass range $0.25~{\rm GeV} < M_{Z_{D}} < 10~{\rm GeV}$. BaBar performed the search in the $ e^{-}e^{+} \to {Z_{D}} \gamma$ channel, while assuming that the ${Z_{D}}$ predominantly decays invisibly. 

In the current analysis, our scanned set of parameter space points have $M_{Z_{D}}$ in between $2~{\rm GeV}$ and $60~{\rm GeV}$, while $\epsilon_{g}$ has been scanned up to $10^{-4}$. Thus, the current constraints on $\epsilon_{g}$ derived from dark vector boson searches do not affect our parameter space. In addition to the $Z_{D}$ searches by LHCb and BaBar, there are numerous beam dump experiments which have also probed a light vector gauge boson. However, the searches by the beam dump experiments mostly concentrate in the mass range of the order $O(MeV)$. Some of these experiments are KLOE~\cite{ARCHILLI2012251},  MAMI~\cite{Merkel:2014avp}, NA-64~\cite{Banerjee:2017hhz}, E141~\cite{PhysRevLett.59.755}, E774~\cite{PhysRevLett.67.2942}, KEK~\cite{PhysRevLett.57.659}, HADES~\cite{Agakishiev:2013fwl}, and MiniBooNE~\cite{Aguilar-Arevalo:2017mqx}. Some of the upcoming experiments like DarkLight~\cite{Balewski:2014pxa} and APEX~\cite{Beacham:2013ka}, are expected to improve upon the sensitivity to $\epsilon_{g}$ by $\sim 2$ orders of magnitude.

\begin{figure}
\begin{center}
\includegraphics[scale=0.3]{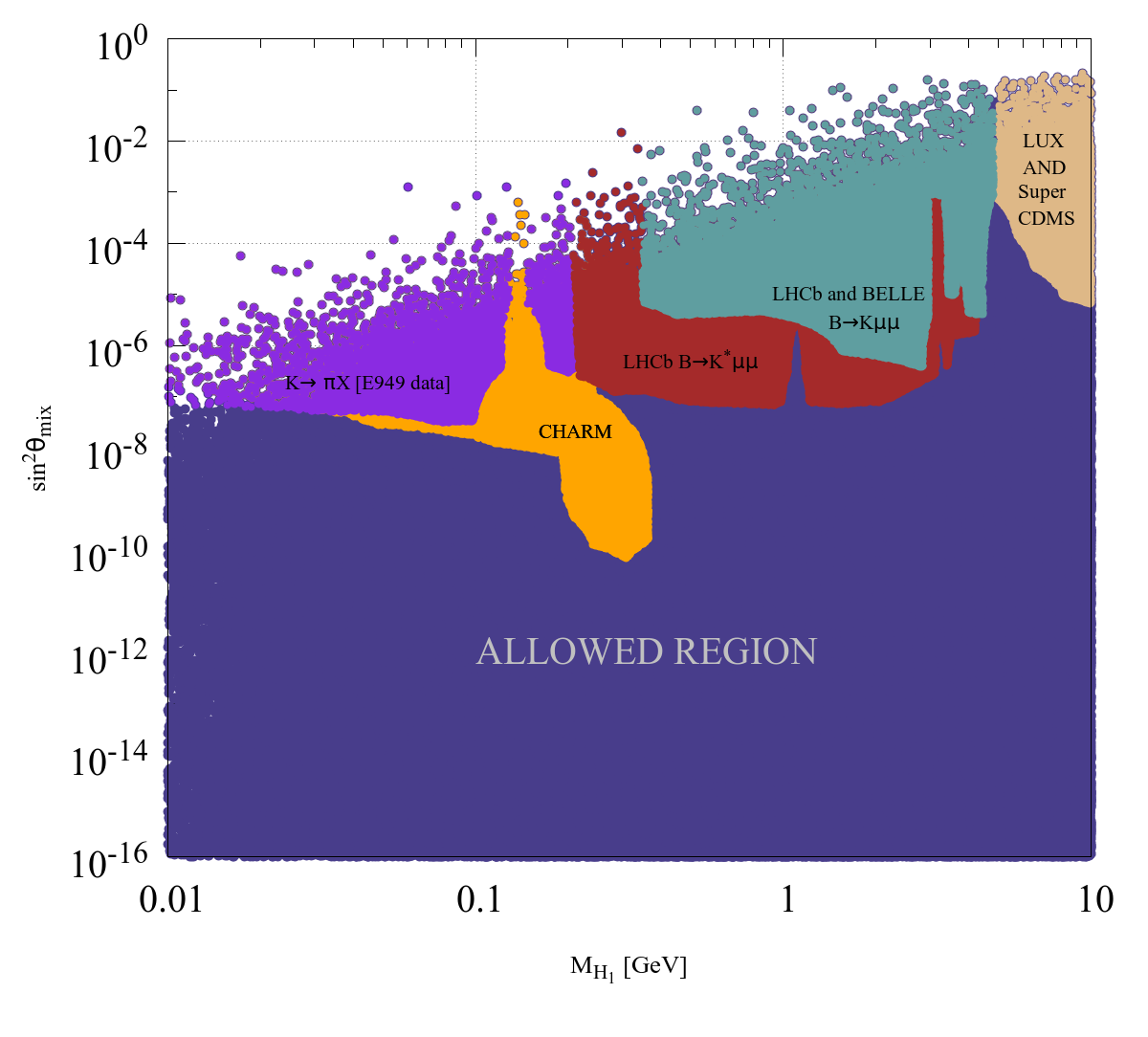}
\caption{Summary of parameter space region excluded by the various beam dump~\cite{Artamonov:2009sz,Bergsma:159811,Agnese:2014aze} and flavor physics experiments~\cite{Aaij:2015tna,Wei:2009zv,Chen:2007zk}.}
\label{fig:flav_beamdump_current}
\end{center}
\end{figure}

Light scalar boson searches in B-factory and beam dump experiments also yield exclusion contours on $\sin\theta_{mix}$ as a function of $M_{H_{1}}$. The E949 experiment~\cite{Artamonov:2009sz} probed the kaon decay process, $K^{+}\to \pi^{+}\nu\bar{\nu}$, in the pion momentum range $140~{\rm MeV} < p_{\pi} < 199~{\rm MeV}$. These search limits have been translated to the $M_{H_{1}}-\sin^{2}{\theta_{mix}}$ plane in \cite{Lanfranchi:2243034} by re-interpreting the analysis scheme of \cite{Flacke:2016szy}. In the context of our analysis, we use the corresponding exclusion contour shown in Fig.~1 of \cite{Lanfranchi:2243034} and show the excluded parameter space points in purple color in Fig.~\ref{fig:flav_beamdump_current}. The CHARM collaboration ~\cite{Bergsma:159811} performed a search for axion like particles using a $400~{\rm GeV}$ proton beam from CERN-SPS dumped onto a copper target. The corresponding limits have also been been translated and presented as exclusion contours in the $M_{H_{1}}-\sin{\theta_{mix}}$ plane in \cite{Lanfranchi:2243034}, which we directly use in our analysis, and the excluded parameter space points have been shown in yellow color in Fig.~\ref{fig:flav_beamdump_current}. We would like to note that the sensitivities of CHARM and E949 experiments overlap in the $M_{H_{1}} \lesssim 250~{\rm MeV}$ region with E949 exerting more stringent constraints below $M_{H_{1}} \lesssim 40~{\rm MeV}$. Results from the search for weakly interacting massive particles by the SuperCDMS collaboration~\cite{Agnese:2014aze} has also been extracted from Fig.~1 of \cite{Lanfranchi:2243034}, and excludes the parameter space region corresponding to $M_{H_{1}} \gtrsim 5~{\rm GeV}$ and $\sin^{2}\theta_{mix} \gtrsim 10^{-5}$. These excluded points have been shown in light brown color in Fig.~\ref{fig:flav_beamdump_current}. The B-factories exert the strongest constraints on $\sin^{2}\theta_{mix}$ in the intermediate light Higgs mass region, $400~{\rm MeV} \lesssim M_{H_{1}} \lesssim 5~{\rm GeV}$. The search for $H_{1}$ performed by LHCb collaboration~\cite{Aaij:2015tna} in the decay process : $B^{0} \to K^{+}\pi^{-}H_{1}$, with the $H_{1}$ eventually decaying into a di-muon pair, excludes $\sin^{2}\theta_{mix} \gtrsim 10^{-7}$ in the mass range of $214~{\rm MeV} \lesssim M_{H_{1}} \lesssim 4~{\rm GeV}$. The corresponding exclusion contour has been taken from Fig.~1 of \cite{Lanfranchi:2243034} and the excluded parameter space points have been shown in red color in Fig.~\ref{fig:flav_beamdump_current}. The measurements in $B \to K H_{1}$ channel by BELLE~\cite{Wei:2009zv,Chen:2007zk} and LHCb have also been translated into limits on $\sin^{2}\theta_{mix}$ in \cite{Lanfranchi:2243034}. The parameter space points excluded by those are shown in sky blue color in Fig.~\ref{fig:flav_beamdump_current}. It can be observed from Fig.~\ref{fig:flav_beamdump_current} that $\sin\theta_{mix}$ values above $\sim 3\times 10^{-3}$ are roughly excluded over the $M_{H_{1}} \lesssim 5~{\rm GeV}$ region. The CHARM sensitivity extends out till $\sin\theta_{mix} \gtrsim 10^{-5}$ over a small range of $M_{H_{1}}$, $200~{\rm MeV} \lesssim M_{H_{1}} \lesssim 400~{\rm MeV}$. The parameter space points which are still allowed by the flavor physics and beam dump experiments have been shown in deep blue color.

\section{Dark Matter Aspects} 
\label{sec:DM_aspect}
 
Having constrained a substantial part of the parameter space from flavour physics, beam dump and collider experiments, we now turn to dark matter phenomenology, which, as we shall see, will help us to put more stringent limits on our model thereby enhancing its predictability.
  
\subsection{Prospects of Direct Detection}
The direct detection experiments pose severe constraints on 
interactions of DM with nucleons. In the present context, we 
will assume $H_1$ to be very light, as required to enhance the 
self-interaction cross-section of the DM. Further, scenarios 
with both heavy DM and light DM ($\chi_-$) will be discussed.   

Note that, the contribution from only $H_1, H_2$ mediated 
processes are important since $Z_D$ couples to 
$\chi_{-}-\chi_{+}$. Thus, if $\delta M = M_{+}-M{-}$ is greater 
than $\dfrac{1}{2} M_{-} v_{\rm esc}^2$ (for $M_- \simeq 
\mathcal{O}(10)$ GeV, $\delta M \simeq \mathcal{O}(100)$ keV), 
where $v_{\rm esc} \simeq 544 ~{\rm km/s}$ denotes the escape 
velocity of our galaxy, the incoming DM particle ($\chi_{-}$) 
would not have enough kinetic energy to excite the heavier state, 
leading to kinematic suppression of the $Z_D$ mediated $t$-channel 
process.

\begin{figure}[htb]
	\hspace*{-1.25cm}                                                           
	\includegraphics[scale=0.21]{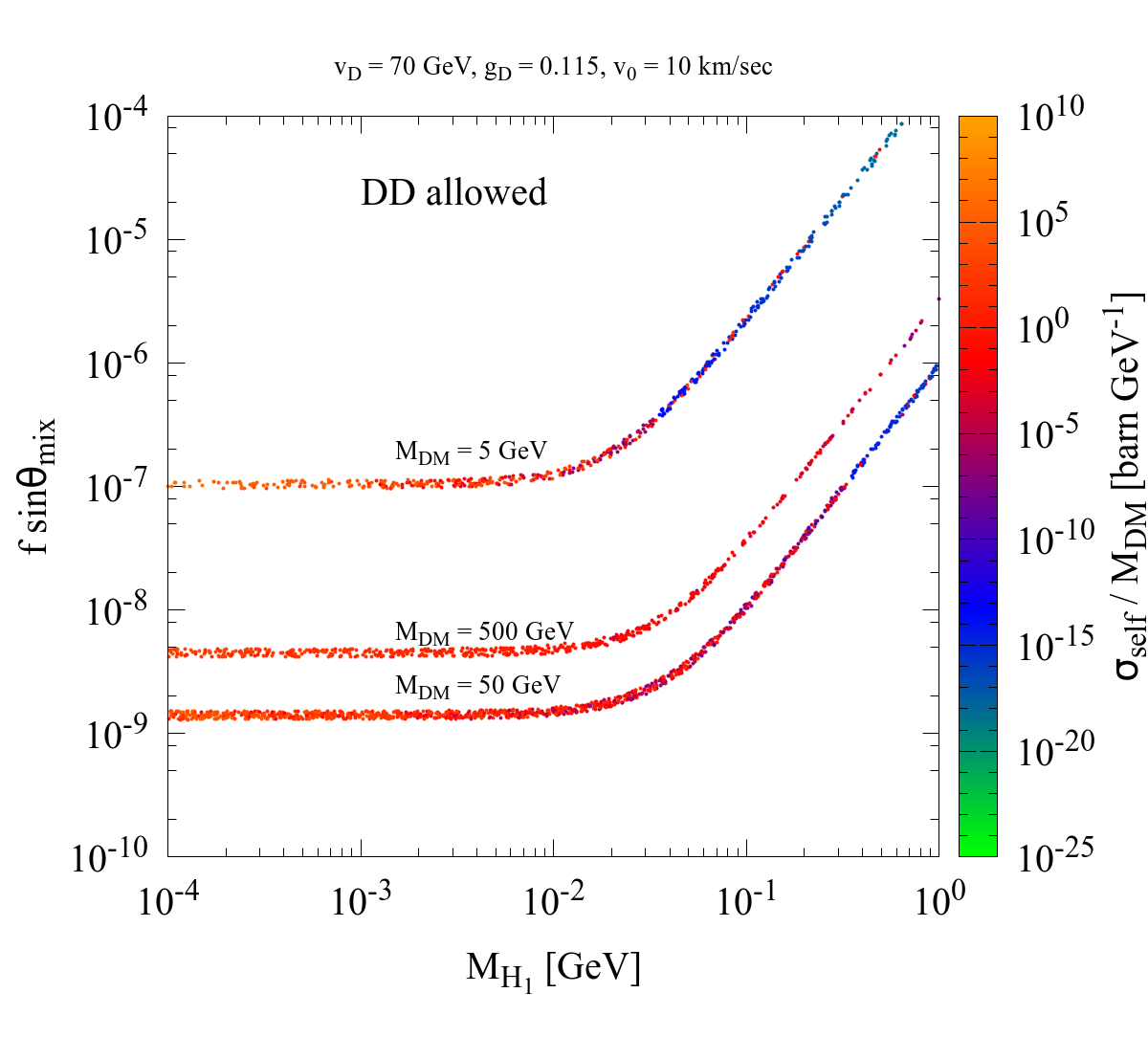}~
	\includegraphics[scale=0.21]{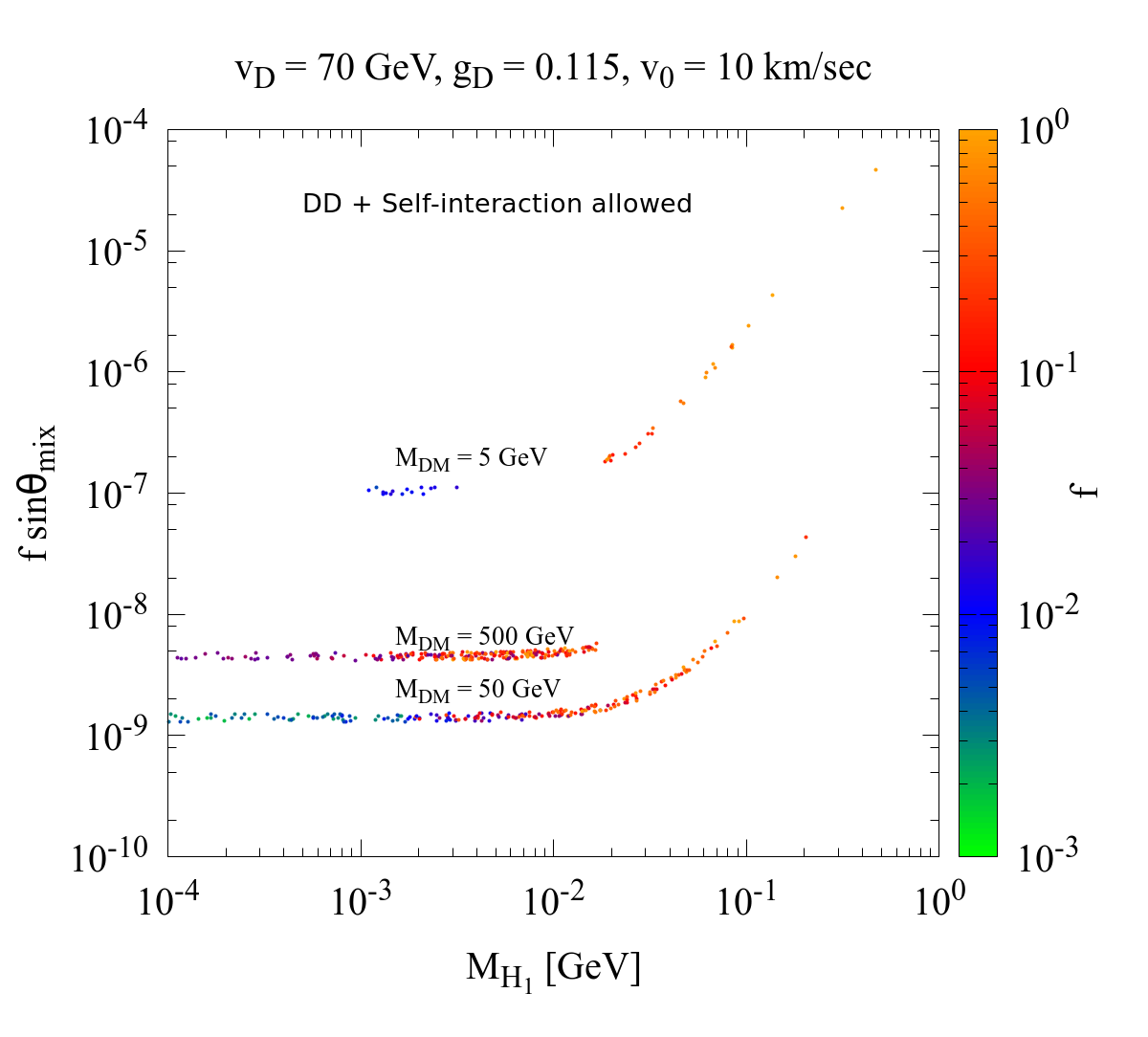}
	\caption{Constraints from direct detection experiment \texttt{Xenon-1T} \cite{Aprile:2017iyp} on $f\, \sin\theta_{\rm mix}$ as a function of dark scalar mass. The degree of self interaction is shown in the colour bar on its side ({\bf left}). Same plot but the plotted points now also allow for a sizable self-interaction. The relevant range of variation of $f$ is shown in the colour palette ({\bf right}).}
	\label{ddcons}
\end{figure}

The differential event-rate at a detector, as a function of the 
nuclear recoil energy $E_R$, is given by, 
\beq
\dfrac{d R}{dE_R} = n_T \dfrac{\rho_{\chi_-}}{M_{-}} \int_{v_{\rm min}} d^3 v f_E(\vec{v}) v \dfrac{d \sigma(v, E_R)}{d E_R}
\eeq
where, $n_T$ is the number of target nuclei in the detector 
material, $\rho_{\chi_-}$ is the local density of DM halo ($\simeq 
0.3 {\rm GeV} {\rm cm}^{-3}$) and $ \dfrac{d \sigma(v, E_R)}{d E_R}$ 
is the scattering cross-section with a nucleus. Further, 
$f_E(\vec{v})$ denotes the velocity distribution of the DM with 
respect to earth and can be related to the velocity 
distribution $f({\vec v})$ of DM in the galactic halo as 
$f_E(\vec{v})=f({\vec v}+
{\vec v}_E)$ where ${\vec v}_E$ denotes the velocity of earth in the 
galactic rest frame. We will assume $f({\vec v})$ to be a Maxwell-
Boltzmann distribution with velocity dispersion $v_0 = 220$ km/s 
and a cut-off set to $v_{\rm esc}$. The minimum velocity 
$v_{\rm min}$ corresponding to a recoil energy $E_R$ to the 
target nucleus is given by, 
\beq
v_{\rm min}(E_R) = \dfrac{m_T  E_R}{2 \mu^2}, 
\eeq
where $m_T$ denotes the mass of the target nucleus and $\mu = m_T 
M_{DM} (m_T +M_{DM})^{-1}$ denotes the reduced mass of the nucleus-DM. 
The interaction with a nucleus with atomic number $A$ and charge 
$Z$, then, is given by, 
\beq
\dfrac{d \sigma (v, E_R)}{dE_R} = \dfrac{m_T}{2 \mu^2 v^2}
\sigma_{\rm SI}^T F^2(2 m_T E_R)
\label{eq:dsigmadd}
\eeq
where $\sigma_{\rm SI}^T =\dfrac{4 \mu^2}{\pi}[Z f_p + (A-Z)f_n]^2$ 
is the $\chi_-$ nucleus cross-section at zero momentum transfer. 
Also, $f_p$ and $f_n$ denotes couplings with $p$ (proton) and $n$ (neutron) respectively. 
We have
\begin{equation}
f_{N}= m_{N} \left( \overset{u,d,s}{\underset{q}{\Sigma}} f^{N}_{q} \frac{\lambda_q}{m_q} + 
\frac{2}{27} \overset{c,b,t}{\underset{Q}{\Sigma}} f_{G} \frac{\lambda_Q}{m_Q} \right); 
N \in \{p,n\}. 
\end{equation}
In the above expression $\lambda_q$ denotes the effective coupling of $\chi_-$ 
with the quark $q$ in the limit of small momentum transfer, and is given by $ f y_q \sin\theta_{\rm mix} \cos\theta_{\rm mix} \left(\dfrac{1}{M_{H_1}^2}- \dfrac{1}{M_{H_2}^2}\right)$, where $y_q$ denotes the Yukawa coupling for quark $q$. $f^{N}_{q}$ denotes the contribution of quark 
$q$ to the mass $m_N$ of nucleon $N$. While the light quarks contribute to the nucleon 
masses directly, the heavy quark contributions to $f^N$ appears through the loop-induced 
interactions with gluons. These are given by, 
\begin{equation}
f^N_q = \frac{1}{m_N} \langle N |m_q \bar{q}{q}| N\rangle,
~f_G = 1 - \overset{u,d,s}{\Sigma}f^{N}_{q}.
\end{equation}
Note that, since $H_1$ and $H_2$ mediated $t$-channel processes contribute, the $s$ 
quark content of the nucleon can be of significant importance. Following
\texttt{micrOMEGAs4.3.5} \cite{Belanger:2014hqa}, we have used $\sigma_{\pi N} = 34 ~{\rm MeV}$ and $\sigma_s 
= 42 ~{\rm MeV}$ to determine the quark contents of the nucleon. Further, $F(q)$ 
denotes the nuclear form factor corresponding to a momentum transfer $q$. 
However, since we are interested in the presence of a very light $H_1$, as 
required for the Sommerfeld enhancement of the self-interaction cross-section, 
the mediator mass can be comparable to or even smaller than the typical momentum 
transfer of $\mathcal{O}(100) \, {\rm keV}$ (for $\sim \mathcal{O}(100) ~{\rm GeV}$ DM).

In such cases, use of $\sigma_{\rm SI}^T$, as described above, 
overestimates the direct detection constraint on $f \sin\theta_{\rm mix}$. In order to 
account for the same, we have introduced a factor $\dfrac{M_{H_1}^4}{(q^2+M_{H_1}^2)^2}$ 
and multiplied the same to the usual Helm form factor ($F(k)$ in eq \ref{eq:dsigmadd}). 
For $k_{max} \ll M_{H_1}$, this additional term simply reduces to 1, while for 
$k \gtrsim M_{H_1}$ this ensures the correct momentum dependence of the DM-nucleus 
interaction cross-section. Note that depending on the detector threshold, the minimum 
momentum transfer $q_{\rm min}$ is different, while the maximum possible momentum 
transfer $q_{\rm max}$ is set by $v_{\rm esc}$. Typically $q_{min}$ falls well below 1 MeV, 
and assuming $H_1$ thermalizes with the SM particles, $q_{min} \gg M_{H_1}$ would 
contribute to the additional relativistic {\it d.o.f} during BBN and would be 
in tension with the relevant observations.

We have modified the publicly available code \texttt{DDCalc} \cite{Workgroup:2017lvb} to incorporate the 
above.  and computed the constraints on $f \sin\theta_{\rm mix}$ from direct detection experiments. 
{\it   }This is shown in Fig.\,\ref{ddcons} (left) along with the self-interaction cross section in the colour bar. The range of $f$ used in the scan is from $10^{-5}$ to $1$. On the right hand side, we show the same plot but this time only the points with strong self interaction is plotted. The suitable range of $f$ corresponding to this is shown in the adjacent colour bar. The direct detection constraints becomes more and more rigid as we increase the dark matter mass being most tightly constraining at $M_{DM} \sim 50$ GeV. Then the bound weakens gradually as we increase the mass. This fact is also reflected in the figure through the arrangement of the different exclusion lines. The line corresponding to the 50 GeV dark matter rules out the largest volume of the parameter space, whereas the 500 GeV dark matter corresponds to a looser constraint than 50 GeV (but tighter than 5 GeV) as expected.

The improved technique of calculation of direct detection constraints opens up a large part of the parameter space as opposed to the conventional calculation. A comparison between the two methods in shown in Fig.\,\,\ref{fig:ddcomp}. 
\begin{figure}[htb]
	\hspace*{-1.25cm}                                                           
	\includegraphics[scale=0.21]{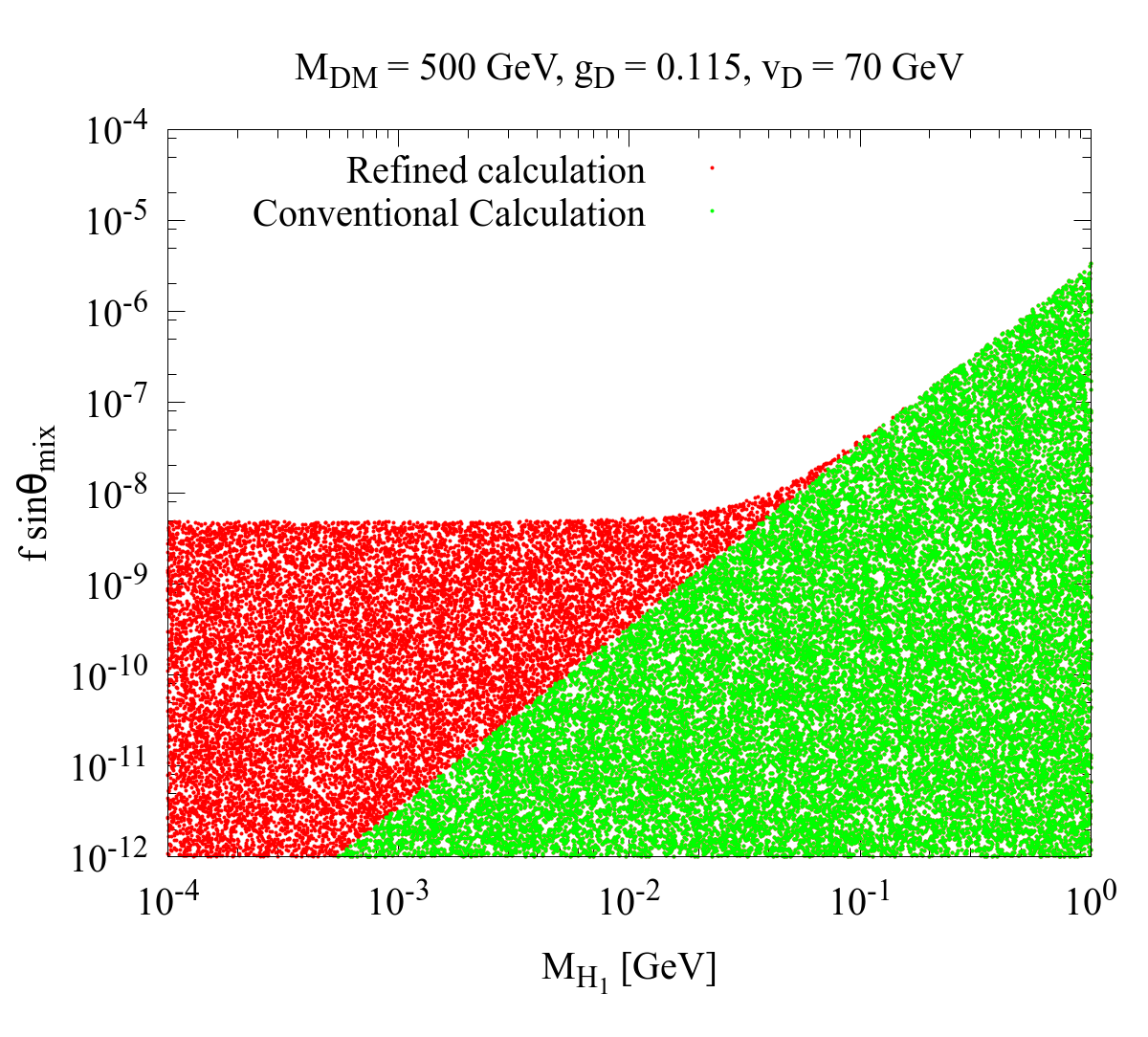}~
	\includegraphics[scale=0.21]{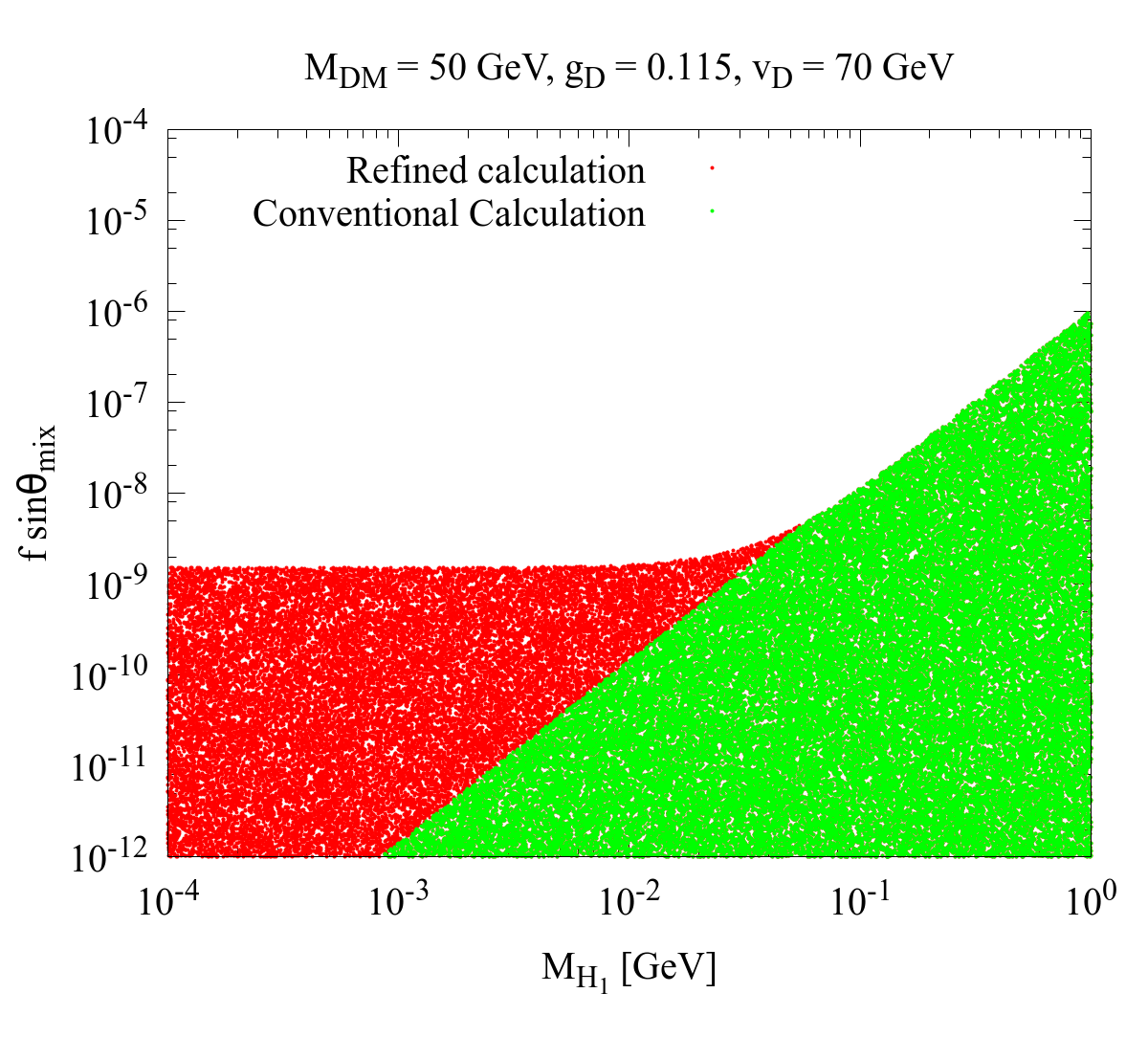}
	\caption{Comparison of calculation of constraints from direct detection experiment \texttt{Xenon-1T} \cite{Aprile:2017iyp} done using the conventional and refined calculation for a 500 GeV dark matter ({\bf left}). Same plot but with a 50 GeV dark matter. Note the slight shift of the point where our refined calculation meets the conventional calculation in this plot when compared with the neighbouring plot. The shift is towards the left for a lighter dark matter as expected ({\bf right}).}
	\label{fig:ddcomp}
\end{figure}
In case of low mass DM ($2 \,{\rm GeV} \lesssim M_{DM} \lesssim 4 \, {\rm GeV}$) 
\texttt{CDMSLite} \cite{Agnese:2017jvy} puts the dominant constraint \footnote {For even lower masses \texttt{CRESST-III} \cite{Angloher:2018fcs} has better sensitivity, however, we have not explored the region $M_{DM} \lesssim 1$ GeV.}. For $M_{DM} \gtrsim 5$ GeV onwards 
\texttt{Xenon-1T} \cite{Aprile:2017iyp} puts forward the most stringent limits \footnote{Note that \texttt{Lux} \cite{Akerib:2016vxi} and \texttt{Panda-II} \cite{Cui:2017nnn} also lead to comparable constraint as \texttt{Xenon-1T} \cite{Aprile:2017iyp}.}. 

Before moving forward some important points are in order. Firstly, till now we were discussing about direct detection experiments which measures nuclear recoil when a dark matter particle hits them. Unfortunately, if dark matter mass goes down below ($\mathcal{O}(300)$ MeV) then nuclear recoils are not an effective method to detect dark matter since recoils energies become pretty low. The effective method is these low mass regions is to measure electron recoils. Experiments like \texttt{SENSEI} \cite{Crisler:2018gci}, \texttt{XENON-10} \cite{2011APh....34..679A}, \texttt{XENON-100} \cite{Aprile:2016swn}, \texttt{SuperCDMS} \cite{Agnese:2018col} and \texttt{DarkSide-50} \cite{Agnes:2018ves} have measured dark matter electron scattering cross sections and have put constraints on the latter. As mentioned in \cite{Essig:2017kqs}, the upper bound on dark matter electron cross section on a $\mathcal{O}(100)$ MeV dark matter is $\sim 10^{-38}$ cm$^2$ (with form factor F$_{\rm DM}$ set to unity). The dark matter electron cross section in our model goes as $\sim f^2\,y_e^2\,\sin^{2}2\theta_{\rm mix}$, where $y_e$ is the electron Yukawa coupling and hence expected to be quite small. For the case of a light mediator (but $M_{H_1} > 10$ MeV to avoid BBN constraints), the dominant contribution to the dark matter electron cross section comes from the t-channel process. For $M_{DM} \lesssim 1$ GeV but greater that $M_{H_1}$ and $M_e$, the DM-electron cross section at small center of mass momentum is approximately given by :
\begin{eqnarray}
\sigma_e &\simeq& f^2 y_e^2 \sin^2\theta_{\rm mix}\biggl(8\ln\left(\dfrac{4M_{H_1}^2}{4M_{H_1}^2+9 M_{DM}^2}\right)\bigl(M_{H_1}^2 M_{DM}^2 - 18 M_{DM}^4 + 4M_{H_1}^4 \bigr) \nonumber \\
&+& 9\left(8M_{H_1}^2 M_{DM}^2 - 7M_{DM}^4\right)\biggr) \bigg/ \left(576\,\pi\, M_{DM}^4\left(4M_{H_1}^2+9 M_{DM}^2\right)\right)
\end{eqnarray}
For a typical $f$ that gives sizable self-interaction {\it i.e.} $f \sim \mathcal{O}(0.1)$ and $\sin\theta_{\rm mix} \sim 10^{-5}$, a 100 MeV dark matter the dark matter has a cross section of $\sim 5.6\times 10^{-50}$ cm$^{2}$ with electrons and hence much below the upper limit set by the DM-electron scattering experiments. 

On the other hand the cross section for an very light dark matter ($M_{DM} \sim 10$ keV, and consequently much lighter than the mediator as well as the electron) is given by the following expression :
\begin{eqnarray}
\sigma_e &\simeq& f^2 y_e^2 \sin^2 2\theta_{\rm mix}\biggl(2M_{H_1}^2\ln\left(\dfrac{4M_{H_1}^2}{M_{H_1}^2+4 M_{DM}^2}\right)+M_{DM}^2\bigl(\dfrac{4M_{H_1}^2}{M_{H_1}^2+4 M_{DM}^2}+4 \bigr) \biggr)\nonumber \\
 &\bigg/& \left(256\,\pi\, M_{DM}^4\right)
\end{eqnarray}
But such low masses are beyond the reach of DM-electron scattering experiments and no limits on this cross section exists as far as the direct detection experiments are concerned.

Secondly, we should be careful about the presence of the neutrino floor \cite{1307.5458}, going below which would render direct detection signals to be meaningless. Hence $f \sin\theta_{\rm mix}$ can not be lowered indefinitely. For a $50$ GeV dark matter and $\mathcal{O}(1)$ GeV $H_1$ mass, we found earlier that $f \sin\theta_{\rm mix} \lesssim 10^{-6}$. This $f \sin\theta_{\rm mix}$ can be further lowered roughly by $\sim 24.5$ times before it hits the neutrino floor for this dark matter and mediator mass.

\subsection{(Thermal) Relic density of dark matter} 
From our discussions in the previous sections, we find that to satisfy constraints from direct detection experiments and at the same time allowing for sizable self interactions, the required value of $\sin \theta_{\rm mix}$ is actually very small ($\simeq 10^{-5}-10^{-7}$, for a suitable $f$, while the latter is fixed from considerations of self-interactions). With such small values of mixing angle, the standard procedure for calculation of relic density (assuming prior thermalization and a subsequent freeze-out) is placed under scrutiny and demands some in depth analysis before proceeding further. In this work, we have already assumed that the kinetic mixing between $Z$ and $Z_{D}$ is very small (denoted by $\epsilon_{g}$). Hence, the scalar mixing angle ($\theta_{\rm mix}$) is the only possibility through which the dark sector can communicate with the standard model sector and eventually can come to equilibrium with the photon bath. The two parameters in our model that can control this thermalization effectively are $\lambda_{\rm mix}$ ($H_2 H_2 \leftrightarrow H_1 H_1$ type of interactions) and $\sin \theta_{\rm mix}$ ($Z Z \leftrightarrow Z_{D} Z_{D}$ type of interactions). Although $\sin\theta_{\rm mix}$ depends on $\lambda_{\rm mix}$, but considering only relevant values of the latter is not sufficient alone to give us an idea about the mixing angle required for thermalization. This is because the processes that depend solely on $\sin\theta_{\rm mix}$ (e.g. $Z Z \leftrightarrow Z_D Z_D$) can also play an important role in equilibrating the dark and standard model sectors. However, before EWSB ($T_\gamma^{\rm EWSB} \sim 153$ GeV), the scalar mixing angle had no significant role as such. Then $\lambda_{\rm mix}$ is the only relevant parameter that can bring the two systems to equilibrium. Before EWSB, at very high temperatures, the scalar masses can be neglected. The cross section of $H_2 H_2 \leftarrow H_1 H_1$ is given by $\sigma \sim \frac{\lambda_{\rm mix}^2}{32 \pi \, s}$. The two systems can thermalize if the rate of annihilation can exceed that of the Hubble expansion at some (sufficiently high) temperatures. Mathematically, the rate of annihilation is given by $\Gamma = n^{\rm eq} \langle \sigma v \rangle$, where $n^{\rm eq}$ is the equilibrium number density of the particle under consideration and $v$ is the relative velocity of the annihilating particles. If at some temperatures $T \gg T_\gamma^{\rm EWSB}$, $\Gamma / H \sim 1$, then we can safely conclude that the dark and the visible sectors did thermalize in the early universe. Here $H$ is the Hubble expansion rate and is given by $1.67\,\sqrt{g_\star} \frac{T^2}{ M_{\rm Pl}}$, where $g_\star$ is the degree of freedom and $M_{\rm Pl}$ is the Planck mass. In our discussion we have taken $T \sim 10^5$ GeV. We find, that for the two sectors to be in equilibrium at such temperatures $\lambda_{\rm mix} \gtrsim 10^{-5}$ (Fig.\,\ref{therm1}). 
\begin{figure}
		\hspace*{-0.05cm}                                                           
	\includegraphics[scale=0.9]{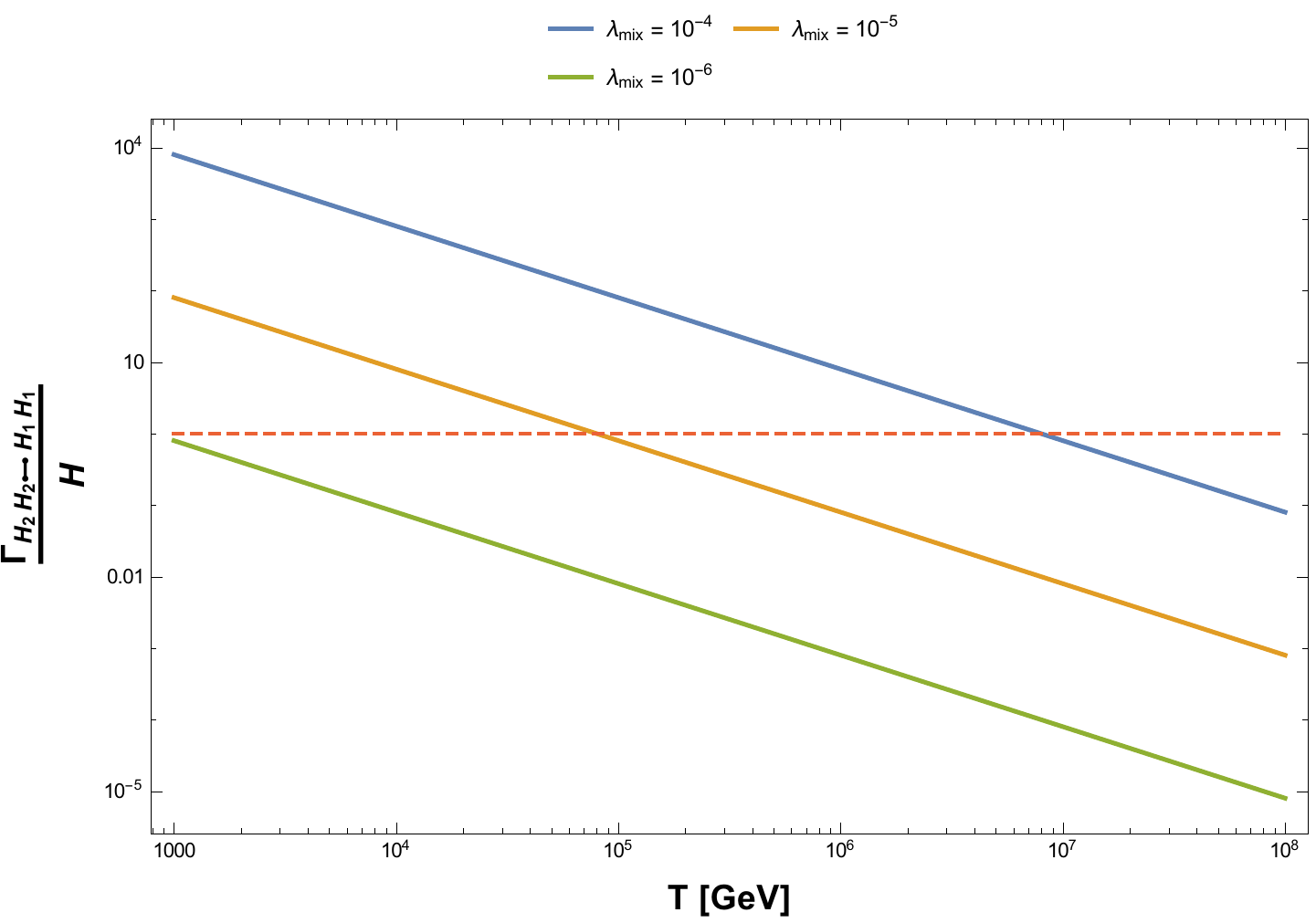}
	\caption{Comparison of the rate of $H_2 H_2 \leftrightarrow H_1 H_1$ vs the Hubble expansion rate in the early universe for different values of the quartic coupling $\lambda_{\rm mix}$. The red dashed line signifies the temperature at which they were equal. The dark and the standard model sectors were in thermal contact before this epoch.}
	\label{therm1}
\end{figure}
After EWSB, $H_2 H_2 \leftrightarrow H_1 H_1$ annihilations can compete with the Hubble rate if $\lambda_{\rm mix} \gtrsim 7 \times10^{-6}$. However not all interactions are dependant on $\lambda_{\rm mix}$ exclusively. For example $Z Z \leftrightarrow Z_D Z_D$ depends only on $\sin\theta_{\rm mix}$. So even if $\lambda_{\rm mix}$ is small, the dark and the standard model sector can come to equilibrium through the aforementioned interaction, and this is in turn can constrain $\sin\theta_{\rm mix}$. For temperatures $\sim 100$ GeV, we find for $\Gamma (ZZ \leftrightarrow Z_D Z_D) \gtrsim H$, $\sin\theta_{\rm mix} \gtrsim 7.5\times10^{-6}$. Hence for mixing angles below this value, the two sectors will fail to thermalize through this channel. Thus in the case where the dark and SM sectors fail to thermalize in the early universe ({\it i.e} both $\lambda_{\rm mix}$ and $\sin\theta_{\rm mix}$ are small) usual techniques of calculation of relic density by freeze-out mechanism will lead to incorrect results. However, as discussed above, keeping one of them large can in principle lead to successful thermalization of the two sectors.

In the previous section, we derived constraints on $f \sin\theta_{\rm mix}$ from direct detection experiments and plotted points which were simultaneously allowed by it and also can lead to sizeable self-interactions. The range of $f$ found from such considerations was $10^{-3} \lesssim f \lesssim 1$. This in turn constrains the scalar mixing angle (depending on the dark matter mass). But, we know,
\begin{equation}
\sin2\theta_{\rm mix} = \dfrac{\lambda_{\rm mix}\left(M_{\rm Z_D}/g_D\right)v_h}{M_{H_2}^2 - M_{H_1}^2}
\label{sint}
\end{equation}
Hence, we can try to visualize the part of the parameter space (in $M_{Z_D}-\lambda_{\rm mix}$ plane, say) that can give rise to a thermal relic. Two values of $f$ and two different dark matter masses were chosen for demonstration purpose. It is evident from Fig.\,\ref{therm2}, that with increasing $f$ the allowed region decreases for a given dark matter mass. On the whole, light dark matters are more favoured with respect to the heavier ones in the thermal scenario. The value of the dark gauge coupling $g_D$ was however arbitrarily fixed to 0.2 in these plots. If we lower $g_D$, the allowed parameter space decreases as expected. In principle, $g_D$ can be fixed from considerations of correct thermal relic (for {\it e.g} see Table.\,\ref{tablerelic}). The constraint of $f\, \sin\theta_{mix}$ from direct detection is dependent on the value of the mass of the light scalar. For conservative estimates we have taken the minimum value of $M_{H_1}$ allowed from BBN {\it i.e. $\sim \mathcal{O}(10)$} MeV. For higher values of the mediator mass more allowed regions are expected to open up.
\begin{figure}[htb]
	\hspace*{-0.7cm}                                                           
	\includegraphics[scale=0.41]{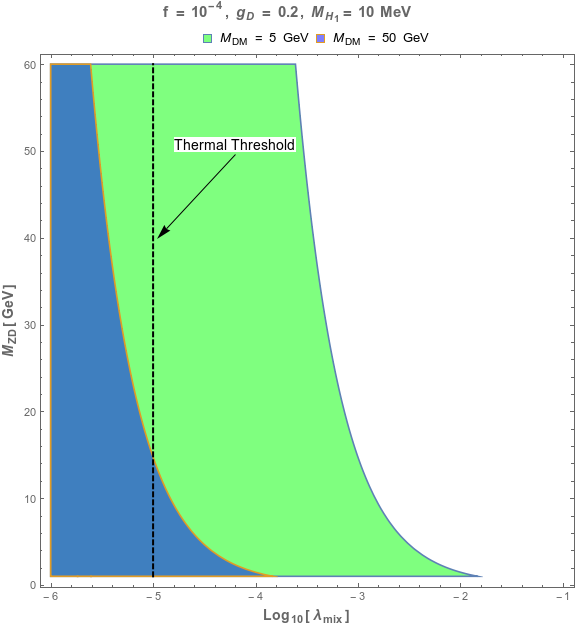}~
	\includegraphics[scale=0.41]{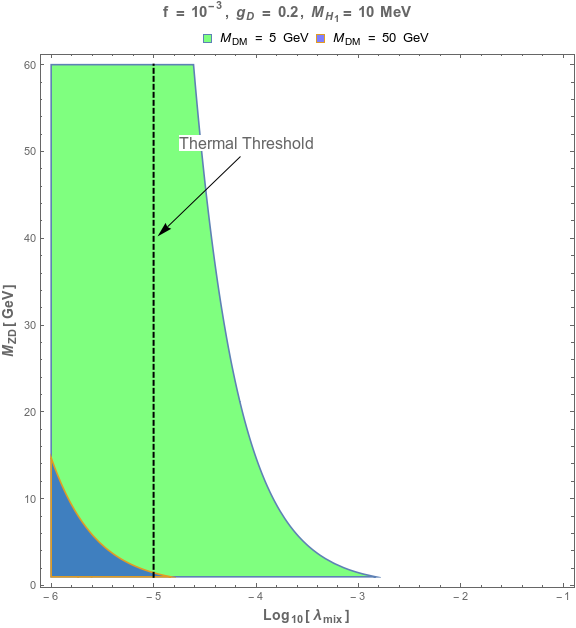}
	\caption{Plots showing the allowed region for accommodating a thermal dark matter in $M_{Z_D}-\lambda_{\rm mix}$ plane. The region on the right from the black dashed line is the region that can support thermal dark matter ($\lambda_{\rm mix} \gtrsim 10^{-5}$). }
	\label{therm2}
\end{figure} 

Finally, analogous to the plot in the previous section, we show the range of variation of relic density alongside that of $\lambda_{\rm mix}$ for the points allowed by direct detection experiments and at the same time giving rise to sizeable self-interaction in Fig.\,\ref{therm3}. From this, we can easily conclude that light dark matter is our best bet if wish to stick to the (usual) regions of thermal relic density.

\begin{figure}[htb]
	\hspace*{-0.7cm}                                                           
 	\includegraphics[scale=0.21]{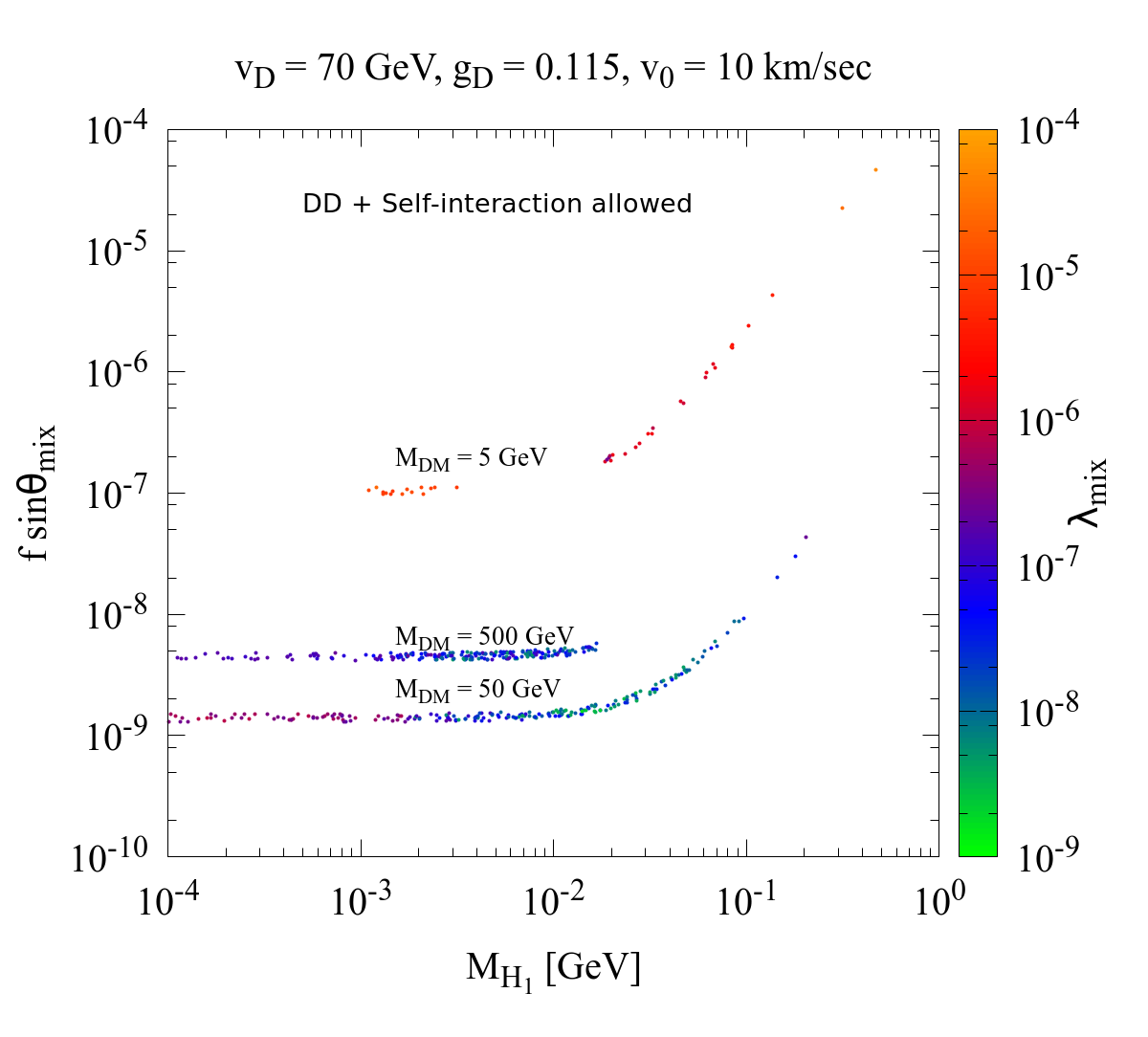}~
 	\includegraphics[scale=0.21]{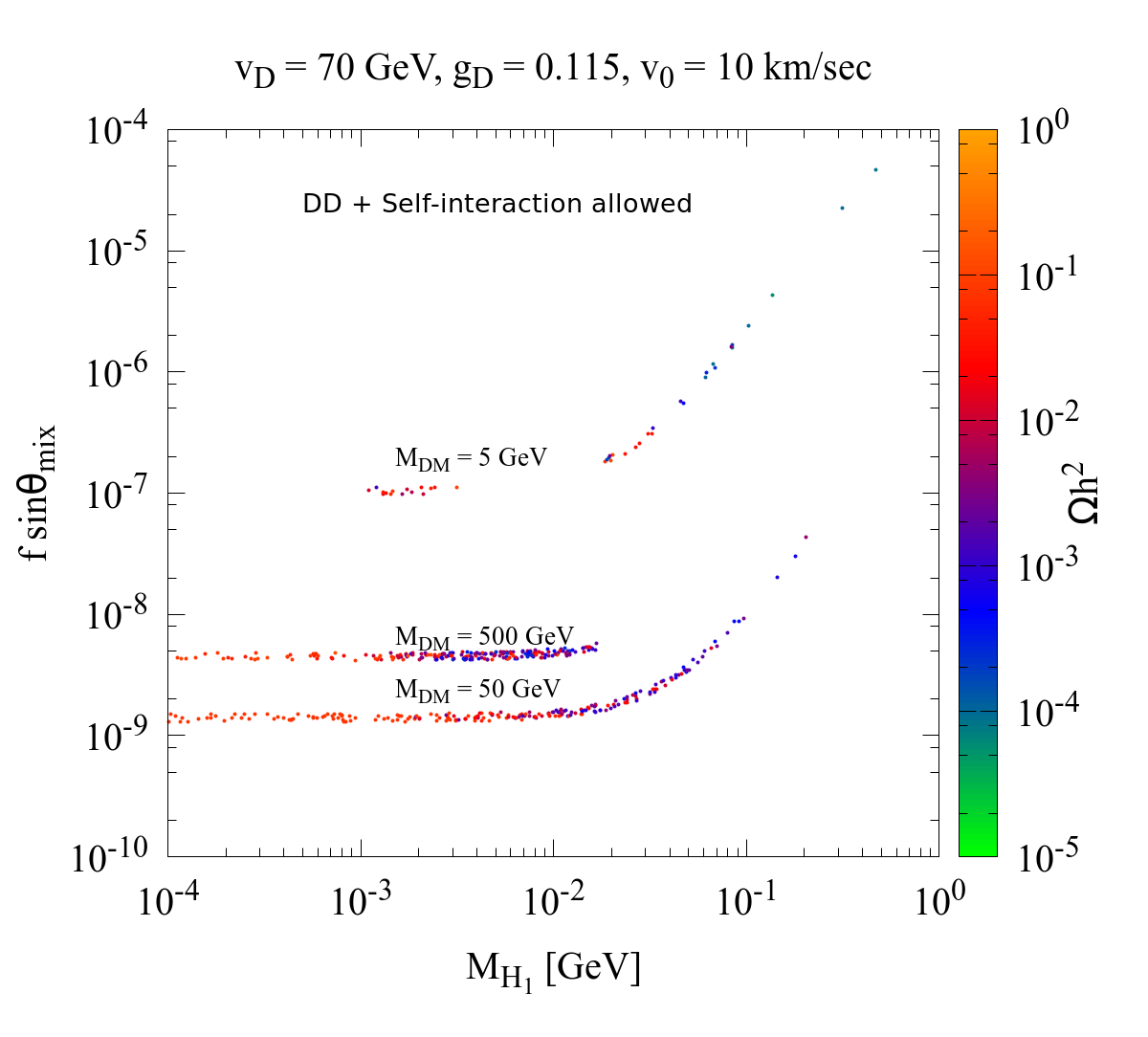}
 	\caption{Plots showing the variation of of relic density along with the quartic coupling, $\lambda_{\rm mix}$ in the colour bar for points allowed by direct detection constraints and also having sizeable self interactions.}
 	\label{therm3}
 \end{figure} 

The dominant contributing channel to the thermal relic density can be broadly classified into three classes :

\begin{itemize}
\item $\chi_1 \chi_1 \leftrightarrow H_1 H_1$ : The cross section solely depends on $f$ ($\sigma \sim f^4$). This type of channel can in general be present in simple extensions of SM (for example in cases where we have only a scalar portal mediating the dark sector and the visible sector). We have seen that the requirement of strong self-interaction among the dark matter particles pushes $f$ towards higher values ($f \sim \mathcal{O}(0.1)$). So, this may lead to under abundance in usual circumstances. But in our model since the dark matter $\chi_-$ is a Majorana particle, this type of annihilations are CP-odd and hence p-wave suppressed consequently satisfying thermal relic density. The sole controlling parameter is the Yukawa coupling $f$ which also allows for a sizable self interaction cross section. But for general dark matter masses, correct relic density and large self-interactions cannot always be satisfied by adjusting this single parameter $f$, and leads to over abundance.

%
%
%
%

\item $\chi_1 \chi_1 \leftrightarrow Z_D Z_D$ : To obtain the correct relic abundance we then have to resort to a new channel employing the extra dark gauge boson $Z_D$. This type of channels are also a generic feature of minimal extensions of SM by a new local gauge group. The cross section is now governed by both $f$ and $g_D$ ($\sigma \sim f^2 g_D^2$). The dark gauge coupling can be tuned to adjust the present day relic density to the observed value. Since $g_D$ does not contribute to the self-interaction of dark matter, this parameter can be freely chosen to fix the relic, while $f$ is fixed to a value that gives strong self-interaction between the dark matter particles. The mass of the dark matter particle is also constrained now such that $M_{DM} > M_{Z_D}$. 

%
%
%
%

\item $\chi_1 \chi_2 \leftrightarrow$ SM SM (co-annihilation) : Thermal relic density in principle can also be dominated by co-annihilation  cross sections of $\chi_{-}$ (DM) and $\chi_{+}$. The impact of co-annihilation is determined by the mass difference of the two dark states. It is strongest if the mass splitting is small. The mass splitting is given by $\Delta m_\chi = \sqrt{2}f\,v_D$. But, from our earlier discussions we came to the conclusion that to stick to the thermal scenarios, its best to adhere to low mass dark matter. On the other hand, self-interaction demands for large values of the Yukawa coupling $f$. Thus co-annihilation fails to be the dominant contributor to the relic density of dark matter if the mass of the latter is taken to be small. This can also be understood clearly as follows : the principle term controlling the thermally averaged co-annihilation cross section is the Boltzmann factor given by $\sim e^{-\frac{\Delta m_\chi}{T_{FO}}}$ which is $\sim e^{-\frac{20\,\Delta m_\chi}{M_{DM}}}$. Here, $T_{FO}$ is the freeze out temperature of the dark matter. Hence higher the dark matter mass and lower the mass splitting, the more important are these co-annihilation channels in their contribution towards the dark matter relic density. From Fig.\,\,\ref{ddcons}, we find that for massive dark matters the upper bound on $f \sin\theta_{\rm mix}$ is quite stringent. But for sizable self interactions $f$ cannot be very small. This forces $\sin\theta_{\rm mix}$ towards smaller values (from direct detection constraints) and consequently demands a small $v_D$ (since $\lambda_{\rm mix} \gtrsim 10^{-5}$ for a thermal scenario). But very small $v_D$ will lead to very light $Z_D$ thereby making them unsuitable for probing via colliders. Hence no such points exists in the parameter space where co-annihilation is the dominant contributor to the relic density and at the same time, it also respects all the other requirements (like direct detection, sizable self interaction, thermal dark matter and not too light $Z_D$).
\end{itemize}

\begin{table}
	\centering
	\begin{tabular}[htbp]{|p{0.8cm}|p{0.8cm}|p{0.8cm}|p{0.8cm}|p{1.3cm}|p{1.3cm}|p{0.8cm}|p{1.5cm}|p{0.8cm}|p{0.8cm}|p{1.2cm}| }
		
		\hline
		\multicolumn{11}{|c|}{Thermal Benchmark} \\
		\hline 
		$M_{\rm DM}$ & $M_+$ & $M_{Z_D}$ & $M_{H_1}$ & $\lambda_{\phi}$ & $\lambda_{\rm mix}$ & $g_D$& sin$\theta_{mix}$& $f$ & $\Omega h^2$ & $\frac{\sigma_{\rm self}}{ M_{\rm DM}}$ \\ [0.5ex] 
		\hline\hline
		
		2 GeV & 5.7 GeV & 2.4 GeV  & 20 MeV & $2.2\times 10^{-7}$  & $4.5\times 10^{-5}$  & $0.08$ & $2.04\times 10^{-5}$ & 0.09 & 0.12 & 0.11 barn GeV$^{-1}$ \\\hline
		
		\hline
		
	\end{tabular}
	\caption{Table demonstrating a benchmark where the thermal relic scenario can be realised. It satisfies all the constraints as well has sizable self-interactions at virialised velocities of $v_0 \sim \mathcal{O}(10)$ km sec$^{-1}$. The dominant channel contributing to the thermal relic density is $\chi_1 \chi_1 \leftrightarrow H_1 H_1$. Since the dark matter mass is beyond the reach of \texttt{Xenon-1T} \cite{Aprile:2017iyp}, this specific benchmark is checked against experiments which are sensitive light dark matters {\it} i.e. \texttt{CRESST-II} \cite{Schieck:2016nrp} and \texttt{CDMSlite (run-2)} \cite{Agnese:2015nto}.}
	\label{tablerelic}
\end{table}

To summarize, thermal dark matters which are strongly self interacting and also respects constraints from direct detection experiments as well as have sizeable collider signatures, should be light {\it i.e.} $\lesssim \mathcal{O}(5)$ GeV. To get a better feel of the numbers let us consider a {\it wimp-ish} dark matter of order $\mathcal{O}(500)$ GeV. From Fig.\,\ref{ddcons} we find that $f \sin\theta_{\rm mix} \lesssim 6\times10^{-9}$ for $M_{H_1} \sim \mathcal{O}(10)$ MeV. The mediator should be light to allow for large self-interaction cross section but not too light so as to violate bounds from BBN ($\gtrsim \mathcal{O}(10)$ MeV). Also, we found that for sizeable self interaction at these dark matter and mediator masses $f \sim \mathcal{O}(0.1)$. Taking $v_D$ to be as low as $\sim \mathcal{O}(1)$ GeV, we find $\lambda_{\rm mix} \lesssim 7.7\times 10^{-6}$ and that falls below our derived limit of thermal threshold. Hence for high mass dark matter, we are almost out of points in the parameter space that satisfies all of our aforementioned desired criteria. Light dark matters are hence more favourable in our scenario. Even for a $\mathcal{O}(5)$ GeV dark matter, direct detection experiments force $f\,v_D \lesssim 0.25$. For reasons previously mentioned, with $f \sim 0.1$ we find $v_D \lesssim 2.5$. Along with this, requirement of $\lambda_{\rm mix} \gtrsim 10^{-5}$ (thermalisability) and not too small $Z_D$ mass restricts our allowed parameter space quite heavily.

Next, as a concrete example, we present a specific benchmark in Table.\,\ref{tablerelic} which illustrates that for specific choice of parameter values one can indeed have a perfectly thermal scenario satisfying all the constraints and also having a sizable self-interactions although only with small dark matter masses.\footnote{This benchmark is compliant with constraints from BBN since the mass of the mediator $M_{H_1} \gg 1$ MeV consequently rendering it to be non relativistic during the time of BBN. Furthermore, the lifetime of $H_1$ is also less than 1 sec ({\it i.e.} onset of BBN).} In the next section we will discuss about the future detection prospects of our scenario in colliders.

\section{Future searches}
\label{subsec:future}

From the discussions in the perspective of dark matter phenomenology, we have thus been able to convince ourselves that for a thermal dark matter with large self interactions we need very light $H_1$ as the mediator ( $\sim \mathcal{O}(1 ~{\rm GeV})$). In the following sections we will investigate about the future of prospects and detectability of our model from a collider perspective. 
We probe the prospect of future collider experiments and estimate their effect on the parameter space which survives the current constraints as discussed in Sec.~\ref{sec:all_constraints}. We begin by a discussion about the future search for the light scalar mediator $(H_1)$.

\subsection{Future prospects of $H_{2} \to H_{1} H_{1} $}

We have already discussed in Sec.\ref{sec:LHC_direct_constraints} that ATLAS and CMS have performed different searches 
for the Higgs boson decaying into two spin-zero particles in various final state using 
Run-I and Run-II datasets : $4\tau$ \cite{Khachatryan:2017mnf}, $2\mu 2b$ \cite{Khachatryan:2017mnf,Aaboud:2018esj}, 
$2\mu 2 \tau$ \cite{Khachatryan:2017mnf,Aad:2015oqa,Sirunyan:2018mbx}, $4\mu$ \cite{Khachatryan:2015wka,CMS-PAS-HIG-18-003}, 
$4\tau$ \cite{Khachatryan:2015nba,CMS-PAS-HIG-14-022}, $2\tau 2b$ \cite{CMS-PAS-HIG-14-041,Sirunyan:2018pzn}, $4b$ \cite{Aaboud:2018iil}. 
Apart from these analyses, CMS has also looked for direct production of a light pseudoscalar Higgs boson 
in the dimuon decay channel using LHC 7 TeV data \cite{Chatrchyan:2012am} and the limits 
are not very strong yet. For lower values of $M_{A}$ or $M_{H_1}$ (in the range 1 to 8 GeV) much 
stronger bounds have been obtained by ATLAS and CMS in $4\mu$ channel\cite{Aaboud:2018fvk}. Rest of the channels like $2\mu 2b$\cite{Khachatryan:2017mnf,Aaboud:2018esj}, 
$2\mu 2 \tau$  \cite{Khachatryan:2017mnf,Aad:2015oqa,Sirunyan:2018mbx}, $2\tau 2b$ \cite{CMS-PAS-HIG-14-041,Sirunyan:2018pzn}, etc. are sensitive for 
higher mass range (typically  15 GeV $ < M_{A} < $ 62.5 GeV).

The possibility to probe a light pseudoscalar particle from Higgs decays at the 
14 TeV LHC has been studied in \cite{Lisanti:2009uy} for $2\mu 2 \tau$ final state. 
This search covered the mass range : $2 M_\tau < M_A < 2 M_b$. Using the upper limit 
obtained from this analysis, expected future reach for 14 TeV LHC with integrated 
luminosity $\lum=$ 300 $\ifb$ have been translated in \cite{Bhattacherjee:2013jca} 
(see Fig. 7 of \cite{Bhattacherjee:2013jca}). The projected upper limit on  $H_2 \to H_1 H_1$ branching ratios 
is roughly 5$\%$ \cite{Bhattacherjee:2013jca}. For our case we find that $H_2 \to H_1 H_1$ branching ratios 
lie in the range $10^{-3}$ to $10^{-6}$, which will be beyond the reach of HL-LHC in $2\mu 2 \tau$ final state.

In \cite{Carena:2007jk}, the expected future reach at the 14 TeV LHC has been studied for $H_2 \to AA \to 4b$ and 
$H_2 \to AA \to 2b 2\tau $ final states, where the Higgs boson is produced in association with a W or Z boson.  
This analysis is sensitive for $M_A > 10$ GeV and 4$b$ final state is more promising than the  $2b 2\tau $ channel. 
The potential of an exotic Higgs decay search for $H_2 \to XX \to 2b 2\mu$ ($X = H_1/A$) in the mass range of
15 to 60 GeV has been presented in \cite{Curtin:2014pda}. It is found that Br($H_2 \to 2X \to 2b2\mu)$ 
can be constrained at the few $\times 10^{-5}$ level at the HL-LHC. Both these analyses \cite{Carena:2007jk, Curtin:2014pda} 
are not sensitive for the parameter space of our interest ($M_{H_2} < 10$ GeV).

\subsection{Future prospects at ILC}


\begin{figure}
\begin{center}
\includegraphics[scale=0.21]{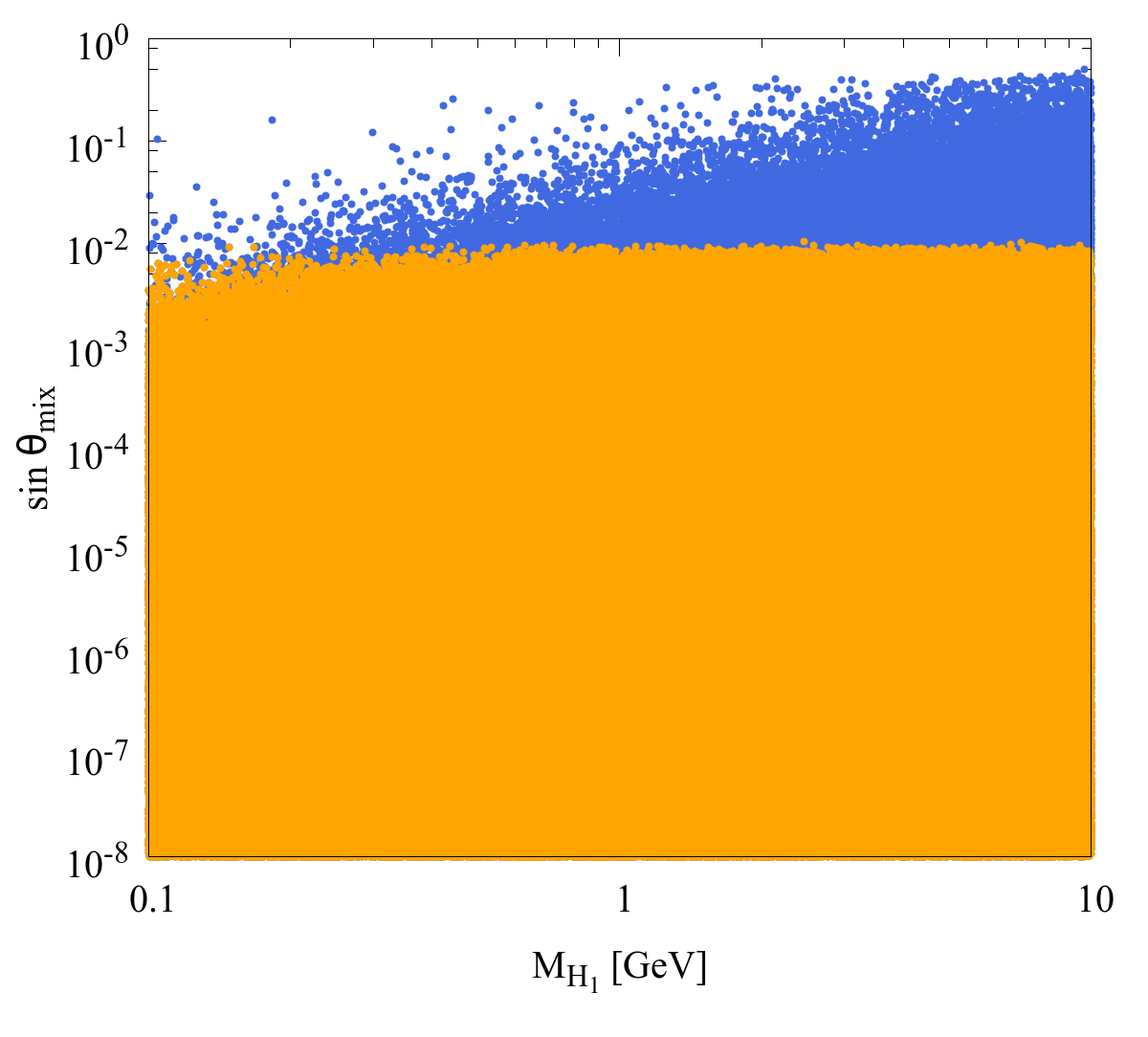}
\caption{Parameter space points in the $M_{H_{1}}-\sin\theta_{mix}$ plane. The blue colored points are excluded by the projected sensitivity of ILC+LHC to probe $HZZ^{*}$ coupling within $0.3\%$. }
\label{fig:ilc_future}
\end{center}
\end{figure} 

Uncertainties in Higgs boson couplings to various SM final states from a combination of global fit of Higgs coupling measurements at the ILC have been presented in \cite{Asner:2013psa}. Combination of the results of Higgs coupling measurements from the $300~{\rm fb^{-1}}$ run of LHC, and $500~{\rm GeV}$ run of ILC, may lead to an error of $0.3\%$ in $HZZ$ coupling~\cite{Asner:2013psa}. Correspondingly, we test the impact of this constraint on our parameter space by choosing those points which generate $\mu^{ggF}_{ZZ^{*}}$ in the range of $0.997-1.003$. Such parameter space points have been shown in orange color in Fig.~\ref{fig:ilc_future}. The blue colored points in Fig.~\ref{fig:ilc_future} correspond to the entire set of parameter space scanned by us. It can be observed that the projected reach of ILC is $\sin\theta_{mix}\gtrsim 7\cdot 10^{-3}$ which is roughly $1$ order of magnitude stronger than the current LEP results.

\subsection{Exclusions from future SHiP and LZ experiments}

Continuing our discussion of Sec.~\ref{sec:beam_dump_current}, we show the projected reach of two relevant future experiments, namely, SHiP~\cite{Anelli:2015pba} and LZ, in Fig.~\ref{fig:Ship_future}. The projected sensitivity of SHiP is remarkably strong when compared to the current experiments. For $M_{H_{1}}\sim 1\,{\rm GeV}$, the current limits exclude $\sin\theta_{mix}$ above $\sim 10^{-3}$, whereas, SHiP is projected to probe until $\sin\theta_{mix} \sim 10^{-5}$. LZ is expected to gain effectiveness in the $O(GeV)$ region, and is expected to improve upon the existing sensitivity by around an order of magnitude. 

\begin{figure}
\begin{center}
\includegraphics[scale=0.21]{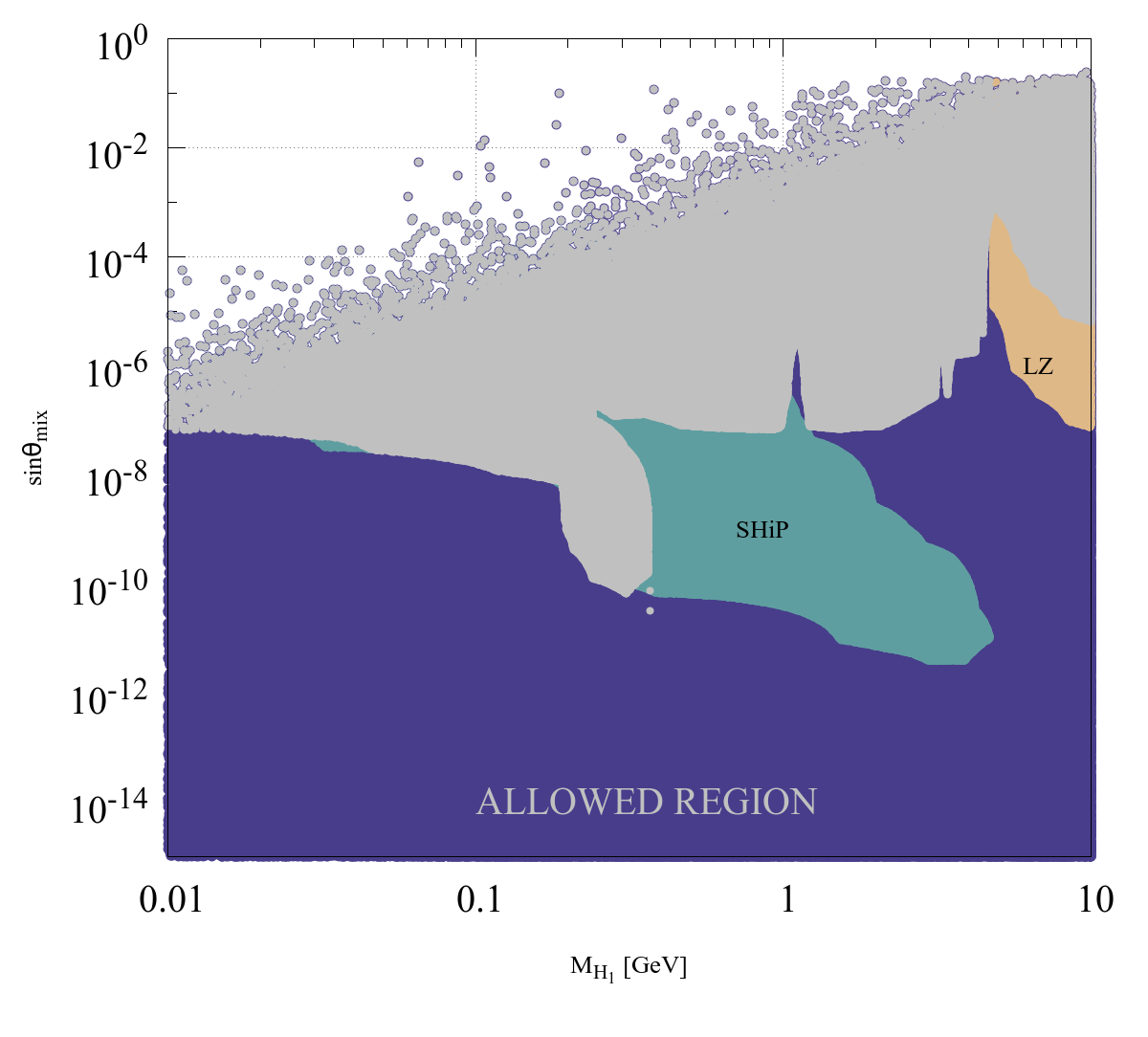}
\caption{Parameter space points in the $M_{H_{1}}-\sin\theta_{mix}$ plane. The grey colored points are excluded by the current search limits from beam dump experiments and flavor physics experiments (Fig.~\ref{fig:flav_beamdump_current}). The green colored points will be within the projected reach of SHiP experiment while the brown colored points could be probed by the proposed LZ. }
\label{fig:Ship_future}
\end{center}
\end{figure} 

As seen from the findings of the preceding three subsections, a hunt for a light scalar mediator in future colliders seem to be challenging. Our model however also possesses an $\mathcal{O}(1)$ GeV $Z_D$ and dark matter. Hence, we can instead try searching for this dark photon in future colliders. It is important to note that the analysis performed in the next two subsections would comply only in the case of a promptly decaying $Z_{D}$ boson. The total decay width of $Z_{D}$ is proportional to $\epsilon_g^{2}$. A value of $\epsilon_g^{2} \gtrsim 10^{-12}$ results in a lifetime which is compatible with the regime of prompt decay. Smaller values of $\epsilon_g$ ($\lesssim 10^{-6}$) results in decay lengths of the order of $\sim 10^{-5}~{\rm m}$ or higher, resulting in a late decaying phenomena. The generic collider features of a late decaying $Z_{D}$ has been discussed in Sec.~\ref{subsec:LLP_zd}.

\subsection{$H_{2} \to Z_{D}Z_{D} \to 4\mu$ at HL-LHC}
\label{sec:coll_zdzd}

A detailed search analysis for $Z_{D}$ is presented ($M_{Z_{D}} = 2-12~{\rm GeV}$), using the process $pp \to H_{2} \to Z_{D} Z_{D} \to 4\mu$ final state, in the context of $14~{\rm TeV}$ run of LHC at an integrated luminosity of $3000~fb^{-1}$. At the LHC, the Higgs boson is dominantly produced via the gluon fusion mode ($ggF$). The $ggF$ mode overshadows the other modes of Higgs production, such as vector boson fusion ($VBF$), associated production with $b$ quarks ($b\bar{b}H_{2}$) and associated production with top quarks ($t\bar{t}H_{2}$), and therefore, in the current analysis, we consider only the $ggF$ mode of Higgs production. 

The signal sample involves the process : $ gg \to H_{2} \to Z_{D} Z_{D} \to 4\mu$. Dominant backgrounds arise from electroweak $4l$ production, and the $H_{2} \to ZZ^{*}$ decay process. The $gg \to H_{2}$ samples have been generated using ${\rm MadGraph5}$~\cite{Alwall:2014hca}, with $M_{H_{2}}$ fixed at $125~{\rm GeV}$. The subsequent decay process, followed by showering and hadronization is performed by \texttt{Pythia-6}~\cite{Sjostrand:2001yu}. The $4l$ background has been generated by ${\rm MadGraph5}$ at the partonic level, while \texttt{Pythia-6} has been used for showering. 

\begin{figure}
\begin{center}
\includegraphics[scale=0.16]{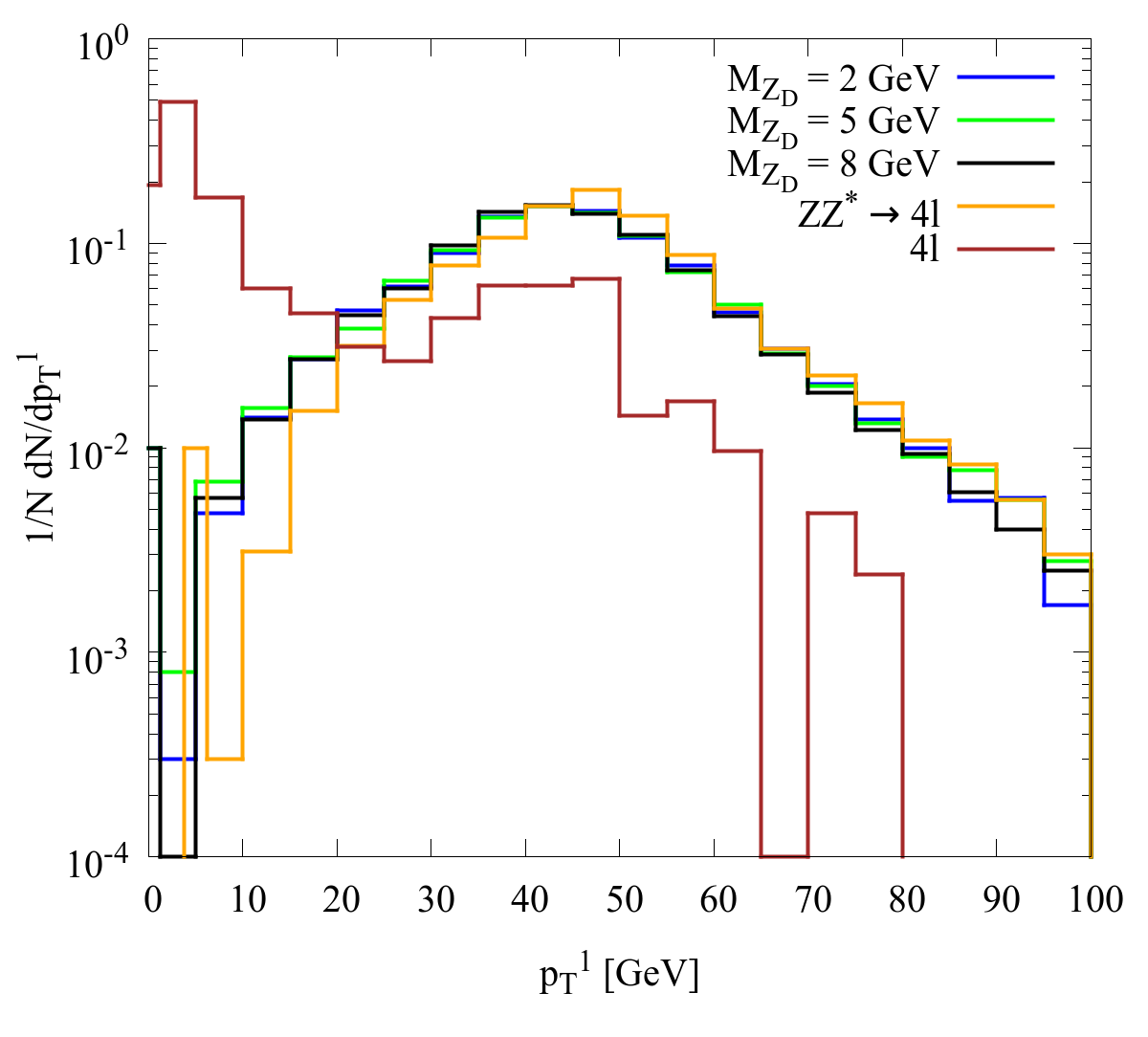}\includegraphics[scale=0.16]{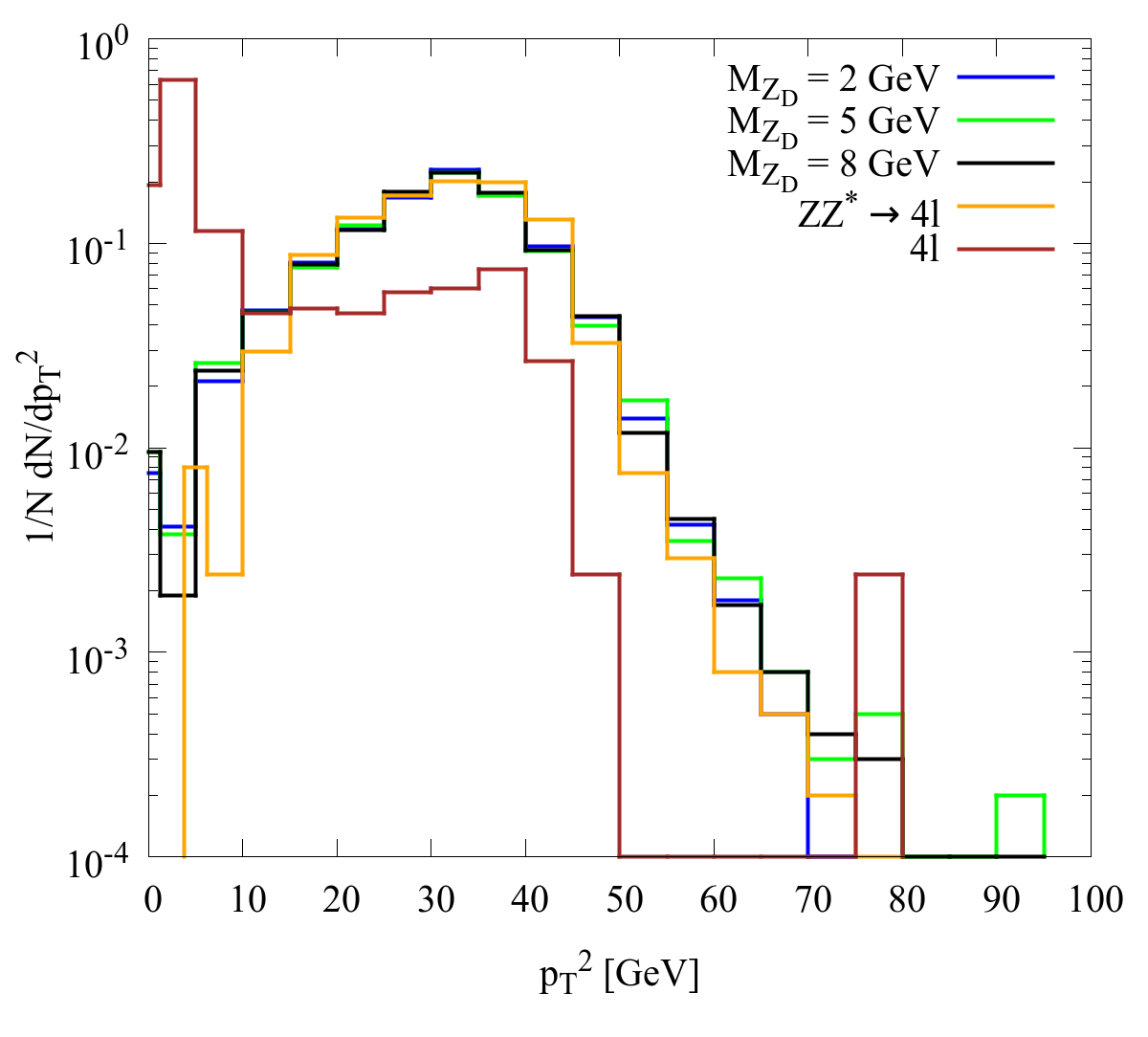}\\
\includegraphics[scale=0.16]{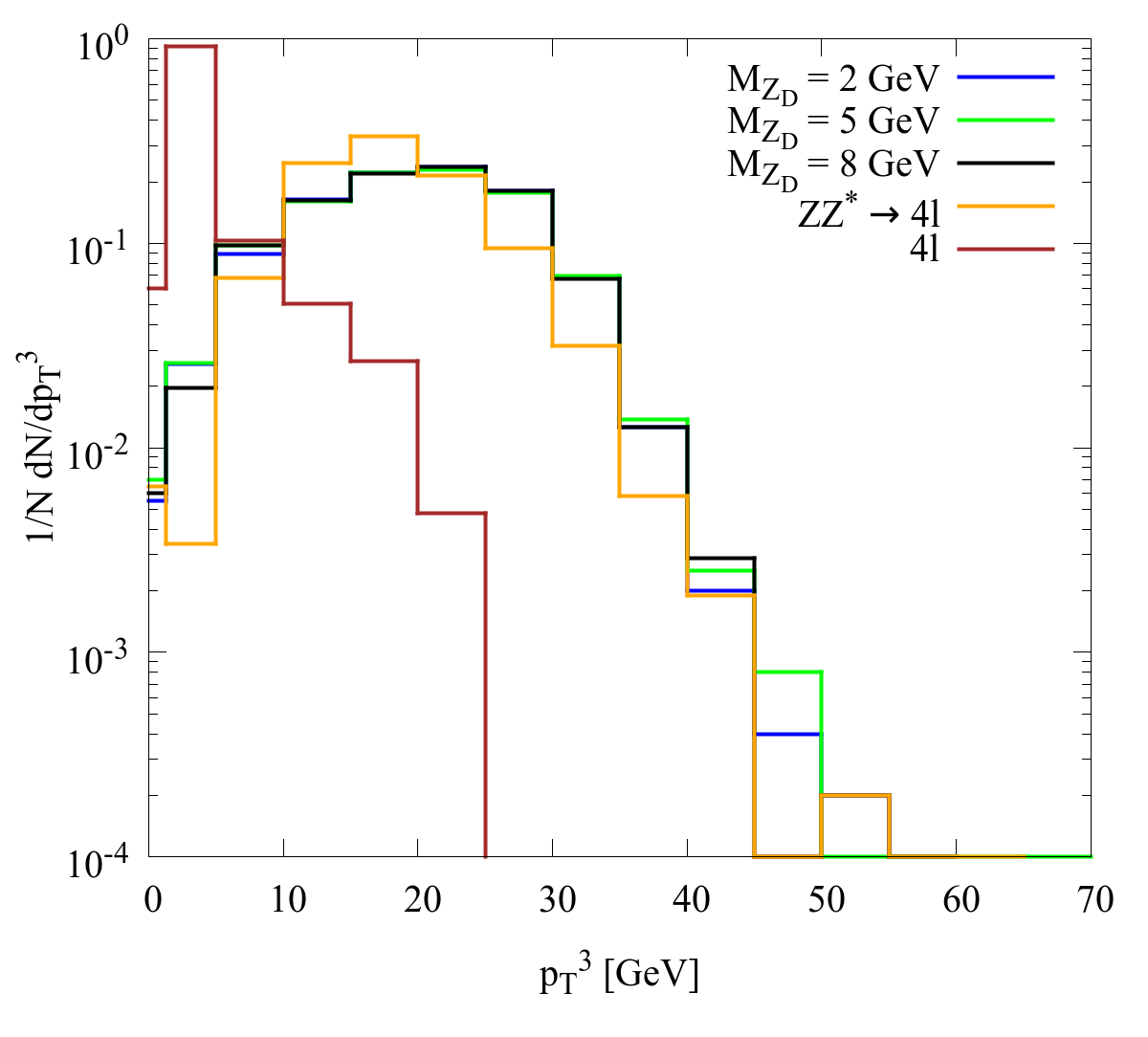}\includegraphics[scale=0.16]{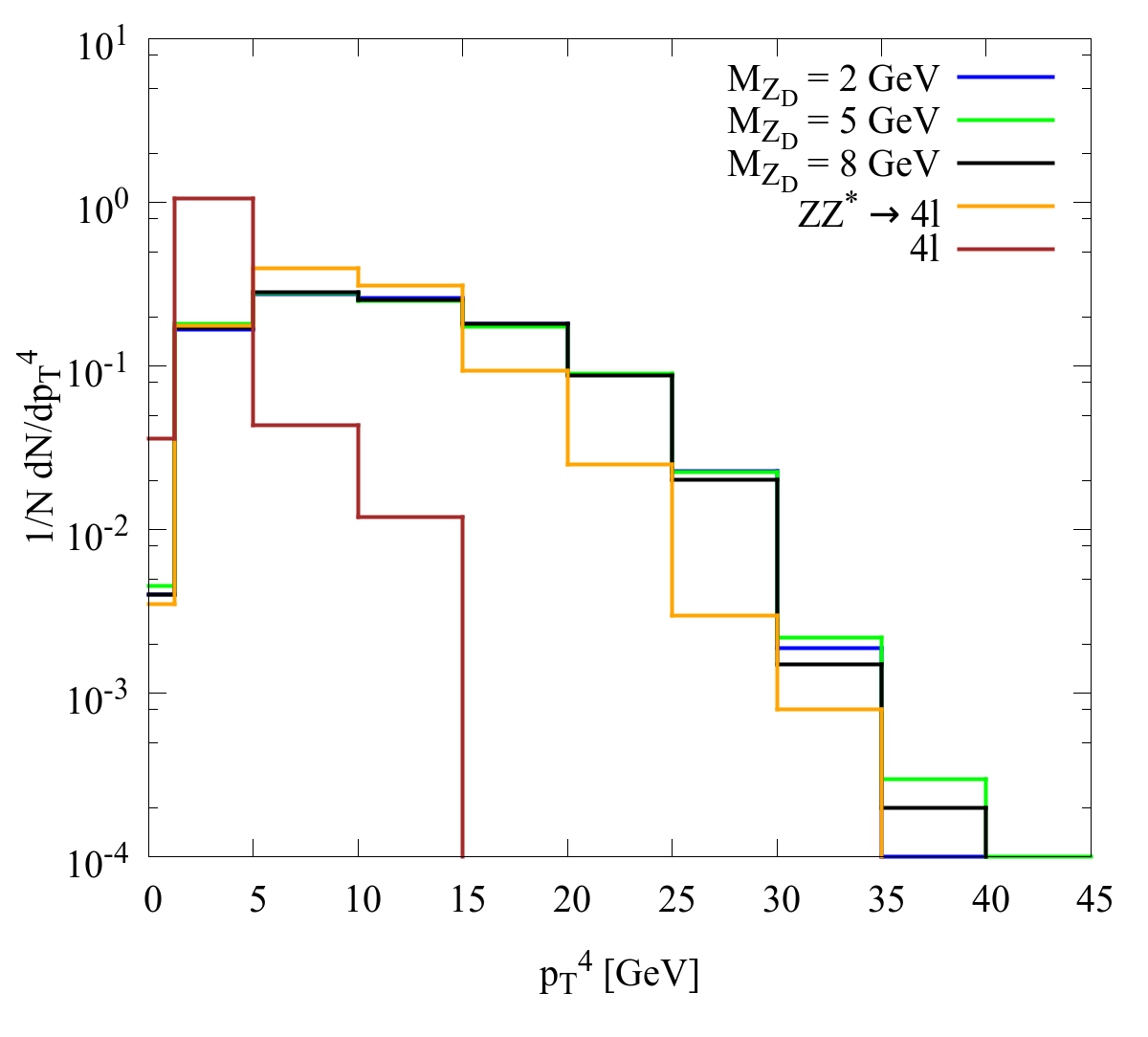}
\caption{Transverse momentum ($p_{T}$) distribution of the final state muons at the partonic level. The blue, green and black colored lines correspond to the signal event generated with different $Z_{D}$ masses, $M_{Z_{D}}=2,~5$ and $8~{\rm GeV}$, respectively. The orange and the brown colored line represents the $p_{T}$ distribution of the $ZZ^{*} \to 4l$ and electroweak $4l$ background.}
\label{fig:future_coll_pt1}
\end{center}  
\end{figure}

The final state muons are required to have transverse momentum, $p_{T} > 2.6~{\rm GeV}$ and must lie within a pseudorapidity range of $|\eta| < 4.0$. Isolation condition requires the sum of transverse momentum of other tracks, excluding the leading four muons, within a cone of radius $\Delta R = 0.2$ around the muon to be less than $30\%$ of the $p_{T}$ of the muon. Furthermore, if the $\Delta R$ between the muon and the reconstructed jet is less than $0.4$, then the muon is removed. Event selection requires exactly $4$ muons to be present in the final state. Before moving on to discuss the choice of selection cuts, we discuss the kinematic distribution of the four muons in the final state. In this context, we choose three different values of $M_{Z_{D}} = 2,~5,~8~{\rm  GeV}$ to generate the signal. The $p_{T}$ distribution of the four muons from the background and the benchmark signals are shown in Fig.~\ref{fig:future_coll_pt1}, where, the $4l$ and $H_{2} \to ZZ^{*}$ backgrounds have been described by brown and orange colored lines, respectively. We would like to mention that the muons have been $p_{T}$ ordered, and therefore, $p_{T}^{1}$ corresponds to the highest $p_{T}$ muon while $p_{T}^{4}$ represents the lowest $p_{T}$ muon. The blue, green and black colored lines correspond to the $p_{T}$ distribution of the signal samples with $M_{Z_{D}} = 2,~5,~8~{\rm GeV}$, respectively. 

\begin{figure}
\begin{center}
\includegraphics[scale=0.17]{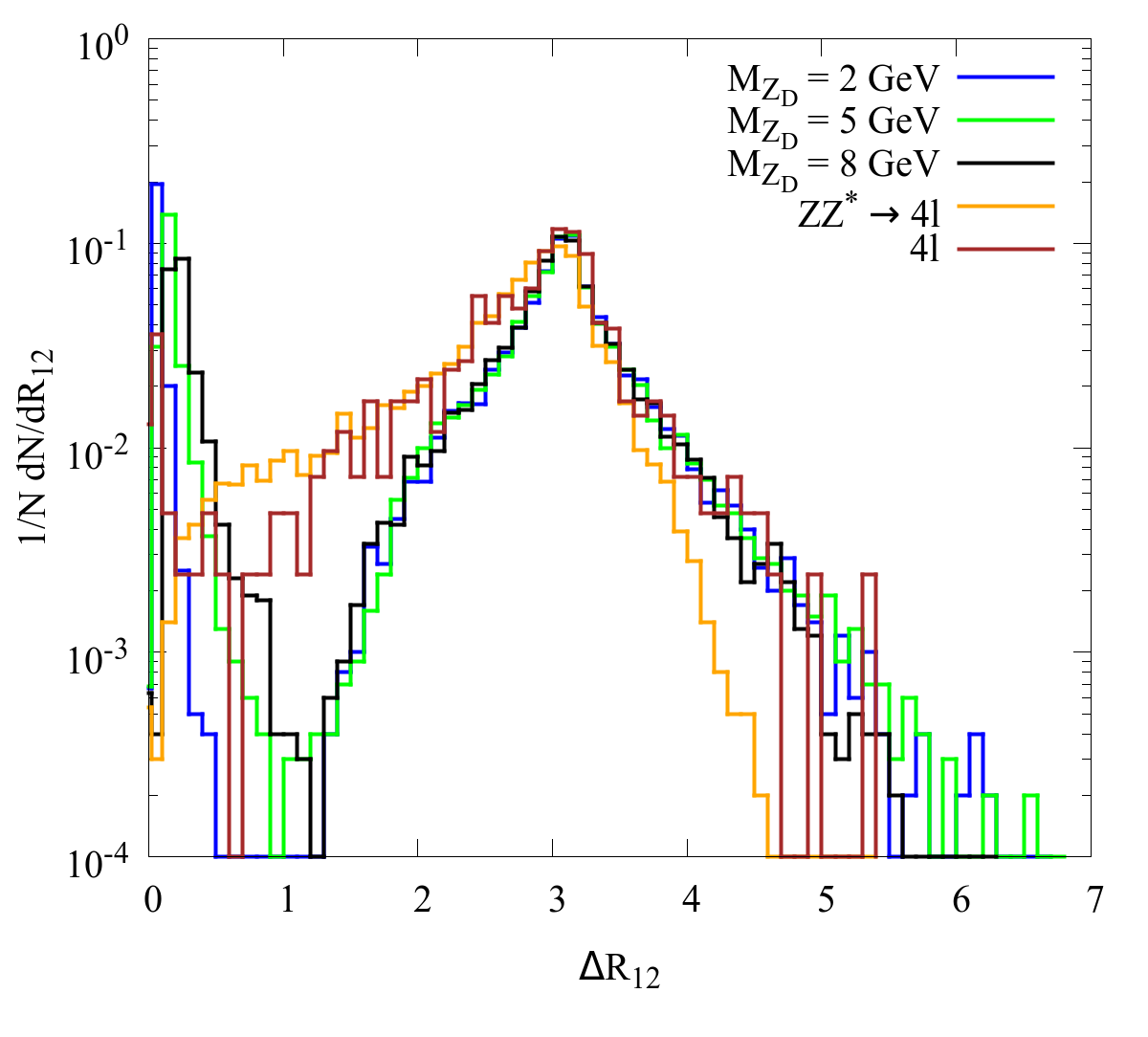}\includegraphics[scale=0.17]{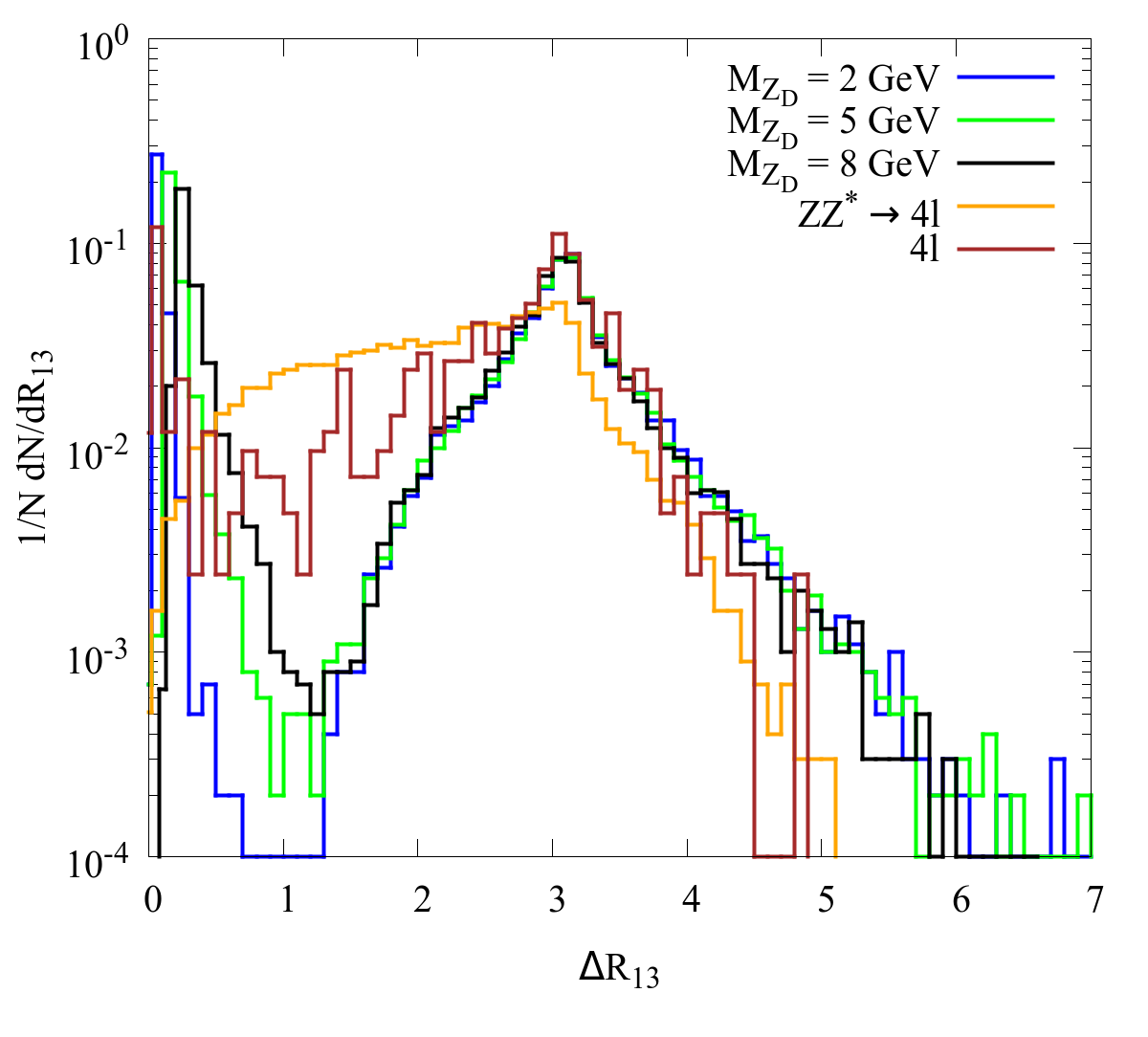}\\ 
\includegraphics[scale=0.17]{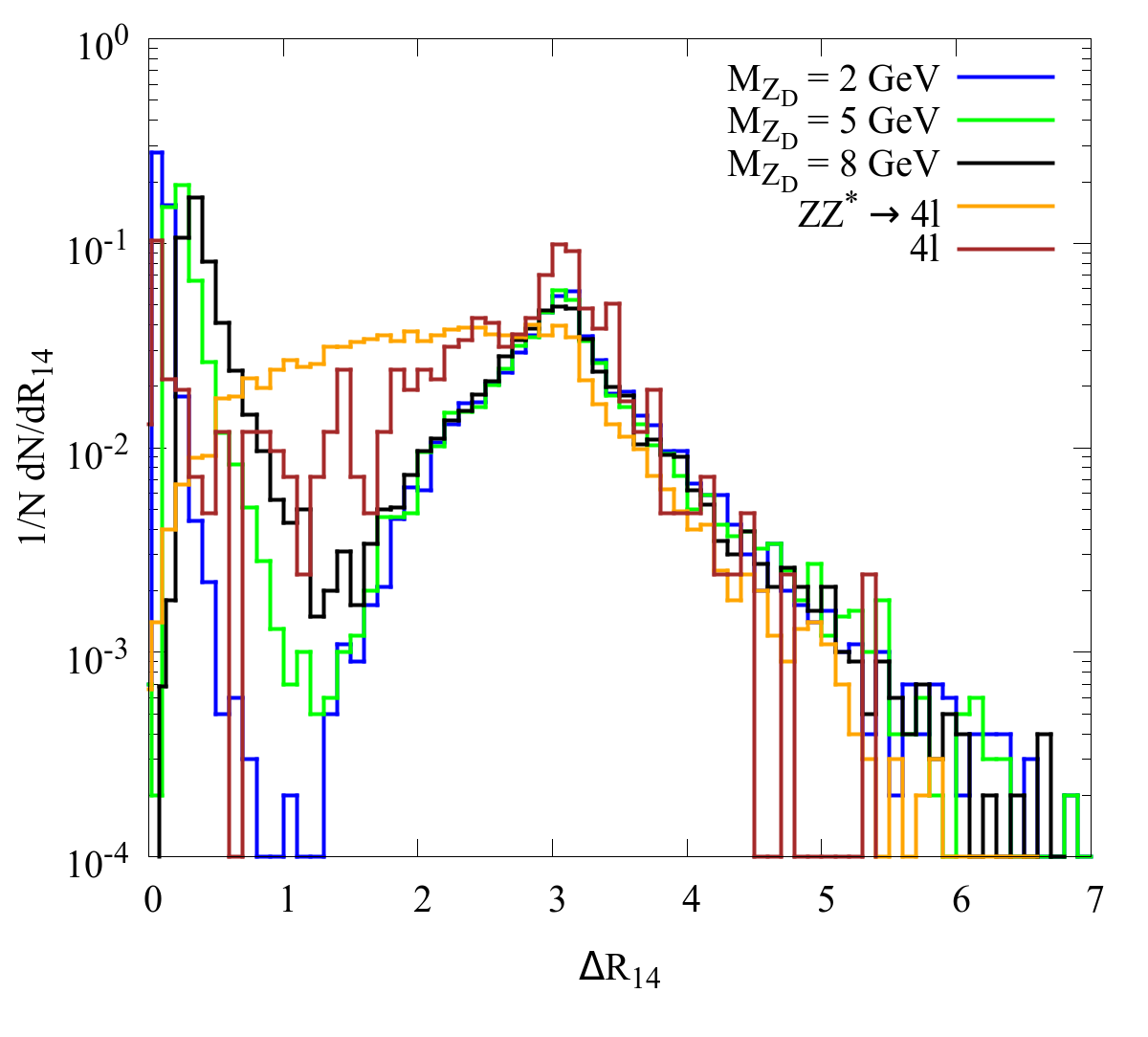}
\includegraphics[scale=0.17]{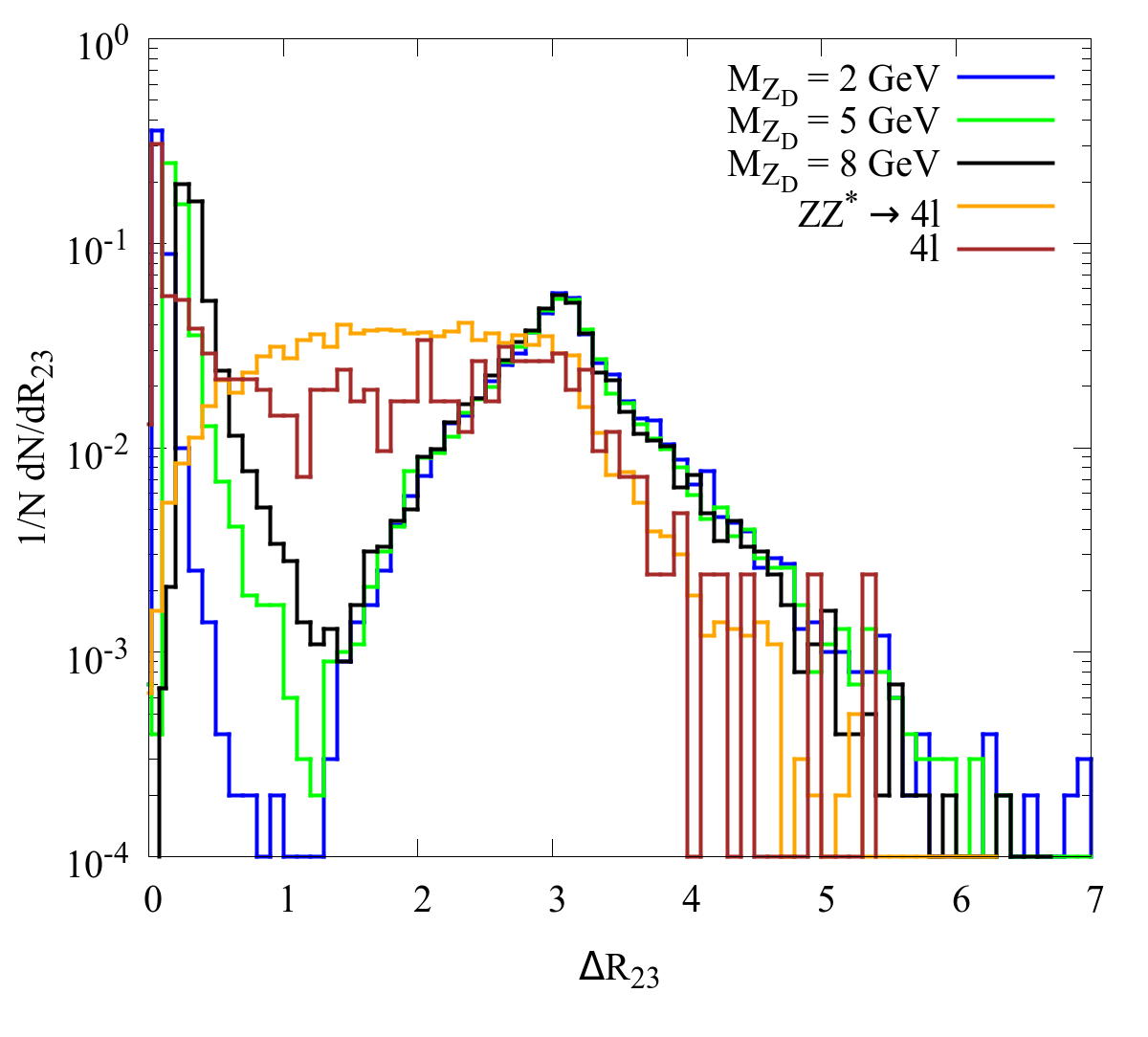}
\caption{Normalised distribution showing $\Delta{R}$ between various final state muon pairs. The color codes are the same as that of Fig.~\ref{fig:future_coll_pt1}.}
\label{fig:future_coll_delr1}
\end{center}  
\end{figure}

\begin{table}
\begin{center}
\begin{tabular}{C{16cm}}
\hline \hline 
Selection cuts \\ 
\hline 
(a). Exactly $4$ muons in final state. \\
(b). $p_{T}^{\mu_{1}} > 20~{\rm GeV}$, $p_{T}^{\mu_{2}} > 15~{\rm GeV} $ \\
 $p_{T}^{\mu_{3}} > 10~{\rm GeV}$, $p_{T}^{\mu_{4}} > 2.6~{\rm GeV}$  \\
$|\eta| < 4$ \\
(c). $\Delta(R)({\mu_{1}\mu_{3}},{\mu_{1}\mu_{4}},{\mu_{2}\mu_{3}},{\mu_{2}\mu_{4}}) > 2$ \\
(d). $0.88~{\rm GeV} < M_{12}^{inv},M_{34}^{inv} < 20~{\rm GeV}$\\
(e). Event veto if : $(M_{J/\Psi} - 0.25~{\rm GeV}) < M_{12,34} < (M_{\Psi(2s)} + 0.30~{\rm GeV})$ \\
                 $(M_{\Upsilon} - 0.70~{\rm GeV}) < M_{12,34} < (M_{\Upsilon(3s)} + 0.75~{\rm GeV})$ \\
(f). $M_{34}^{inv}/M_{12}^{inv} > 0.85$ \\
(g). $120~{\rm GeV} < M_{4\mu}^{inv} < 130~{\rm GeV} $ \\
\hline 
\end{tabular}
\caption{Selection cuts for the cut-based analysis in the $4\mu$ final state, following ~\cite{Aaboud:2018fvk}.}
\label{tab:sel_cut_zdzd4mu}
\end{center}
\end{table}

\begin{table}
\begin{center}
\begin{tabular}{||C{1.4cm}|C{1.2cm}|C{1.2cm}|C{1.2cm}|C{1.4cm}|C{1.2cm}|C{1.2cm}|C{1.0cm}|C{1.0cm}||} \hline\hline
\multirow{2}{*}{} & \multicolumn{6}{c|}{Fraction of events surviving selection cuts from Table.~\ref{tab:sel_cut_zdzd4mu}} & \multicolumn{2}{c|}{U.L. [fb]} \\\cline{2-7}
 & $(a)$ & $(b)$ & $(c)$ & $(d)+(e)$ & $(f)$ & $(g)$ & \multicolumn{2}{c|}{$\sigma_{H_{2}\to Z_{D}Z_{D}}$}  \\ \hline\hline
$M_{Z_{D}}$ & \multicolumn{6}{c|}{\multirow{2}{*}{Signal}} & \multirow{2}{*}{$5\sigma$} & \multirow{2}{*}{$2\sigma$} \\
$ [{\rm GeV}]$ & \multicolumn{6}{c|}{}  &  & \\ \cline{1-7} \cline{1-7}\hline
$2$ & $0.838$ & $0.752$ & $0.665$ & $0.664$ & $0.651$ & $0.637$ & $ 0.197 $ & $ 0.078 $  \\ 
$2.80$ & $0.831$ & $0.747$ & $0.674$ & $0.674$ & $0.659$ & $0.643$ & $ 0.166 $ & $ 0.066 $\\
$4$ & $0.829$ & $0.746$ & $0.662$ & $ 0.528$ & $0.522$ & $ 0.522$ & $ 0.426 $ & $ 0.170 $ \\
$6$ & $0.821$ & $0.737$ & $0.651$ & $0.648$ & $0.636$ & $0.614$ &  $0.465 $ & $ 0.186 $ \\
$8.75$ & $0.814$ & $0.734$ & $0.640$ & $0.640$ & $0.620$ & $0.598$ & $ 0.489 $ & $ 0.196 $ \\
$11.11$ & $0.815$ & $0.731$ & $0.637$ & $0.406$ & $0.397$ & $0.396$ & $ 0.738 $ & $ 0.295 $ \\
$12$ & $0.812$ & $0.729$ & $0.621$ & $0.580$ & $0.573$ & $0.568$ & $ 0.515 $ & $ 0.206 $ \\\hline\hline
 & \multicolumn{6}{c|}{\multirow{2}{*}{Background yield ($14~{\rm TeV}~3000~{\rm fb^{-1}}$)}} & \multicolumn{2}{c|}{Cross section} \\
 & \multicolumn{6}{c|}{}  &  \multicolumn{2}{c|}{before cut (a)}\\ \cline{1-7} \cline{1-7}\hline
\multirow{2}{*}{$4l$} & \multirow{2}{*}{$2\cdot 10^{5}$} & \multirow{2}{*}{$1\cdot 10^{5}$} & \multirow{2}{*}{$1\cdot 10^{4}$} & \multirow{2}{*}{$773$} & \multirow{2}{*}{$169$}  & \multirow{2}{*}{$6.87$} & \multicolumn{2}{c|}{\multirow{2}{*}{$458~{\rm fb}$}} \\
 & & & & & & & \multicolumn{2}{c|}{}  \\\hline 
$H_{2} \to ZZ^{*}$ & \multirow{2}{*}{$2836$} & \multirow{2}{*}{$2583$} & \multirow{2}{*}{$588$} & \multirow{2}{*}{$42.9$} & \multirow{2}{*}{$13.4$} & \multirow{2}{*}{$11.8$} &  \multicolumn{2}{c|}{\multirow{2}{*}{$1.0~{\rm pb}$}} \\
 & & & & & & & \multicolumn{2}{c|}{}  \\\hline
 \multirow{2}{*}{$bbll$} & \multirow{2}{*}{$2\cdot 10^{5}$} & \multirow{2}{*}{$4\cdot 10^{4}$} & \multirow{2}{*}{$7\cdot 10^{3}$} & \multirow{2}{*}{$0$} & \multirow{2}{*}{$0$} & \multirow{2}{*}{$0$} &  \multicolumn{2}{c|}{\multirow{2}{*}{$249.6~{\rm pb}$}} \\
 & & & & & & & \multicolumn{2}{c|}{}  \\\hline\hline 

\end{tabular}
\caption{The cut flow table showing the signal efficiency on successive application of selection cuts specified in Table.~\ref{tab:sel_cut_zdzd4mu}, for $M_{Z_{D}} = 2,~2.8,~4,~6,~8.75,~11.11,~12~{\rm GeV}$. The background yields corresponding to $14~{\rm TeV}$ $3000~{\rm \ifb}$ LHC are also shown. The expected upper limits at $2\sigma$ level and the discovery reach at $5\sigma$ level, on $\sigma(gg \to H_{2} \to Z_{D}Z_{D})$, are also presented.}
\label{tab:cut_flow_zdzd4mu}
\end{center}
\end{table}

The signal samples and the $H_{2} \to ZZ^{*}$ background yields a similar $p_{T}$ distribution, while the $4l$ background peaks at very low values of $p_{T}$. Taking motivation from a similar search analysis by the ATLAS collaboration~\cite{Aaboud:2018fvk}, and the $p_{T}$ distribution of the $4l$ background, we demand the highest $p_{T}$ muon to have $p_{T} > 20~{\rm GeV}$. The second and third leading $p_{T}$ muons are required to have $p_{T}^{2} > 15~{\rm GeV}$ and $p_{T}^{3} > 10~{\rm GeV}$, respectively. The angular distributions of the final state muons also provide an additional control to filter out the signal events. In this context, we show the $\Delta R$ distribution between all some of the possible final state muon pairs in Fig.~\ref{fig:future_coll_delr1}. Before specifying the selection cuts on the $\Delta R$ variables, it is important to discuss the criteria for formation of same flavor opposite sign (SFOS) muon pairs. The four muons in the final state are required to form two SFOS pairs. Each quadruplet of muon per event can result in two separate combinations of di-SFOS pairs. The di-SFOS pair which results in smaller difference of SFOS muon invariant masses is chosen to be the correct di-SFOS pair. The SFOS pair with invariant mass closest to the $Z$ boson mass is referred to as the leading SFOS pair and its invariant mass will be represented by $M_{12}$., while $M_{34}$ represents the invariant mass of the sub-leading SFOS pair. Henceforth in this subsection, we will refer to the muons in the leading SFOS pairs as $\mu_{1}$ and $\mu_{2}$, while muons in the other SFOS pair will be referred to as $\mu_{3}$ and $\mu_{4}$. Now, we go back to the discussions on the $\Delta R$ distributions shown in Fig.~\ref{fig:future_coll_delr1}. It can be observed that the backgrounds mostly peak around $\Delta R \sim 1-2.5$ region, while the signal events peak at two distinct regions, $\Delta R \sim 0 $ and $\Delta R \sim 3$. The peak in the $\Delta R \sim 0 $ arise from muon pairs which originate from the same $Z_{D}$ pair, and therefore, mostly correspond to the same SFOS pair, while the $\Delta R \sim 3$ peak manifests from the muons belonging to different SFOS pairs. Deriving motivation from this observation, we impose a lower cut on $\Delta R$ between muon pairs from separate SFOS pairs, $\Delta R_{13},~\Delta R_{14},~\Delta R_{23},~\Delta R_{24} > 2$.

In addition, the invariant mass of the SFOS pairs, $M_{12}$ and $M_{34}$ are required to lie within the range $0.88 - 20~{\rm GeV}$, and the ratio $M_{34}/M_{12}$ is required to be greater than $\sim 0.85$, following \cite{Aaboud:2018fvk}. The invariant mass of the four isolated muons ($M_{4\mu}$) is also required to lie within $M_{4\mu} = 120-130~{\rm GeV}$. We would like to mention that events with any of the SFOS pairs having invariant mass in the range of $J/ \Psi$ resonance ($M_{OS} = 2.846 - 3.9861~{\rm GeV}$) or $\Upsilon$ resonance ($M_{\Upsilon(2s,3s,4s)} = 8.7603 - 11.1052~{\rm GeV}$), are vetoed. The selection cuts are summarized in Table.~\ref{tab:sel_cut_zdzd4mu}. The $t\bar{t}$ and $b\bar{b}ll$ backgrounds were also evaluated, and no events survived the selection cuts. 

\begin{figure}
\begin{center}
\includegraphics[scale=0.30]{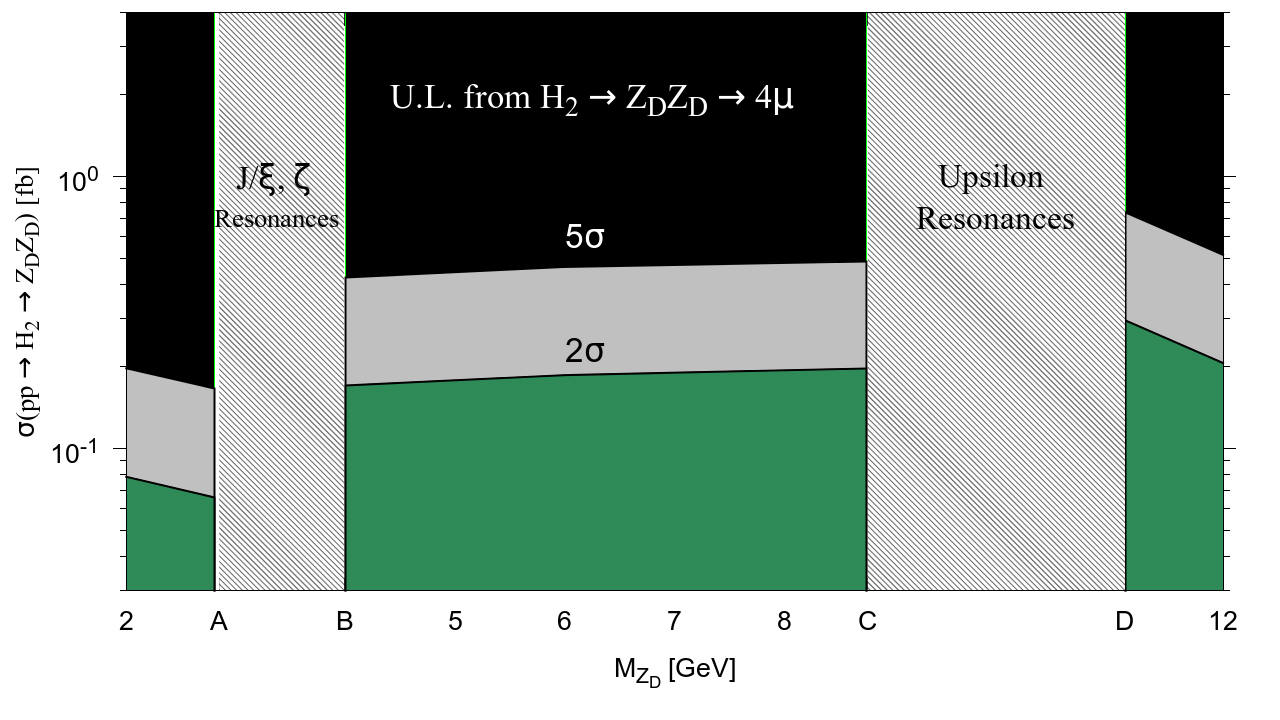}
\caption{Upper limits on $\sigma_{ggH_{2}} \times Br(H_{2} \to Z_{D} Z_{D})$ derived from $Z_{D}$ search in the $4\mu$ final state for $14~{\rm TeV}$ LHC corresponding to an integrated luminosity of $3000~{\rm fb^{-1}}$.}
\label{fig:coll_zdzd_4mu}
\end{center}
\end{figure}

We perform the search for $Z_{D}$ for different values of $M_{Z_{D}} = 2,~2.85,~4,~6,~8.75$, $11.11,~12~{\rm GeV}$, in the context of $14~{\rm TeV}$ LHC corresponding to an integrated luminosity of $3000~{\rm fb^{-1}}$. The cut-flow table showing the signal and background yields are shown in Table.~\ref{tab:cut_flow_zdzd4mu}. Upper limits are obtained on the production cross-section of SM-like $H_{2}$ through $ggF$ mode times the branching fraction $H_{2}\to Z_{D}Z_{D}$ at $5\sigma$ and $2\sigma$, assuming zero systematic uncertainty, for the case of $14~{\rm TeV}$ LHC operating an at integrated luminosity of $3000~{\rm fb^{-1}}$, as shown in Fig.~\ref{fig:coll_zdzd_4mu}. The branching ratio for $Z_{D} \to \mu^{+}\mu^{-}$ has been taken from \cite{Berger:2016vxi}. The upper limits for HL-LHC presented in Fig.~\ref{fig:coll_zdzd_4mu}  is roughly $\sim 10-15$ times stronger than the current bounds shown in Fig.~\ref{fig:LHC_zdzd}.

We conclude this subsection by evaluating the prospect of exclusion/discovery of the allowed benchmark point (BP1) of Table.~\ref{tablerelic} by HL-LHC using the search limits derived in this subsection. BP1 furnishes a value of $Br(H_{2} \to Z_{D}Z_{D}) \sim 5 \cdot 10^{-6}$ for $M_{Z_{D}} \sim 2.4~{\rm GeV}$, while $\sigma(gg \to H_{2})$ attains a value of $\cos^{2}{\theta_{mix}} \cdot \sigma(gg \to H_{2})_{SM} $, where, $\sigma(gg \to H_{2})_{SM} $ corresponds to the SM value of $H_{2}$ production cross-section in the $ggF$ mode. At NNLO+NNLL level, $\sigma(gg \to H_{2})_{SM} = 39.56^{+7.32\%}_{-8.38\%}$ fb~\cite{deFlorian:2013jea,deFlorian:2015moa,hhtwiki} for the 14 TeV run of LHC. Thus, $\sigma(gg \to H_{2} \to Z_{D}Z_{D})$ attains a value of $\sim 0.2$ for $\sqrt{s} = 14~{\rm TeV}$. A comparison with the corresponding upper limit derived in Fig.~\ref{fig:coll_zdzd_4mu} reveal that BP1 could be marginally probed at HL-LHC\footnote{The reach can be further improved upon inclusion of $4e$ and $2e2\mu$ channels.}. These searches have the potential to yield strong exclusion limits in the context of future high energy/ high luminosity colliders. Directly produced $Z_{D}$ process, $pp \to Z_{D} \to \mu\mu$,  could be another viable mode of probing $Z_{D}$. However, the signal is marred by a huge continuum Drell-Yan background. Consequently, the search strategies would involve an efficient treatment of low $p_{T}$ muons. Some recent studies have focused on the case of such low $p_{T}$ muons through scouting techniques \cite{zd_mu_scouting_2,zd_mu_scouting_1}. A more precise understanding of the background and derivation of further improved scouting techniques might help in alleviating $pp \to Z_{D} \to \mu\mu$ as an important probe of $Z_{D}$ in the future runs of LHC.

\subsection{$H_{2} \to \chi_{+}\chi_{+} \to Z_{D} \chi_{-}Z_{D}\chi_{-}$ at HL-LHC}

The SM-like Higgs boson, $H_{2}$, can also undergo decay into a pair of $\chi_{+}\chi_{+}$, which can result in a $4\mu + \met$ final state, through cascade decay via $\chi_{+}\chi_{+}\to (Z_{D} \to \mu^{+}\mu^{-})\chi_{-}(Z_{D}\to \mu^{+}\mu^{-})\chi_{-}$. In this case, $\chi_{-}$, constitutes a source of $\met$. 

In this subsection, we present a search strategy for $Z_{D}$ in the $H_{2} \to \chi_{+}\chi_{+} \to Z_{D} \chi_{-}Z_{D}\chi_{-}$ channel, in context of a $14~{\rm TeV}$ LHC with $\lum = 3000~{\rm fb^{-1}}$. Similar to the case of Sec.~\ref{sec:coll_zdzd}, we only consider the $ggF$ mode of Higgs production. $gg \to H_{2} \to \chi_{+}\chi_{+} \to Z_{D} \chi_{-}Z_{D}\chi_{-}$ constitutes the signal, where, $gg\to H_{2}$ samples have been generated with \texttt{MadGraph}~\cite{Alwall:2014hca}, while \texttt{Pythia-6}~\cite{Sjostrand:2001yu} has been used to perform the cascade decay and showering. The benchmark point shown in Table.~2, with $M_{\chi_{+}}=5.8187~{\rm GeV}$ and $M_{\chi_{-}}=2.4~{\rm GeV}$, has been chosen to perform this search.  

\begin{figure}
\begin{center}
\includegraphics[scale=0.16]{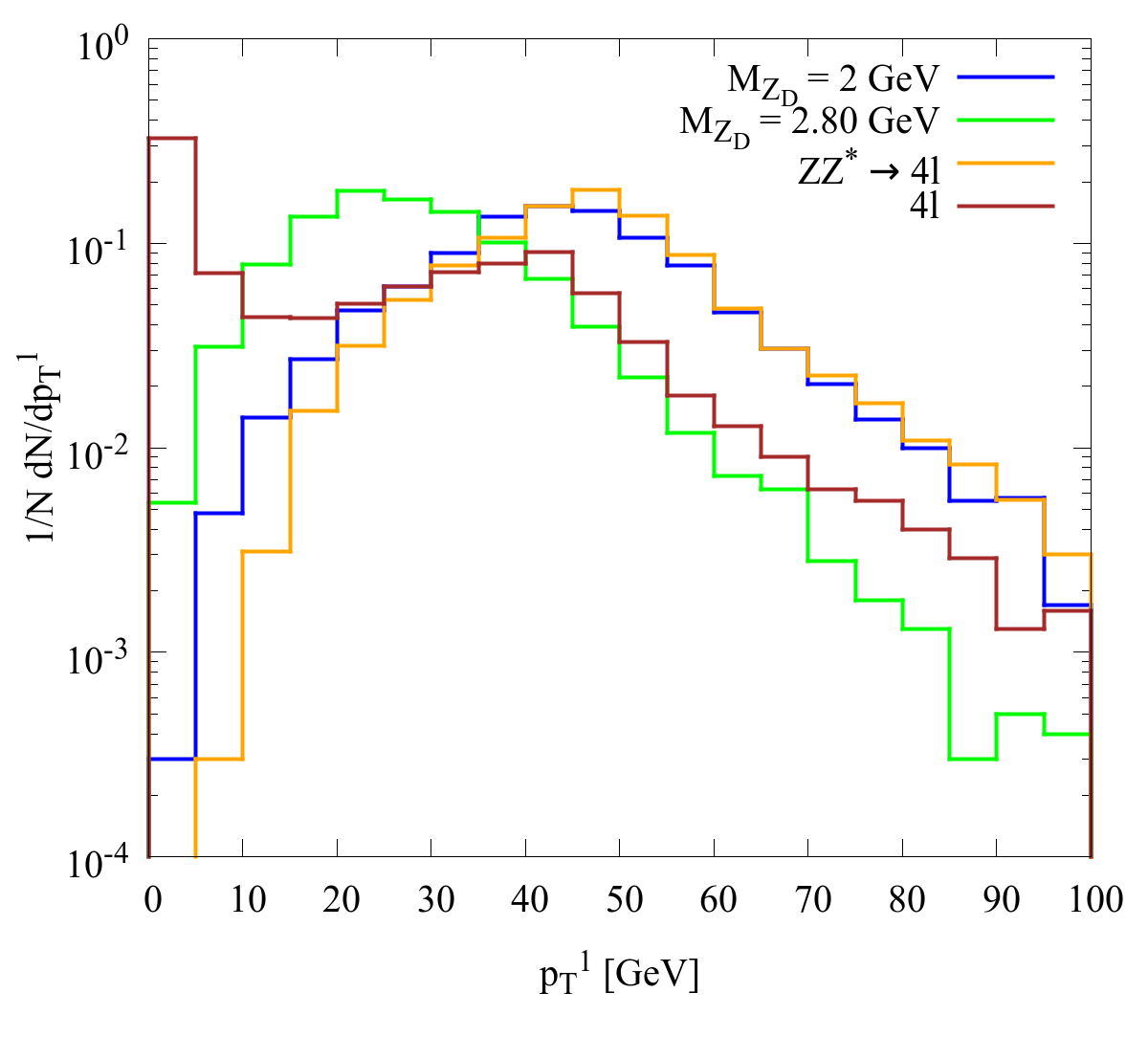}\includegraphics[scale=0.16]{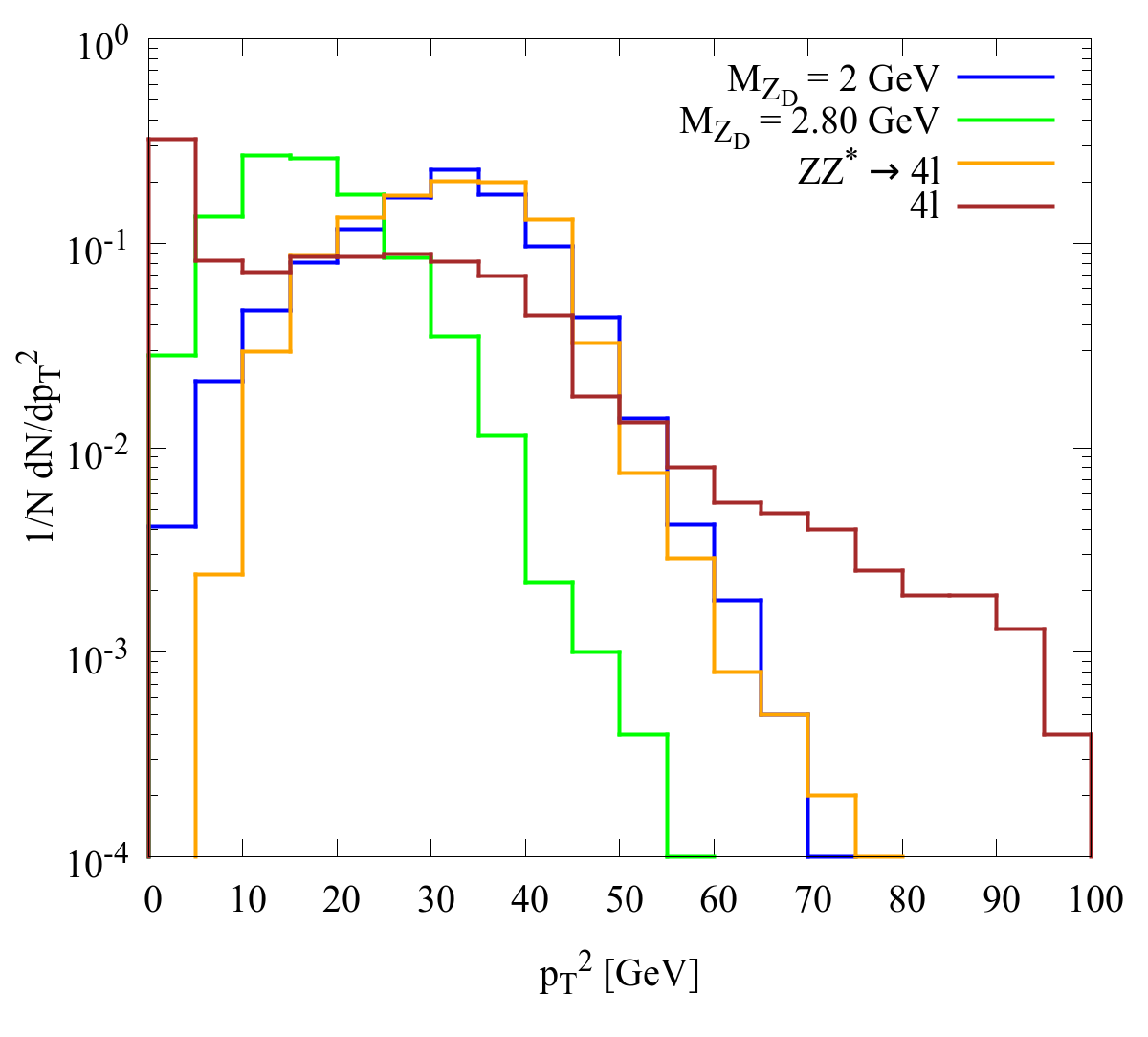}\\
\includegraphics[scale=0.16]{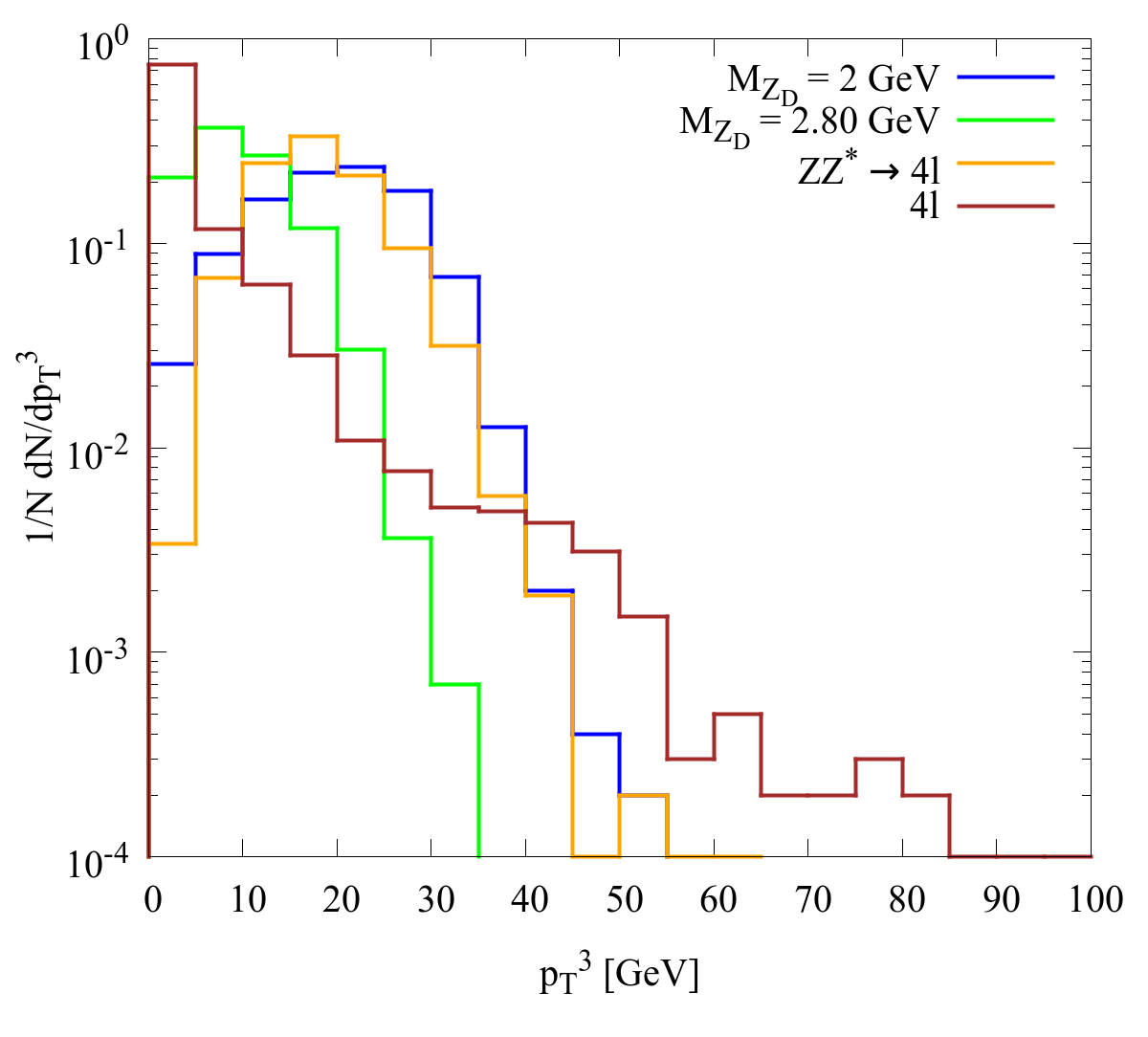}\includegraphics[scale=0.16]{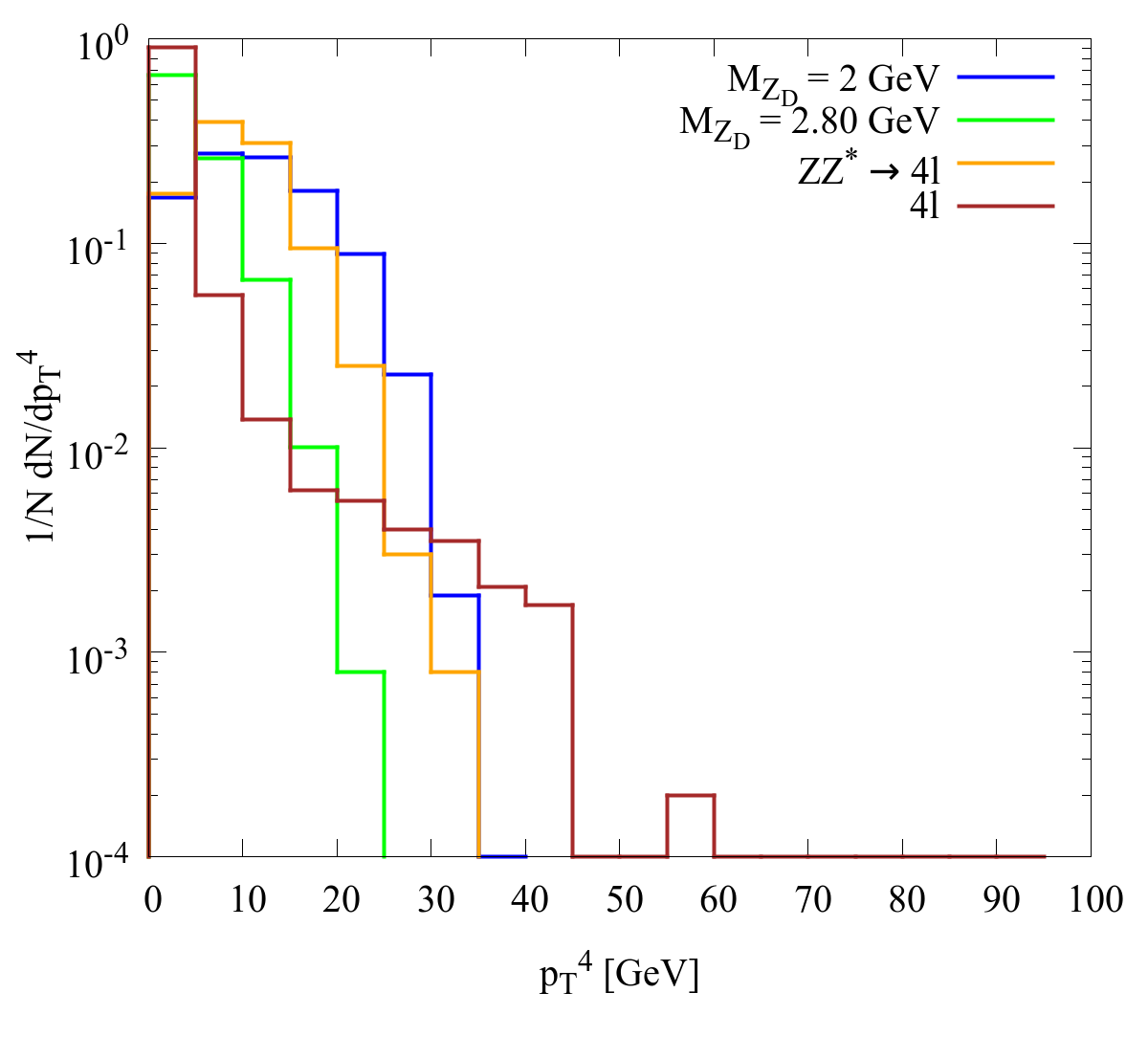}
\caption{Transverse momentum ($p_{T}$) distribution of the final state muons at the partonic level. The blue and green colored lines correspond to the signal event generated with different $Z_{D}$ masses, $M_{Z_{D}}=2,~$ and $2.80~{\rm GeV}$, respectively. The orange and the brown colored line represents the $p_{T}$ distribution of the $ZZ^{*} \to 4l$ and electroweak $4l$ background.}
\label{fig:future_coll_pt1_met}
\end{center}  
\end{figure}


\begin{figure}
\begin{center}
\includegraphics[scale=0.17]{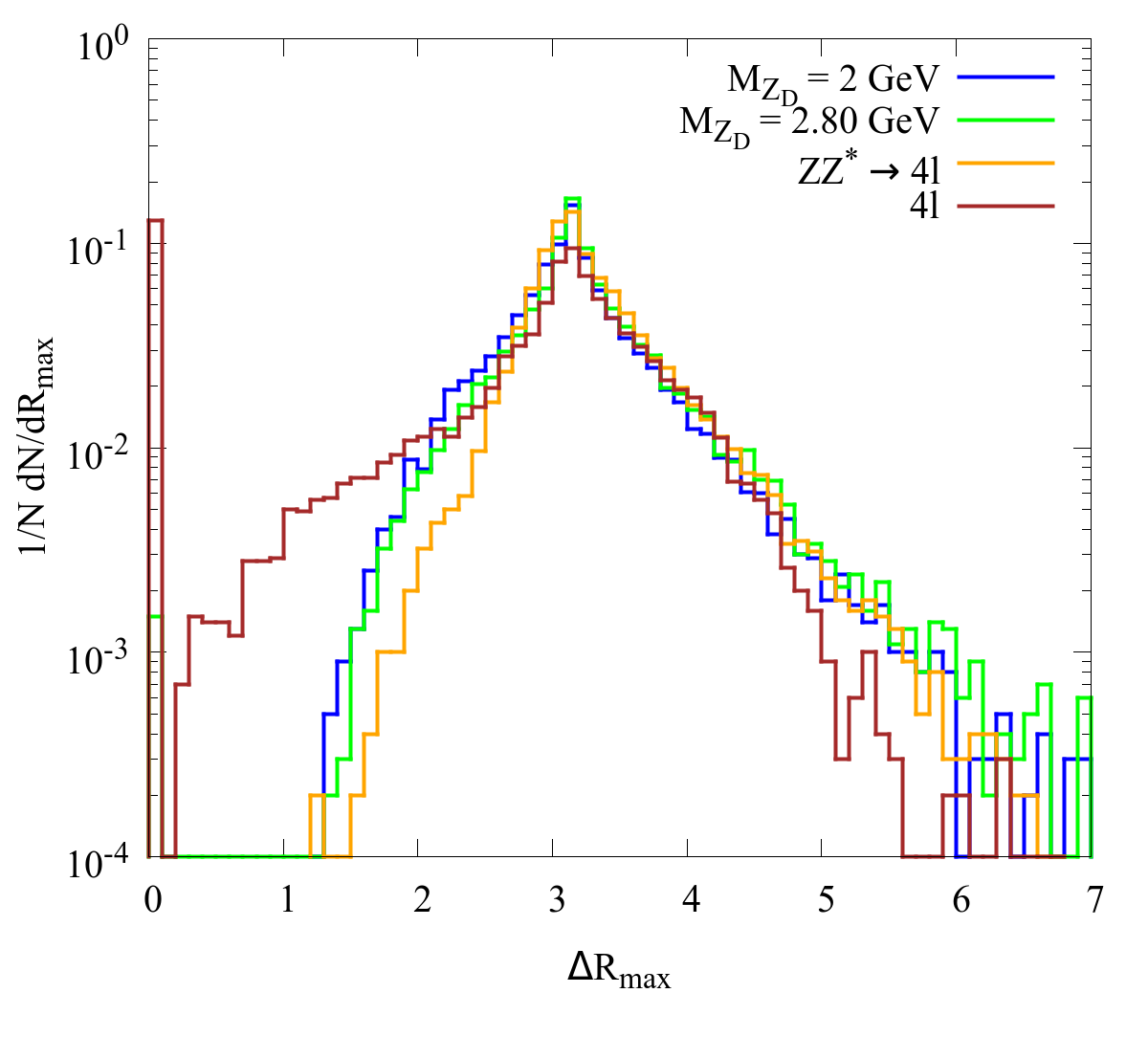}\includegraphics[scale=0.17]{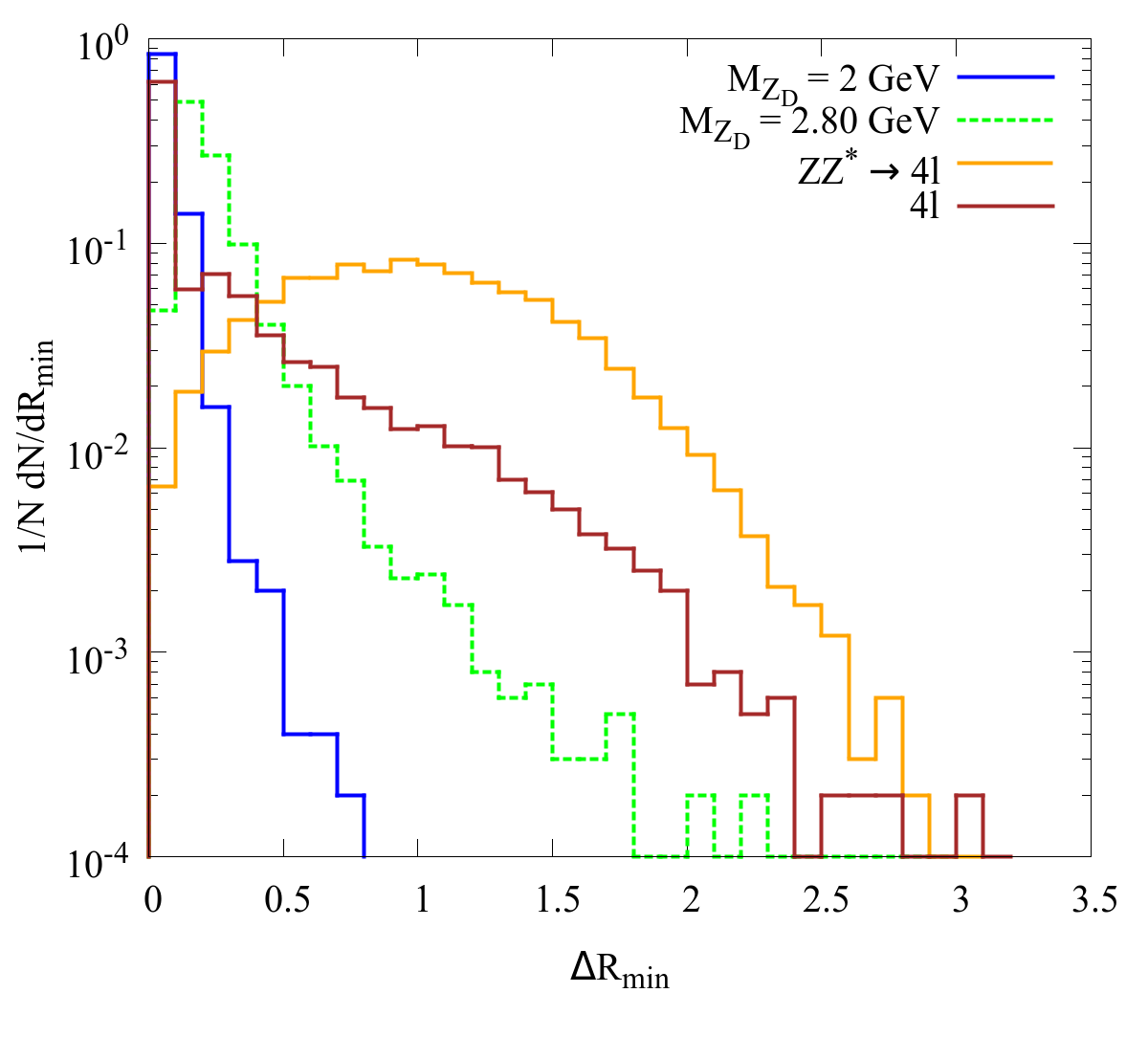}
\caption{Normalised distribution showing $\Delta{R_{max}}$ and $\Delta{R_{min}}$. The color codes are the same as that of Fig.~\ref{fig:future_coll_pt1_met}.}
\label{fig:future_coll_delr1_met}
\end{center}  
\end{figure}

The muon isolation and selection criteria specified in Sec.~\ref{sec:coll_zdzd} is applied here as well, and an event is required to have exactly four isolated muons in the final state. We show the $p_{T}$ distribution of the final state muons, corresponding to the relevant backgrounds (same color code as Fig.~\ref{fig:future_coll_pt1}) and two benchmark signal events ($M_{Z_{D}} = 2.0~{\rm GeV}$ and $M_{Z_{D}}=2.8~{\rm GeV}$), in Fig.~\ref{fig:future_coll_pt1_met}, where, $p_{T}^{1}$ represents the highest $p_{T}$ muon and $p_{T}^{4}$ represents the lowest $p_{T}$ muon. When compared to the case of Sec.~\ref{sec:coll_zdzd}, the $p_{T}$ distribution of the signal events peak at a relatively smaller value on account of a residual $\met$. Consequently, we apply slightly weaker trigger cuts on the muon $p_{T}$. The muon with the highest $p_{T}$ is required to have $p_{T} > 15~{\rm GeV}$, while the second (third) leading muon is required to have $p_{T} > 10~{\rm GeV}(5~{\rm GeV})$. Here again, the angular variables, $\Delta R$, provides an additional control in improving upon the signal significance. In this context, we show the $\Delta R_{max}$ and $\Delta R_{min}$ distributions in Fig.~\ref{fig:future_coll_delr1_met}, following the color code of Fig.~\ref{fig:future_coll_pt1_met}. $\Delta R_{max}$ ($\Delta R_{min}$) represents the maximum (minimum) value of $\Delta R$ between all possible muon pairs. It can be observed from Fig.~\ref{fig:future_coll_delr1_met} that the $\Delta R_{max}$ distribution for the $4l$  background falls off earlier than the signal samples. Consequently, we impose $\Delta R_{max} > 2.0$. In addition, a lower limit on $\Delta R_{min} > 0.5$ is also imposed. The construction of SFOS pair is done following the strategy specified in Sec.~\ref{sec:coll_zdzd}. An event is required to have exactly two SFOS pairs, with the invariant mass of the leading and sub-leading SFOS pairs required to be within the range $0.88~{\rm GeV} < M_{12},M_{34} < 20~{\rm GeV}$. The ratio of invariant masses of the sub-leading and leading SFOS pair is required to be within $\dfrac{M_{34}}{M_{12}} > 0.85$. The $J/\Psi$ and $\Upsilon$ resonance vetoes have also been applied in this analysis. A summary of selection cuts is presented in Table.~\ref{tab:sel_cut_zdzd4mumet}. The HL-LHC upper limits on $\sigma(gg \to H_{2}) \times Br(H_{2} \to \chi_{+}\chi_{+} \to \chi_{-}Z_{D}\chi_{-}Z_{D})$ corresponding to a signal significance of $5\sigma$ and $2\sigma$ are shown in Table.~\ref{tab:zdzd_met_signi}. For the sake of comparison, the upper limits on $\sigma(gg \to H_{2} \to Z_{D}Z_{D})$, derived in Sec.~\ref{sec:coll_zdzd}, are also shown in the same table.

\begin{table}
\begin{center}
\begin{tabular}{C{16cm}}
\hline \hline 
Selection cuts \\ 
\hline 
(a). Exactly $4$ muons in final state. \\
(b). $p_{T}^{\mu_{1}} > 15~{\rm GeV}$, $p_{T}^{\mu_{2}} > 10~{\rm GeV} $ \\
 $p_{T}^{\mu_{3}} > 5~{\rm GeV}$, $p_{T}^{\mu_{4}} > 2.6~{\rm GeV}$  \\
$|\eta| < 4$ \\
(c). $\Delta R_{max} > 2.0 $ and $\Delta R_{min} > 0.5 $\\
(d). $0.88~{\rm GeV} < M_{12}^{inv},M_{34}^{inv} < 20~{\rm GeV}$\\
(e). Event veto if : $(M_{J/\Psi} - 0.25~{\rm GeV}) < M_{12,34} < (M_{\Psi(2s)} + 0.30~{\rm GeV})$ \\
                 $(M_{\Upsilon} - 0.70~{\rm GeV}) < M_{12,34} < (M_{\Upsilon(3s)} + 0.75~{\rm GeV})$ \\
(f). $M_{34}^{inv}/M_{12}^{inv} > 0.85$ \\ 
\hline 
\end{tabular}
\caption{Selection cuts for the cut-based analysis in the $4\mu + \met$ final state.}
\label{tab:sel_cut_zdzd4mumet}
\end{center}
\end{table}

\begin{table}
\begin{center}
\begin{tabular}{||C{2.0cm}||C{2.0cm}|C{2.0cm}||C{2.0cm}|C{2.0cm}||} \hline\hline
\multirow{2}{*}{$M_{Z_{D}}$ [GeV]} & \multicolumn{2}{c|}{$H_{2} \to \chi_{+}\chi_{+} \to 4\mu + \met $} & \multicolumn{2}{c|}{$H_{2} \to Z_{D}Z_{D} \to 4\mu$} \\\cline{2-5}
 & $5\sigma$ [fb] & $2\sigma$ [fb] & $5\sigma$ [fb] & $2\sigma$ [fb] \\ \hline \hline
$2.40$ & $1.03$ & $0.41$ & $0.181$ & $0.072$ \\\hline
\end{tabular}
\caption{Upper limits corresponding to $5\sigma$ and $2\sigma$ signal significances on $\sigma(gg \to H_{2}) \times Br(H_{2} \to \chi_{+}\chi_{+} \to \chi_{-}Z_{D}\chi_{-}Z_{D})$ for HL-LHC. The upper limits derived in Sec.~\ref{sec:coll_zdzd} for $\sigma(gg \to H_{2} \to Z_{D}Z_{D})$ are also listed.}
\label{tab:zdzd_met_signi}
\end{center}
\end{table}

The case of $H_{2} \to Z_{D}Z_{D}$ (discussed in Sec.~\ref{sec:coll_zdzd}) , furnishes stronger limits as compared to the current case, and the reason can be attributed to the possibility of invariant mass reconstruction of $H_{2}$ ($M_{4\mu}$) in the previous case. We would like to note that in the case of $H_{2} \to Z_{D}Z_{D}$, we had imposed a selection cut on $M_{4\mu}$ and had restricted it within $120-130~{\rm GeV}$, which was extremely efficient in filtering out the background.

\subsection{The curious case of a late decaying $Z_{D}$ boson at $14~{\rm TeV}$ and $27~{\rm TeV}$ high luminosity LHC}
\label{subsec:LLP_zd}

In the recent times, non-traditional search methodologies for beyond Standard Models (BSM) have started garnering popularity. Search for the late decaying long-lived particles (LLP) which feature a characteristic secondary vertex has emerged as one such avenue. In general, a particle is said to be long-lived if its proper decay length exceeds $c\tau > 10^{-4}~m$. $Z_{D}$ is one such viable LLP candidate within the framework of the $U(1)_{D}$ model considered in this analysis. The proper decay length of $Z_{D}$ has an inverse square dependence on the kinetic mixing factor $\epsilon_g$, and can become a potential LLP candidate for smaller values of $\epsilon_g$. For example, at $M_{Z_{D}}\sim 2~{\rm GeV}$, a value of $\epsilon_g \gtrsim (10^{-5}-10^{-6})$ will result in the $Z_{D}$ to become a LLP (see Fig.~2 of \cite{Curtin:2014cca}). 

Before proceeding ahead with the details of the analysis, we lay down a brief description of the existing segmentation of the ATLAS detector. The ATLAS detector can be broadly classified into four different segments, viz., the tracker region, the electromagnetic calorimeter (ECAL), the hadronic calorimeter (HCAL) and the muon spectrometer. The tracker region can be further subdivided into three major segments. The first segment involves the Pixel detector, which contains three sub-layers at a radii of $33.25~{\rm mm}$, $50.5~{\rm mm}$, and $88.5~{\rm mm}$. The next segment within the tracker is the SCT which can be further classified into three segments with radii $299~{\rm mm}$, $371~{\rm mm}$, and $443~{\rm mm}$. The final segment in the tracker is the TRT which spans the radii $554~{\rm mm}-1082~{\rm mm}$. The ECAL segment lies roughly in between a radii of $\sim 1300~{\rm mm}-2100~{\rm mm}$, while the HCAL scans across a radii of $\sim 2285~{\rm mm} - 3815~{\rm mm}$. The final segment of the ATLAS detector, the muon spectrometer, extends from a radii of $\sim 4100~{\rm mm}$ all the way up to $10000~{\rm mm}$. 

\begin{figure}
\begin{center}
\includegraphics[scale=0.26]{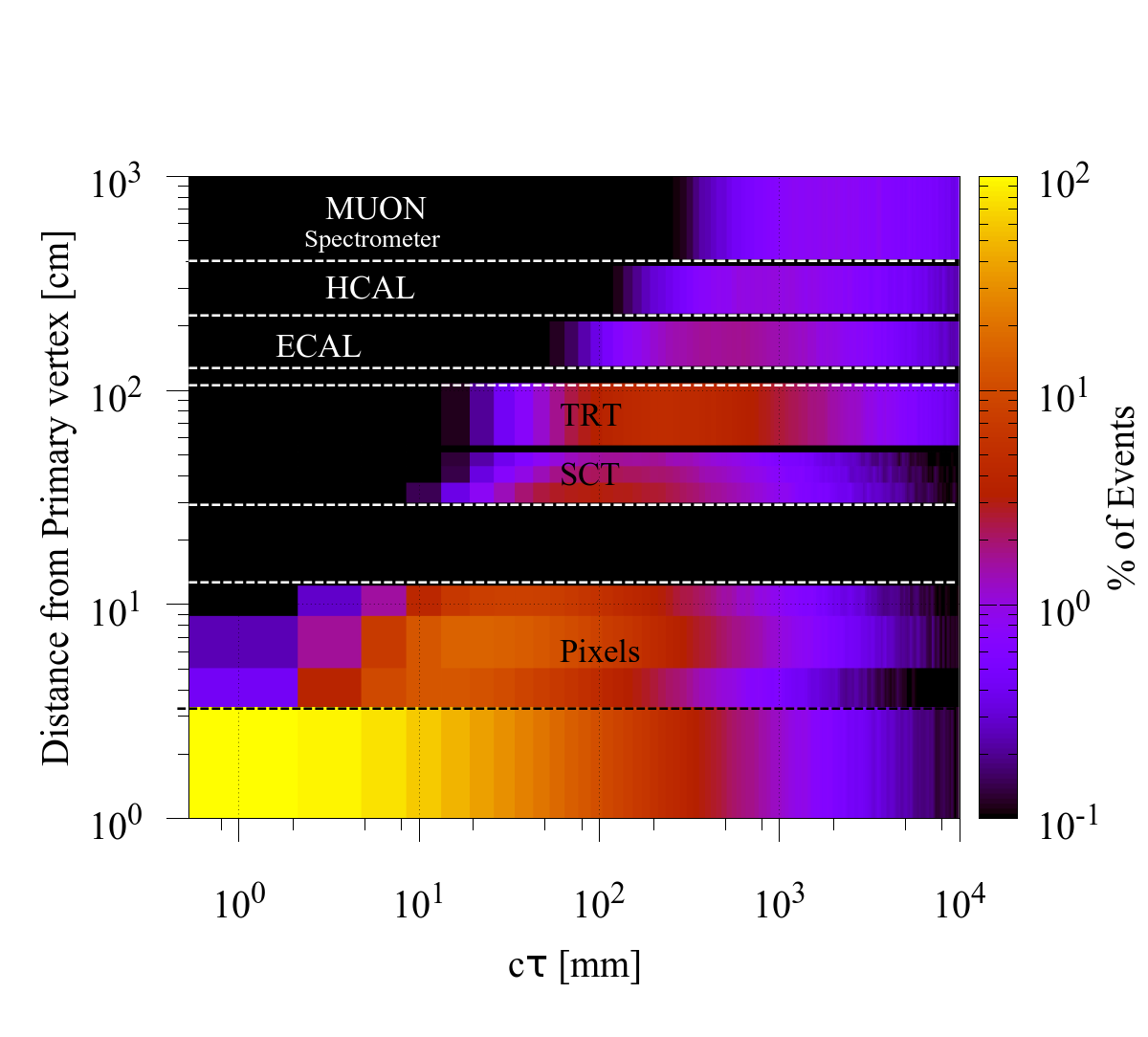}
\caption{The vertical and horizontal axes represent the radii from the beam axis and proper decay length, respectively, of the LLP. The color palette corresponds to the percentage of events which will undergo decay in the corresponding segments, for the case of $14~{\rm TeV}$ LHC with an integrated luminosity of $3000~{\rm fb^{-1}}$. The segmentation along the vertical axis corresponds to the existing geometry of the ATLAS detector.}
\label{fig:LLP_14}
\end{center}
\end{figure}

The LLP can undergo decay in different segments of the LHC, based on its kinematic distribution and its decay length. The decay length is given by $l_{d}=\beta\gamma c\tau$, where, $\beta$ is the boost of the particle and is defined as the ratio of its velocity ($v$) to the speed of light ($c$): $\beta = v/c$, whereas, $c\tau$ is the proper lifetime of the particle in its rest frame, and $\gamma$ is the relativistic factor defined as : $\gamma = \dfrac{1}{\sqrt{1-\beta^{2}}}$. It can be inferred from the form of the decay length that identification of the secondary vertex within the detector (which corresponds to the point of decay of the LLP) and measurement of the boost factor, can also be used to estimate the proper lifetime of the LLP. Within a typical detector, if $N_{0}$ be the number of long-lived particles produced with a proper life time $\tau_{i}$ and a mean life-time of $\tau$, then the exponential distribution gives the total number of decay events, $N = N_{0}e^{-\tau_{i}/\tau}$. Consequently, the fraction of long-lived particles undergoing decay in different segments of the detector will be a characteristic reflection of the proper decay length. 

\begin{figure}
\begin{center}
\includegraphics[scale=0.26]{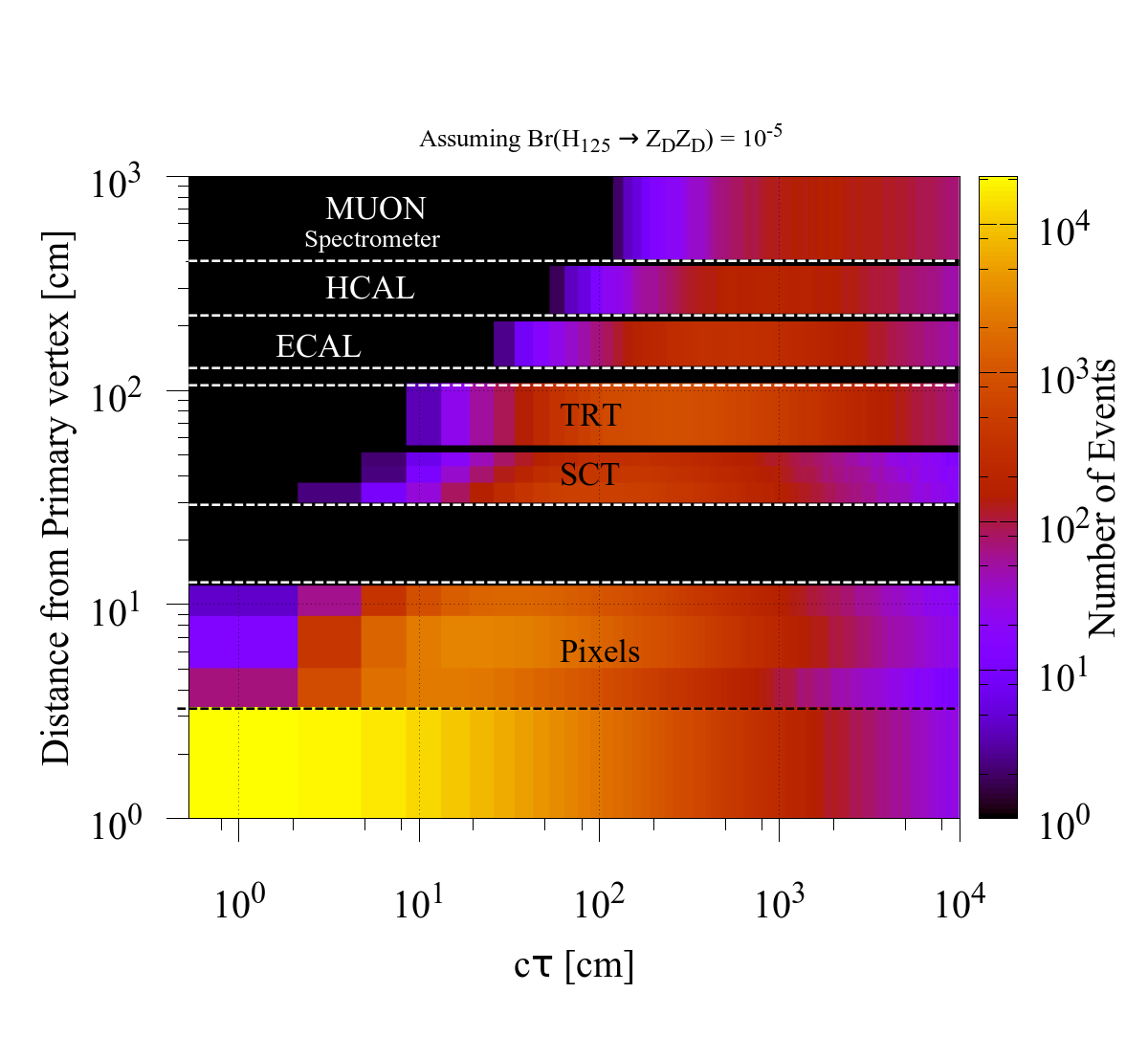}
\caption{The vertical and horizontal axes represent the radii from the beam axis and proper decay length, respectively, of the LLP. The color palette corresponds to the total  of events which will undergo decay in the corresponding segments, for the case of $27~{\rm TeV}$ LHC at an integrated luminosity of $15~{\rm ab^{-1}}$. Here, we assume $Br(H_{2} \to Z_{D}Z_{D})=10^{-5}$. The segmentation along the vertical axis corresponds to the existing geometry of the ATLAS detector.}
\label{fig:LLP_27}
\end{center}
\end{figure}

To visualize this effect, we used \texttt{Pythia-6} to simulate the signal $pp \to H_{2} \to Z_{D}Z_{D}$ with $Z_{D}$ being a LLP. Event samples were generated by varying the decay length of $Z_{D}$ over the range from $0.5~{\rm mm}$ to $10000~{\rm mm}$, and $20,000$ events were generated for specific choice of decay length. We show the proper decay length (in mm) along the horizontal axis and the radii (in cm) along the vertical axis in Fig.~\ref{fig:LLP_14} (following the ATLAS detector geometry), with the color palettes being representative of the percentage of events decaying within the concerned segment. The black colored regions correspond to the void regions within the detector and also represents those segments which do not register an event decay. The event samples for Fig.~\ref{fig:LLP_14} have been simulated for the case of $14~{\rm TeV}$ LHC. It can be observed that particles with proper decay length up to $\sim 10~{\rm cm}$ mostly decay before reaching the Pixel detector. Particles with large proper decay length can be observed to have a fairly uniform probability of decaying throughout the detectors. We would like to note that, in order to correctly simulate the effect of a smeared distribution, we generated events with proper decay lengths up to $20000~{\rm mm}$ and then truncated the horizontal range in Fig.~\ref{fig:LLP_14} at $10000~{\rm mm}$. Observation of the LLP at different segments of the detector can be instrumental in motivating the design of segment-specific search analyses. Reconstruction of the same type of particle at the tracker and the ECAL involves vastly different strategies, and different search analyses would be required at these two different segments. Under such circumstances, having a generic idea about the probability of decay within a certain segment of the detector can prove instrumental in performing more specific and focused searches.

\begin{table}
\begin{center}
\begin{tabular}{||C{1.6cm}|C{1.3cm}|C{1.3cm}||C{1.3cm}|C{1.3cm}||C{1.3cm}|C{1.3cm}|} \hline\hline
\multirow{3}{*}{Segment} & \multicolumn{6}{c|}{Percentage of events undergoing decay} \\ \cline{2-7}
 & \multicolumn{2}{c|}{$c\tau~=13.2~{\rm mm}$} & \multicolumn{2}{c|}{$c\tau~=305.2~{\rm mm}$} & \multicolumn{2}{c|}{$c\tau~=1025~{\rm mm}$} \\ \cline{2-7}
 & $ggF$ & $ZH_{2}$ & $ggF$ & $ZH_{2}$ & $ggF$ & $ZH_{2}$ \\ \hline  
 Prompt & $52.84$ & $48.70$ & $3.720$ & $3.325$ & $1.127$ & $1.010$ \\
 Pixel & $32.84$ & $31.18$ & $7.117$ & $6.320$ & $2.275$ & $2.230$\\
 SCT & $0.347$ & $0.685$ & $4.908$ & $4.320$ & $2.050$ & $1.770$\\
 TRT & $0.020$ & $0.085$ & $4.452$ & $3.795$ & $2.523$ & $2.090$ \\
 ECAL & $0.002$ & $0.000$ & $1.442$  & $1.215$ & $1.290$ & $1.030$ \\
 HCAL & $0.000$ & $0.000$ & $0.485$  & $0.560$ & $0.942$ & $0.755$ \\
 $\mu$ spec. & $0.000$ & $0.000$ & $0.117$ & $0.194$ & $0.667$ & $0.580$ \\ \hline 
\end{tabular}
\caption{Percentage of events undergoing decay in different segments of the detector corresponding to the $ggF$ and $ZH_{2}$ production modes of $H_{2}$, for the case of $14$ TeV LHC. The difference is the manifestation of different kinematics for the two cases.}
\label{tab:LLP}
\end{center}
\end{table}

We perform a similar study for the case of a $27~{\rm TeV}$ LHC machine as well, assuming the current geometry of ATLAS detector, and the corresponding results are represented in Fig.~\ref{fig:LLP_27}. Here, we show the total number of $gg \to H_{2} \to Br(H_{2} \to Z_{D}Z_{D})$ events, assuming $\sigma(gg \to H_{2})=140~{\rm pb}$, $Br(H_{2} \to Z_{D}Z_{D})=10^{-5}$, and an integrated luminosity of $15~ab^{-1}$, in the color palette. The horizontal and vertical axes in Fig.~\ref{fig:LLP_27} correspond to the similar quantities in Fig.~\ref{fig:LLP_14}. A higher boost in the case of $27~{\rm TeV}$ collision results in an upward shift of color pattern in Fig.~\ref{fig:LLP_27} as compared to the previous case.

The preceding discussion assumed the Higgs production in the $ggF$ channel. The production of Higgs in other production channels will result in alteration of the segment-wise decay fraction, owing to different boosts in the transverse direction. In this respect, we briefly explore the case of $ZH_{2}$ production, and the LLP being produced from decay of $H_{2}$, in the context of $14$ TeV LHC. We present a comparison between the fraction of events undergoing decay within various segments of the detector, for the case of $H_{2}$ being produced via $ggF$ mode and in $ZH_{2}$ mode, in Table.~\ref{tab:LLP}. 

\section{Summary and Conclusion}
\label{sec:conclusion}
In this work, we have explored an $U(1)_D$-gauge extension of Standard Model from a phenomenological perspective. Our model has two Majorana fermions to render the theory anomaly free. The lightest of them serves as a dark matter candidate. As generic to any gauge theory, our model has an extra neutral Z like boson ($Z_D$) and since we have also agreed to employ Higgs mechanism to make the particles massive, there also exists an extra neutral scalar ($H_1$) besides the usual Standard Model Higgs as well. Motivation of studying strong dark matter self interactions led us to restrict ourselves only to light scalar mediators ($10$ MeV $\lesssim M_{H_1} \lesssim 10$ GeV). Here we have studied the specific case where Sommerfeld enhancement is mediated only through one light mediator ($H_1$). In general, both the $Z_D$ and $H_1$ could have been light and led to large self interactions, but very light $Z_D$ would have been devoid of any collider signatures. 

For simplicity, throughout this work we have kept the kinetic mixing between $Z$ and $Z_D$ to small values. However it plays a role in the study of collider signatures arising from the prompt decay of $Z_D$. The primary goal of this work was to investigate if there exists a substantial parameter space for a thermal dark matter with large self-interactions. The latter is desirable since it solves some small scale structure issues as already mentioned earlier. The range of $M_{H_1}$ was hence motivated by the requirement of large self-interaction cross sections. We have systematically checked constraints on scalar mixing angle (as a function of $M_{H_1}$) arising from  LEP data, B-factories and beam dump experiments. We found that LEP data rules out $\sin\theta_{\rm mix} \gtrsim 0.2$ for the range of $M_{H_1}$ considered in this work. Higgs signal strength measurements on the other hand constrained $\sin\theta_{\rm mix}$ to $\lesssim 0.1$ in our parameter space of interest. Beam dump and flavour physics experiments put a tighter bound of $\sin\theta_{\rm mix} \lesssim 10^{-4}$ on almost the whole of our parameter space. LHC analyses mainly dealt with Standard Model Higgs decaying to 4$l$ channels and provided us with a probe to the otherwise small gauge kinetic mixing ($\epsilon_g$).

From the point of view of dark matter phenomenology we however have the most stringent constraints on the scalar mixing angle (for a given Yukawa coupling $f$). The Yukawa coupling $f$ is the sole controlling parameter that gives rise to sizable self interactions while direct detection experiments tend to put limits on $f\sin\theta_{\rm mix}$. Due to the presence of light mediators ($H_1$), the standard calculation of DM-nucleon cross section had to be refined using momentum dependent propagators. On the other hand, we have shown that condition of thermalization of the dark and visible sectors sets a lower limit to $\lambda_{\rm mix}$. This competes with the upper bound on $\lambda_{\rm mix}$ that arises from the direct detection experiments (for a given $v_D$ and a suitable $f$ which will give rise to sizable self interactions). Our findings suggest that such points on the parameter space satisfying both the limits are very rare and automatically pushes us to low dark matter masses ($\sim \mathcal{O}(1)$) GeV where constraints from direct detection is considerably weaker. Alternatively, we can also try to do away with the usual {\it thermal} dark matter scenario and probe into other exotic mechanisms like dark freeze-out \cite{Bernal:2015ova} and freeze-in \cite{Hall:2009bx,Biswas:2016bfo,Biswas:2016iyh} to help us get the correct relic density. $\lambda_{\rm mix}$ will have no lower limit in these cases and hence a large portion of the parameter space can be recovered. 

As discussed earlier, we found that it is very difficult to probe the MeV scale dark $H_1$ in our model via future collider experiments. Our model however also predicts an extra dark gauge boson $Z_D$ which may be probed at future HL-LHC. We have presented the reach of $Z_D$ from $H_2$ decay via 4$\mu$ final state and obtained projected upper limits on $\sigma_{H_2} \times Br(H_2 \rightarrow Z_D Z_D)$ for HL-LHC. We have also looked for the scenario where $H_2$ decays into a pair of $\chi_+$ ($\chi_+ \rightarrow Z_D \chi_-$) and finally results into 4$\mu +$ MET signal 
and found that this channel gives weaker limit compared to $H_2 \rightarrow Z_D Z_D \rightarrow 4\mu$ channel. For very small values of $\epsilon_g$, it is observed that $Z_D$ becomes a long lived particle (LLP) and different search strategies are required for the LLP scenarios. This non-traditional prospects of LLP ($Z_D$) at 14 TeV and 27 TeV high luminosity LHC have also been discussed.

Thus, solely from a data driven perspective, in a general and minimal $U(1)$-gauge extension of Standard Model we were able to restrict the presence of a {\it heavy} {\it thermal} dark matter candidate and this motivates (quantitatively) future explorations along these uncharted avenues.

\section*{Acknowledgements}
The work of BB is supported
by the Department of Science and Technology (DST), Government of India, under the Grant Agreement number IFA13-PH-75 (INSPIRE Faculty Award). AC acknowledges support from DST, India, under grant number IFA15-PH-130 (INSPIRE Faculty Award).

\newpage

\bibliographystyle{JHEP}
\bibliography{bibdm}
\end{document}